\documentclass[a4paper,fleqn]{cas-dc}
\usepackage[numbers]{natbib}

\usepackage{float}
\usepackage{pgfplots}
\pgfplotsset{compat=1.18}
\usepackage{placeins}

\usepackage{tikz}
\usetikzlibrary{positioning, calc, shapes.geometric, decorations.pathreplacing, shadows}

\def\tsc#1{\csdef{#1}{\textsc{\lowercase{#1}}\xspace}}
\tsc{WGM}
\tsc{QE}
\tsc{EP}
\tsc{PMS}
\tsc{BEC}
\tsc{DE}

\begin{document}
\let\WriteBookmarks\relax

\shorttitle{Energy-Aware Computing in 2026}

\shortauthors{R.N. Tchakoute and C. Tadonki}

\title [mode = title]{Energy-Aware Computing in the Year 2026}                      



%
\author[1]{Roblex Nana Tchakoute}[type=editor,orcid=0009-0009-8636-8466]

\cormark[1]

\fnmark[1]




\affiliation[1]{organization={Mines Paris - PSL University\\Centre de Recherche en Informatique},
    addressline={35 rue Saint-Honoré}, 
    city={Fontainebleau},
    postcode={77300}, 
    country={France}}

\author[1]{Claude Tadonki}

\cortext[cor1]{Corresponding author}
\cortext[cor2]{Principal corresponding author}



\begin{abstract}
High-Performance Computing (HPC) has recently entered the Exascale era, and considerable efforts are underway to fully harness this power for large-scale applications, such as cutting-edge generative AI (training and exploitation). The corresponding energy consumption is very high, and forecasts are alarming, making this metric a critical systemic bottleneck. Addressing this issue presents a genuine challenge for the entire cloud-edge-HPC continuum at all scales, from low-power IoT microcontrollers to multi-megawatt data centers. Beyond financial costs, green computing is driven by considerations related to climate change and environmental concerns such as carbon footprint ($CO_2e$), as well as constraints on energy production and supply, leading to a real need to regulate {\em information and communication technology} (ICT) activities. This article presents a comprehensive overview of energy-efficient computing, incorporating the most recent and significant contributions. Based on this exploration of the state of the art, we design and describe a holistic taxonomy of the aforementioned publications, structured around various perspectives, including {\em hardware and software aspects, measurement instrumentation, software optimizations, dynamic task scheduling, voltage scaling, workload consolidation, federated learning}, and {\em cooling}. Particular emphasis is placed on large-scale AI, which receives significant attention due to its considerable resource requirements. We conclude with an analysis of a forward-looking roadmap that considers the main perspectives of sustainable computing.
\end{abstract}



\begin{keywords}
Energy-Aware Computing \sep Cloud-Edge Continuum \sep High-Performance Computing (HPC) \sep Green AI \sep Carbon Footprint \sep Power Capping \sep Job Scheduling \sep Energy Efficiency
\end{keywords}

\maketitle

\section{Introduction}


The trajectory of computer architecture has long been shaped by two foundational principles: {\em Moore's Law}, which predicted the doubling of transistor density every two years, and {\em Dennard Scaling}, which ensured that total chip power remained roughly constant as transistors shrank. Together, these laws underpinned decades of increasing performance gains at manageable energy costs, which is no longer the case due to the fundamental {\em power wall} that now governs the design and operation of modern computing infrastructures. In addition to theoretical peak performance, an important focus is now on {\em energy production/supply, heat dissipation}, and {\em environmental sustainability} \cite{iea_datacenters_2025}. The Exascale era, currently led by landmark systems including El Capitan ($\approx$30 MW), Frontier ($\approx$25 MW), Aurora ($\approx$39 MW), JUPITER Booster ($\approx$16 MW) \cite{top500_2025}, has pushed facility power draws to unprecedented levels, making electricity and cooling costs the dominant factor in total cost of ownership (TCO). At this scale, even marginal improvements in energy efficiency translate directly into significant financial savings and a noticeable reduction in carbon emissions.

The emergence of Generative Artificial Intelligence (GenAI) and Large Language Models (LLMs) has significantly exacerbated the (projected) energy crisis \cite{meta_llama3_2024}. Training trillion-parameter models requires large-scale clusters that run continuously for several months. High-performance AI is currently powered by high-end accelerators with thermal design powers (TDPs) greater than 1000W per chip (e.g., NVIDIA Blackwell, AMD MI300X) \cite{nvidia_blackwell_2024, amd_mi300_2024}. Thus, mitigating the energy requirement and the carbon footprint associated with large-scale AI has become a top priority \cite{paris_agreement_2015}. 

Intelligent applications are increasingly migrating toward the "Edge" in order to reduce latency. This introduces a two-way energy context: {\em embedded systems} and the {\em Internet of Things} (IoT). Here, energy is a critical resource. Microcontrollers and embedded AI accelerators (e.g., Coral Dev Boards, Nvidia Jetson) must run inference tasks within milliwatt-scale power limits in order to cope with the limited energy budget, since they are battery-powered. Thus, modern computing ecosystems must be evaluated as a unified \textit{Cloud-Edge-HPC continuum}.

Historically, power efficiency was hardware-centric. However, achieving true sustainability requires a cross-stack approach that combines {\em hardware optimizations} with advanced  {\em system-level mechanisms} such as energy-aware job scheduling, dynamic resource allocation, and intelligent power capping \cite{Philipo2025}. Evaluating Sustainable AI inherently requires balancing its predictive capabilities with the ecological consequences of its carbon footprint \cite{LIN2024100989, Hasan2025}. Recent meta-synthesis studies have demonstrated a crucial need for sustainability-related investigations, and most qualitative surveys have focused on the suitability of all-inclusive approaches in which all energy-accountable hardware and software aspects are analyzed as a whole \cite{McGuire2023}. 

\begin{figure*}[htbp]
    \centering
    \includegraphics[width=1\linewidth]{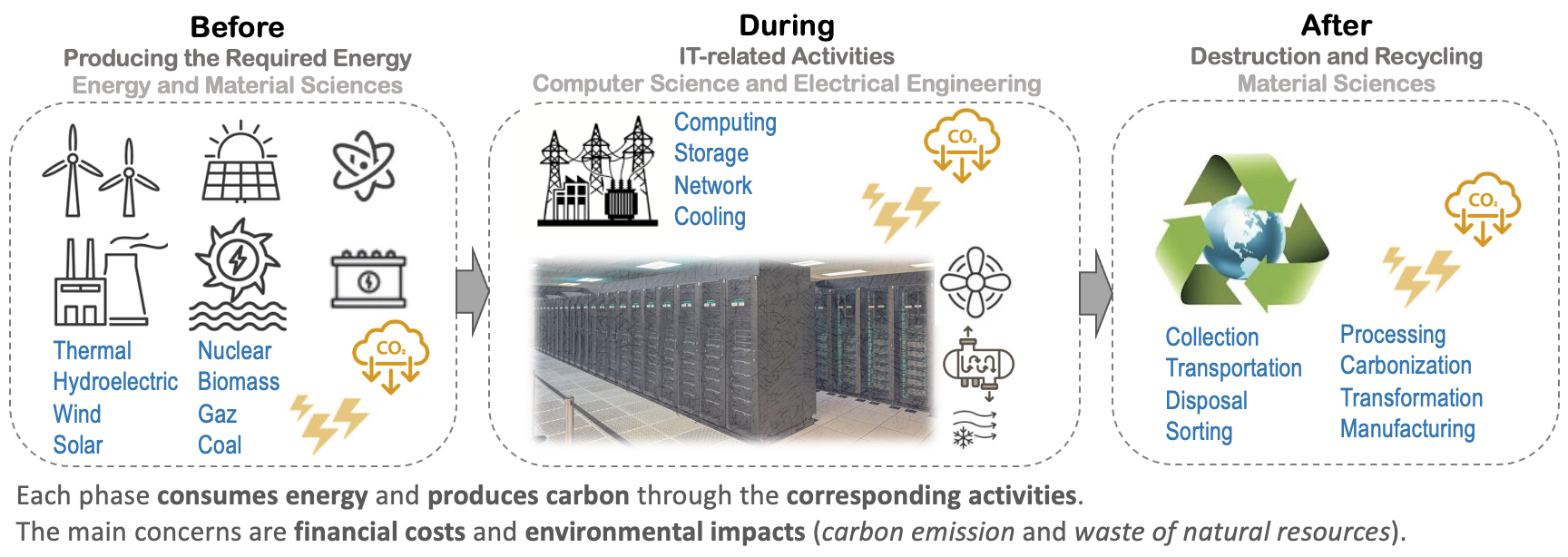}
    \caption{The whole cycle of energy consumption and related consequences.}
    \label{fig:energy_clycle}
\end{figure*}

Concerns related to the energy consumption of IT activities can/should be considered from a broader perspective, including the three main phases: {\em energy production}, {\em IT-related activities}, and {\em disposal/recycling}, as illustrated in Figure \ref{fig:energy_clycle}. Energy-efficient computing, in all its aspects and paradigms, is associated to the intermediate phase only, and this survey describes the landscape of related contributions.

This paper aims to provide a structured and commented view of the landscape of the major scientific and technical contributions within the state-of-the-art of power-aware computing in the \textit{Computing continuum}. We provide and articulate the following viewpoints:
\begin{itemize}
    \item \textbf{Comprehensive Taxonomy:} An up-to-date taxonomy of energy-related studies  spanning {\em edge microcontrollers}, {\em cloud computing infrastructures}, {\em AI accelerators}, and {\em Exascale supercomputers}.
    \item \textbf{Cross-Stack Profiling:} A comparative analysis of hardware and software energy profiling tools, distinguishing between in-band estimations and out-of-band measurements.
    \item \textbf{Code Optimization:} A review of static and dynamic energy optimization strategies, with an emphasis on {\em job scheduling}, {\em power capping}, and {\em workload consolidation}.
    \item \textbf{Sustainable AI:} An explicit focus on the carbon footprint of cutting-edge/large-scale AI through the review of frameworks for tracking and mitigation of energy-related metrics associated to LLM training and inference.
    \item \textbf{Cooling and conventional rules:} An evaluation of liquid cooling technologies and a prospective analysis of conventional rules and technical challenges related to sustainable computing.
\end{itemize}

\begin{figure*}[pos=ht]
    \centering
    \includegraphics[width=1\linewidth]{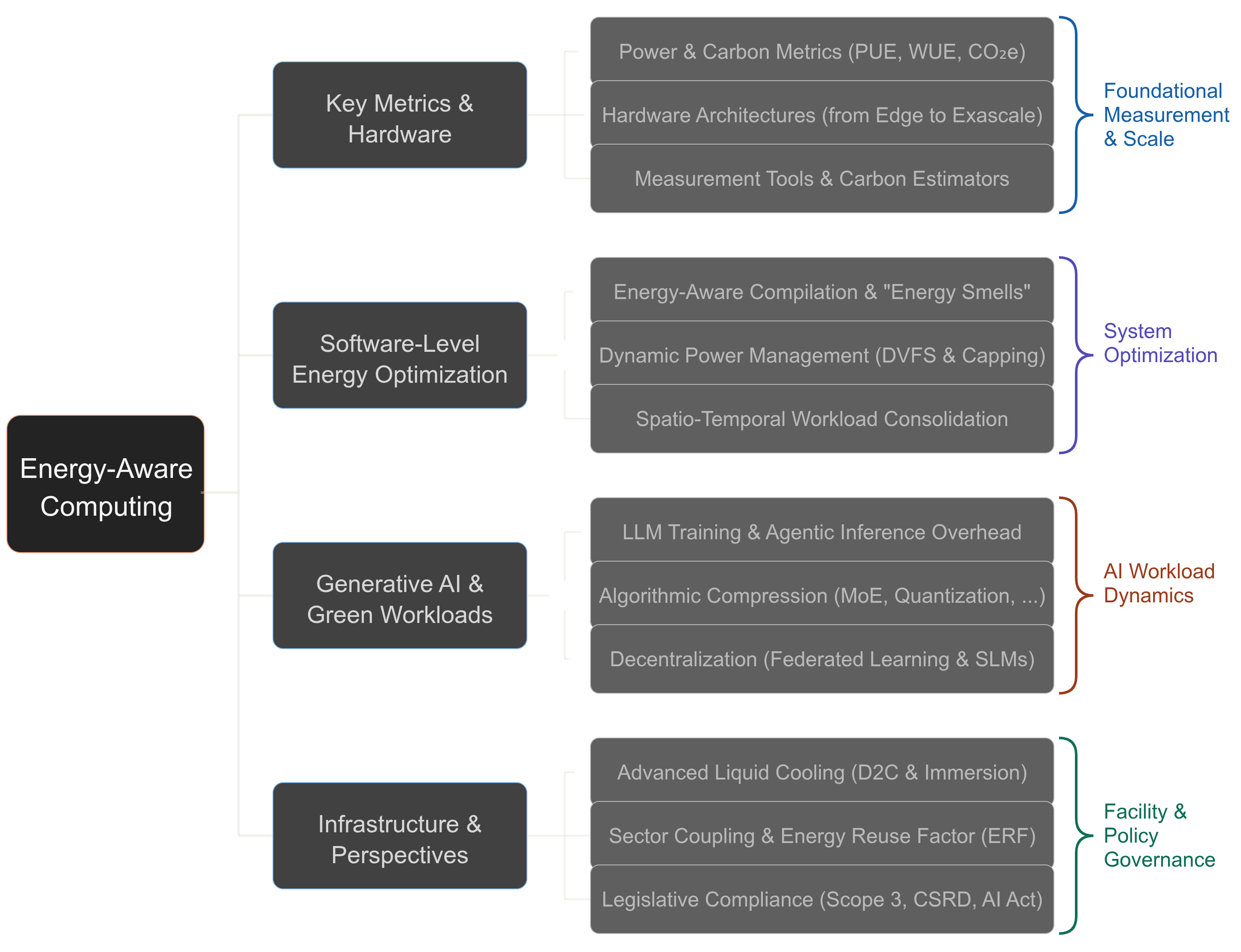}
    \caption{The landscape of energy-aware computing in the Cloud-Edge-HPC continuum.}
    \label{fig:survey_summary}
\end{figure*}

\section{Selection and Methodology}
\label{sec:methodology}

To ensure a rigorous and reproducible overview of the state of the art, we adopt a multi-stage hybrid Systematic Literature Review (SLR) methodology. Given the rapid evolution of AI hardware, relying solely on data from classical databases might lead to outdated configurations. Thus, we chose an objective filtering pipeline that covers the full Cloud-to-Edge continuum, integrating chronological bibliometric dynamics and high-impact industrial reports.

\subsection{Popularity-Driven Filtering and Selection}
Our starting pool is a set of highly cited papers that explicitly focus on {\em sustainable hpc-cloud-Edge computing, cutting-edge AI workloads, power capping}, and {\em hardware telemetry}. Among these contributions, we selected a corpus of roughly 250 pivotal references, prioritizing the most recent ones (preferably post-2023).

We applied the following two rules:
\begin{itemize}
    \item \textbf{Journal/Conference Impact:} We consider mainly Q1 journal papers (via SCImago/JCR) and CORE A/A* conferences (e.g., SC, ISCA, MICRO, ASPLOS, EuroSys), the ranking being considered at the publication date.
    \item \textbf{Empirical Validation Rule:} Regarding works on theoretical energy models or optimization frameworks, we have considered those that provide empirical evaluation on real devices or validated simulators.
\end{itemize}

\subsection{Integration of High-Impact Grey Literature}
In AI and large-scale HPC, the state-of-the-art is usually established by practitioners before being considered (in a more formal way ) in academic journals. Following this observation, we selected a set of non-academic literature:
\begin{itemize}
    \item \textbf{Industrial Reports:} Technical documentation, architectural whitepapers, and carbon footprint reports released by leading vendors (e.g., NVIDIA's Blackwell specifications \cite{nvidia_blackwell_2024}, Google's TPU architectures \cite{google_tpu_trillium_2024}, and Meta's open-source LLM reports \cite{meta_llama3_2024}).
    \item \textbf{Institutional Frameworks:} Macro-environmental data and reports from governmental  institutions, including {\em data center forecasts} of the International Energy Agency (IEA)  \cite{iea_datacenters_2025} and {\em objectives of the Paris Agreement} \cite{paris_agreement_2015}.
\end{itemize}

\subsection{Overview of the Major Selected Contributions}
\begin{table}[pos=htpb]
    \centering
    \begin{tabular}{|p{2.3cm}||p{5.1cm}|}
        \hline
    \textbf{Topic} & \textbf{References} \\
    \hline
        \hline
        Profiling methodology & \cite{Raffin2025};\cite{Yang2024};\cite{Ilsche2015}, \cite{Rajput2024b}, \cite{GULDNER2024402}, \cite{Zhang2025} \\
        \hline
        Profiling tools  & \cite{Rajput2024b}, \cite{ZONG201727}, \cite{Weber2025}, \cite{likwid2010}, \cite{Rajput2024}  \\
        \hline
        Monitoring tools  & \cite{Paipuri2025}, \cite{Terboven2024}, \cite{RAHMANI2026108428} \\
        \hline
        Hardware design  & \cite{Sanni2019}, \cite{WangKe2022}, \cite{Sadi2025}, \cite{LiGuoyu2025}, \cite{Laskar2025}, \cite{Liu2025}, \cite{LiTaixin2025}, \cite{Dipta2023}, \cite{Ebrahimi2024}, \cite{Fayza2025}, \cite{Gupta2023}, \cite{Pashaeifar2019}, \cite{Danopoulos2025}, \cite{Golden2025}, \cite{Kalyanapu2025}, \cite{Abu2017}, \cite{Cochet2024}, \cite{Panteleaki2025}, \cite{Sudarshan2024b}, \cite{DARABI2026101307}, \cite{Spieck2025}, \cite{Shiflett2021}, \cite{LiJiajun2021} \\
        \hline
        Green software/OS \& Code  & \cite{LiXinyi2025}, \cite{Swaraj2026}, \cite{Georgiou2022}, \cite{Sikand2024}, \cite{Cesarini2021}, \cite{Stoico2025}, \cite{Song2024}, \cite{Tunzina2026}, \cite{Martino2025}, \cite{Shah2026}, \cite{McGuire2023}, \cite{Hort2023}, \cite{Ayoola2025}, \cite{Chen2016}, \cite{Kempen2025}, \cite{Cruz2025}, \cite{DeMartino2025}, \cite{Roque2025}, \cite{Maquoi2025}, \cite{Alizadeh2025}, \cite{Zanfardino2025}, \cite{Tundo2024}, \cite{ANDRESLARRACOECHEA2026130046}, \cite{Dhawan2026} \\
        \hline
        Power/perf predictive modeling  & \cite{Marahatta2021}, \cite{RAHMANI2026108428}, \cite{Shim2022}, \cite{Al2020}, \cite{Eeckhout2025}, \cite{Zhou2022}, \cite{GULDNER2024402}, \cite{Conficoni2016}, \cite{Vahabi2025}, \cite{Menear2025}, \cite{Alan2015}, \cite{Venkatesh2015}, \cite{Xia2015}, \cite{Sikal2025}, \cite{Schoonhoven2022}, \cite{Fayza2025}, \cite{Zhou2022}, \cite{Eeckhout2025}, \cite{Gupta2023}, \cite{Pashaeifar2019}, \cite{Kakolyris2025}, \cite{Jiang2025}, \cite{Wang2025b}, \cite{SK2024329}, \cite{CESARIO2026108245}, \cite{DACOSTA2025101106}  \\
        \hline
        Job scheduling \& VM placement & \cite{Sun2022}, \cite{Wajid2016}, \cite{Tarplee2016}, \cite{Menear2025}, \cite{Georgiou2015}, \cite{Wang2017}, \cite{Wang2022}, \cite{Ali2018}, \cite{Sahoo2026}, \cite{Jayanetti2024}, \cite{Xu2020}, \cite{Carvalho2025}, \cite{Beena2025}, \cite{Tang2025}, \cite{Li2025}, \cite{Peng2022}, \cite{Guan2024}, \cite{Chen2022}, \cite{Ding2023}, \cite{Desai2025}, \cite{Qiao2025}, \cite{CHEN2025107760}, \cite{QURESHI2019453}, \cite{HASSAN2020431}, \cite{LI2018887}, \cite{LUO2020119}, \cite{HU2017119}, \cite{DING2020361}, \cite{MENCARONI2025146787}, \cite{KASSAB2021100590}, \cite{WANG2026129008}, \cite{AMINOROAYA2026101327}, \cite{Singh2023}, \cite{SINGHAL2024100985}, \cite{Hidayat2025}, \cite{LI2020789}, \cite{BANERJEE2024376}, \cite{HOSSEINISHIRVANI2023100856}, \cite{BAYDOUN2025101258},  for AI (\cite{Gu2025}, \cite{Chen2025}, \cite{Stojkovic2025b}) \\
        \hline
        Others optimization approaches  & \cite{Dutot2017}, \cite{Shen2023}, \cite{Ana2024}, \cite{Yang2019}, \cite{Fettes2019}, \cite{Elgamal2025}, \cite{Zaw2021}, \cite{Fan2023}, \cite{SINGH2021100463}, \cite{Nan2017}, \cite{Zhao2025}, \cite{Zhong2026}, \cite{Huang2025}, \cite{Son2025}, \cite{Zheng2025}, \cite{Hanafy2024}, \cite{Jayaweera2024}, \cite{KRZYWANIAK2023396}, \cite{VERMA2025101115} \\
        \hline
        Cooling Technologies & \cite{Conficoni2016}, \cite{Abu2017}, \cite{MIRHOSEININEJAD2020174}, \cite{Yang2019}, \cite{KAHIL2025125734}, \cite{epa_pfas}, \cite{Jiang2025}, \cite{Ding2023}, \cite{Wu2025}, \cite{Kilian2025} \\
        \hline
        Special case of AI  & \cite{strubell2019energy} \cite{Hogade2025}, \cite{Nasrin2022}, \cite{Rajput2024b}, \cite{Liu2025}, \cite{Georgiou2022}, \cite{Sikand2024}, \cite{LiTaixin2025}, \cite{YAO2025101220}, \cite{Patel2024}, \cite{Tschand2025}, \cite{Hasan2025}, \cite{Ifath2026}, \cite{John2025}, \cite{Wang2024}, \cite{Stojkovic2025}, \cite{Xia2024}, \cite{Alswaitti2025}, \cite{Barbierato2024}, \cite{Salehi2024}, \cite{Kakolyris2025}, \cite{Shen2025}, \cite{Ifath2026}, \cite{Chung2024}, \cite{Moro_Ragazzi_Valgimigli_2023}, \cite{Tian2026}, \cite{CHADLI2026108218}, \cite{SK2024329}, \cite{Sikand2024}, \cite{Panteleaki2025}, \cite{CASTANO2026101347}, \cite{Shiflett2021} \\
        \hline
    \end{tabular}
    \caption{Overview of Recent Energy-Aware Computing Studies}
    \label{tab:placeholder}
\end{table}
We made a specific methodological choice regarding the selection of the papers for our corpus as previously described. We acknowledge that this choice may have resulted in the omission of genuine/valuable/valued contributions (less recent ones, books, blogs, conferences/journals with unmatched rank, open archives). The reader should be aware of this deliberate choice that we made. However, we think that it is highly likely that most of the aforementioned papers are cited in those selected for this survey.

\subsection{Keyword-based Analysis of our corpus}

\begin{figure}[pos=htpb]
    \centering
    \includegraphics[width=1\columnwidth]{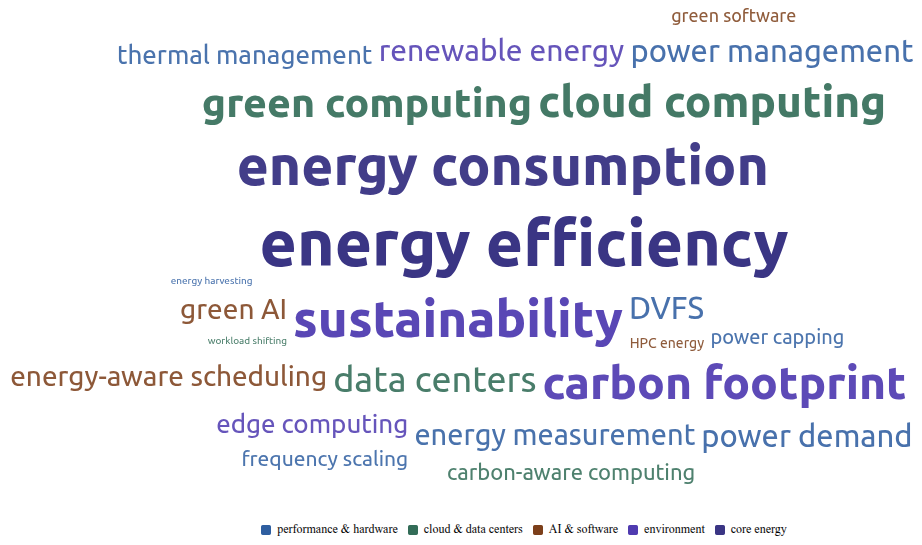}
    \caption{Panorama of the keywords from our corpus.}
    \label{fig:energy_keywords_wordcloud}
\end{figure}

To identify the dominant research themes within the studied corpus, we analyzed the keywords provided by the authors. After processing and normalizing the extracted keywords, we obtained 2,133 terms with redundancies and 949 unique terms (45\%). We also considered the case of semantic equivalence. This aggregation revealed that \textit{Energy Efficiency} was the most prominent topic (99 occurrences), followed by \textit{Energy Consumption} (77), \textit{Sustainability} (67), and \textit{Carbon Footprint} (61). Other recurrent themes included \textit{Cloud Computing} (46), \textit{Green Computing} (41), \textit{Power Demand} (27), \textit{Energy-Aware Scheduling} (26), \textit{Data Centers} (26), and \textit{Power Management} (25). Figure \ref{fig:energy_keywords_wordcloud} provides a visual summary of the keywords according to their magnitude.

\section{Main Energy-Related Metrics}
\label{sec:metrics}

To optimize energy across the Cloud-Edge continuum, researchers rely on fine-grain metrics besides the classical {\em electrical power} (Watts) to encompass {\em thermal dissipation, water consumption}, and {\em carbon emissions} (over the whole life cycle). This extended focus implies adapting traditional HPC benchmarking so as to consider large-scale applications like those from {\em Generative AI} and large-scale infrastructures like {\em Exascale machines}.

\subsection{Key Metrics for Large-Scale Infrastructures}
In extreme-scale computing, facility-level power consumption is the main critical ceiling. The November 2025 TOP500 list illustrates this unprecedented energy budget: \textit{El Capitan} system achieves 1.809 ExaFlops drawing 29.68 Megawatts (MW) \cite{top500_2025}, while \textit{JUPITER Booster} (the first European Exascale system) achieves 1.000 ExaFlops at a noteworthy 63 GFlops/Watt. Table \ref{tab:electricity_cost} provides sample data related to processing performance and energy cost, which illustrate the significant operational expenditure (OpEx) that justifies the need for energy efficiency \cite{Patki2025, BANCHELLI2026108125}. The total hourly cost is calculated using the average price per kWh within [2023;2026] \cite{elect_price_2026}.

Thermal Design Power (TDP), which now routinely exceeds 1000W for AI chips such as the NVIDIA B200 \cite{nvidia_blackwell_2024}, governs the magnitude of cooling requirements. At the individual facility level, Power Usage Effectiveness (PUE) remains the standard \cite{greengrid_pue}, but is no longer sufficient from the isolation point of view. The shift toward extreme-density AI clusters requires a huge amount of chilled water, which promotes {\em Water Usage Effectiveness} (WUE) to the rank of a critical metric. In water-scarce regions, minimizing the volume of water evaporated per kilowatt-hour (L/kWh) is currently prioritized even above PUE \cite{uptime_2025}. Furthermore, optimizing modern data centers requires auditing their direct supply architecture; Power Supply Efficiency (PS-Loss) captures the resistance conversion of Uninterruptible Power Supplies (UPS) \cite{Zhong2026}, where active load-balancing can reduce lifecycle expenditures independently of IT scheduling \cite{Huang2025, Jayanetti2024}.

\begin{table}[htpb]
\centering
\caption{Estimated Hourly Electricity Cost of Exascale Supercomputers (Based on Top500 Data - Nov. 2025)}
\label{tab:electricity_cost}
\resizebox{\columnwidth}{!}{%
\begin{tabular}{|l|l|c|c|c|r|}
\hline
\textbf{Machine} & \textbf{Host Region} & \textbf{Sustained Perf.} & \textbf{Power (MW)} & \textbf{\$ / kwh} & \textbf{Cost / Hour} \\ \hline
EL CAPITAN  & USA (California)  & 1.809 EFLOPS  & 29.68 MW  & 0.148 & \$4,392  \\ \hline
FRONTIER    & USA (Tennessee)   & 1.353 EFLOPS  & 22.78 MW  & 0.148 & \$3,371  \\ \hline
AURORA      & USA (Illinois)    & 1.012 EFLOPS  & 38.69 MW  & 0.148 & \$5,726  \\ \hline
JUPITER     & Germany (Jülich)  & 1.000 EFLOPS  & 15.87 MW  & 0.285 & \$4,522  \\ \hline
\end{tabular}%
}
\end{table}
\subsection{AI-Related Energy-Carbon Data}
The traditional FLOPS metric is no longer the only focus when it comes to large-scale AI workloads. The complexity analysis of GenAI considers specific metrics such as \textit{Energy per Token (Joules/Token)} for inference, and \textit{Training Throughput Efficiency (Tokens/Second/Watt)} for training. Hybrid benchmarks such as \textit{Carbon-Aware Accuracy (Carburacy)} evaluate hyperparameter sweeps simultaneously across cognitive effectiveness (F1/BLEU) and eco-sustainability curves, identifying models that achieve competitive accuracy at the lowest $CO_2$ cost \cite{Moro_Ragazzi_Valgimigli_2023}.

Mostly focusing on raw power consumption obscures the environmental impact, which is quantified through the operational {\em Carbon Footprint} represented by the {\em Carbon Intensity} ($kg\:CO_2e / kWh$). Table \ref{tab:carbon_footprint} provides the hourly footprints of the latest Exascale architectures based on regional grid intensities \cite{elec_carbon}. This dichotomy creates distinct contextual optimization scenarios: \textit{Cap-Ex Power Limits} (maximizing FLOPS/W under strict utility caps), \textit{Renewable-Abundance} (focusing purely on time-to-solution when powered by 100\% renewables, this is the case of the LUMI system), and \textit{Grid-Aware Migration} (report or migration of cloud-based AI training for sustainability consideration).

\begin{table}[htpb]
\centering
\caption{Estimated Hourly $CO_2$ Footprint of Leading Supercomputers}
\label{tab:carbon_footprint}
\resizebox{\columnwidth}{!}{%
\begin{tabular}{|l|c|r|r|r|}
\hline
\textbf{Machine} & \textbf{Power Grid} & \textbf{Power (MW)} & \textbf{$kg(CO_2e)$ / kWh} & \textbf{Total $kg(CO_2e)$ / Hr} \\ \hline
EL CAPITAN (USA) & Mixed Renewable          & 29.68 MW            & \textit{0.220}                   & $\sim 6,529$                      \\ \hline
FRONTIER (USA)   & Regional Average         & 22.78 MW            & \textit{0.360}                   & $\sim 8,200$                      \\ \hline
AURORA (USA)     & High Fossil              & 38.69 MW            & \textit{0.320}                   & $\sim 12,380$                     \\ \hline
JUPITER (DE)     & Mixed European           & 15.87 MW            & \textit{0.380}                   & $\sim 6,030$                      \\ \hline
LUMI (FI)*       & 100\% Hydro/Nuclear      & 7.10 MW             & \textit{0.050*}                  & $\sim 355$                        \\ \hline
\multicolumn{5}{l}{\small \em *LUMI operates on nearly 100\% renewable energy; its lowest carbon footprint illustrates the impact of the carbon factor.}
\end{tabular}%
}
\end{table}
\section{Architectures and Energy Trade-offs}
\label{sec:hardware}

The end of Dennard scaling has led to the emergence of highly specialized architectures within the hardware landscape. Performance is no longer bounded by transistor density, but rather by {\em power delivery} and {\em heat dissipation}. An energy-aware computing investigation requires examining processors across the continuum: from battery-powered embedded systems to power-hungry (megawatt-scale) processors \cite{LASO2025101088}.

\subsection{The Edge and IoT: Milliwatt-Scale Inference}
Within the edge ecosystem (remote sensors, drones, robotics), devices strictly target  {\em space, wattage}, and {\em performance} (SWaP) envelopes. Ultra-low-power microcontrollers (MCUs) natively utilize ratio-based energy pre-assignment to distribute micro-joules proportionally to the priority level of the tasks. In energy-aware systems, schedulers dynamically adapt to intermittent RF/solar availability. By embedding ultra-lightweight adaptive checkpoints into Non-Volatile RAM (NVRAM) a few milliseconds before voltage failure, Edge devices process Deep Neural Networks via transient energy recovery without having to restart \cite{CHEN2026108009}. 

In order to maximize {\em inference-per-watt}, edge platforms leverage {\em Coarse-Grained Reconfigurable Architectures} (CGRAs) and {\em approximate computing}. By modeling hardware error as an acceptable "additive noise", compilers schedule the execution of some blocks in a non-deterministic way on specialized co-processors, thus sacrificing numerical precision constraints in favor of energy reduction \cite{Ebrahimi2024, Pashaeifar2019}. As edge computing is subject to an increasing demand (e.g., Raspberry Pi 5 has a power peak of 15W), Single-Board Computers tend to rely on Heterogeneous Multicore Processors (big.LITTLE architectures). In this context, non-stationary workloads require predictive learning-based orchestration \cite{Zhou2022} and decentralized power-management, thus allowing local components to autonomously regulate operating states \cite{Cochet2024, Sudarshan2024b, Spieck2025}.

A specific case of extreme-edge computing is that of Low Earth Orbit (LEO) satellites, which suffer from high operational thermal throttling. To bypass the limited solar energy budget, LEO architectures employ \textit{Layer-Wise Co-Inference Scheduling}, splitting the neural network layers across satellite radio links and offloading tensor computations on Earth-bound servers \cite{Li2024, Chen2025}. Furthermore, processing software updates on these decentralized sensors requires microservice containerization, yet traditional abstractions like Docker are well-known for being detrimental w.r.t power-constrained silicon. Compiling frameworks that directly target bare-metal devices (e.g., ONNX) or that use low-memory WebAssembly (WASM) yields significantly efficient on-device inference \cite{Hampau2022}. Finally, as Edge AI pushes Multi-Processor System-on-Chips (MPSoCs) to their thermal limits, internal Network-on-Chip (NoC) wiring becomes a primary energy drain. State-of-the-art NoC schedulers (e.g., \textit{IntelliNoC}) deploy continuous Q-learning agents natively on silicon to apply fine-grained DVFS, thus autonomously shutting down hardware routing components to prevent localized overheating \cite{Ali2018, Wang2019, WangKe2022, Fettes2019}.

\subsection{Cloud Infrastructures and Network Overheads}
At the data center level, the energy battle pits traditional x86 architectures against ARM-based hyperscaler chips. To cope with the highly increasing power penalty of high-clock frequencies, x86 architectures (e.g., Intel Xeon 6 Sierra Forest, AMD EPYC Turin) follow the path of heterogeneous design, deploying up to 288 Efficiency Cores (E-cores) per socket \cite{intel_xeon6_2024, amd_turin_2024}. Although highly efficient at the core level, these dense sockets carry 500W TDPs, thus testing the air cooling system. In contrast, driven by the RIKEN Fugaku supercomputer, cloud providers increasingly deploy custom ARM RISC silicon (AWS Graviton4, NVIDIA Grace). By eliminating legacy decode overheads and integrating LPDDR5X memory, ARM architectures maximize predictable performance-per-Watt \cite{aws_graviton}. 

Beyond the basic processors, the internal infrastructure of the Cloud represents a considerable energy vector. Current workflows tend to generate massive "Coflows." To reduce DCN power waste, server-centric networks and Software-Defined Networking (SDN) controllers dynamically map traffic dependencies, intentionally collapsing idle optical paths and packet-switching routing tables so as to minimize the networking footprint by up to 40\% without detrimental packet losses \cite{Alan2015}.

\subsection{AI Super-Accelerators and Exascale Horizons}
LLM training is driving a noteworthy shift in hardware design. While GPUs in data centers show an operational TDP of 300W, modern reticle-limit devices (e.g., NVIDIA Blackwell B200, AMD MI300X) draw from 1000W to 1200W \cite{nvidia_blackwell_2024}. Evaluating raw TDP hides a vulnerability induced by Transformers: \textit{Voltage Emergencies ($di/dt$)}. The highly synchronous, bursty execution of Attention layers currently creates unprecedented microsecond spikes, thus destabilizing power delivery networks and forcing a conservative static guard-band throttling that degrades the overall energy-to-solution margin. 

Domain-specific accelerators (e.g., Google TPUs) address this through fine-grained gating, selectively powering down inactive functional units cycle-by-cycle \cite{Xue2025, google_tpu_trillium_2024}. Novel accelerator architecture such as GCNAX, which tailors the compute engine, buffer structure, and size based on specific Graph convolutional neural networks (GCNs) dataflow, was proposed \cite{LiJiajun2021}. Advances in interconnects (e.g., NVLink) optimize massive die-to-die tensor communication \cite{Laskar2025, Liu2025, ZhaoYingnan2024}, while granular voltage scaling at the level of individual computational operators allows precise adaptation to workload characteristics \cite{Wang2025b, Han2025}. 

\begin{table}[htpb]
\centering
\caption{Current Leading AI Accelerators (Training \& Inference Engines - 2025/2026)}
\label{tab:sota_accelerators}
\resizebox{\columnwidth}{!}{%
\begin{tabular}{|l|c|c|c|p{4cm}|}
\hline
\textbf{Accelerator} & \textbf{TDP (W)} & \textbf{Peak Precision} & \textbf{Memory} & \textbf{Noteworthy Energy Features} \\ \hline
NVIDIA B300 (Ultra) & 1400 W & FP4, FP6, INT8 & 288GB HBM3e & Dynamic tensor scaling; standardizes liquid cooling \\ \hline
AMD MI400X & 1600 W & FP4, FP8, FP16 & 432GB HBM4 & Advanced multi-die packaging isolates logic from I/O \\ \hline
Google TPU v7 & 157 W & BF16, FP8, INT8 & 192GB HBM3e & Systolic arrays minimize logic decode footprint \\ \hline
Intel Gaudi 3 & 900 W & BF16, FP8 & 128GB HBM2e & Integrated RoCE networking eliminates NIC overhead \\ \hline
\end{tabular}%
}
\end{table}

Looking toward 2026 and post-Exascale deployments, systems like  \textit{Alice Recoque} (AMD MI430X/SiPearl Rhea2 - France) and  \textit{Discovery} (ORNL - USA) rely on data-driven operational analytics to strictly enforce {\em facility energy budgets} \cite{Shin2024, eurohpc_alice_recoque}. To power these cloud systems, accelerators have evolved towards "Rack-Scale" paradigms (e.g., the 120 kW NVIDIA GB300 NVL72). Hardware selection requires dealing with a compromise between energy efficiency, precision (FP4 natively drastically reduces energy waste), and software lock-in (ASICs provide superior FLOPS/W but restrict developers to specific compiler stacks).

\subsection{Quantitative Overview of Hardware Trends}
A statistical analysis of recent literature in computer hardware ($N=43$ selected papers) reveals a decisive shift away from traditional CPU. Approximately 42\% of recent research works are dedicated to AI Super-Accelerators (e.g., GPUs, TPUs, Processing-in-Memory, and Wafer-Scale architectures), thus reflecting the industrial response to thermal concerns when processing large-scale workloads like those related to LLMs. The Edge and IoT continuum accounts for 33\% of the literature, with a major emphasis on Sub-Watt NPUs and Battery-Free RF-harvesting chips. General-purpose cloud CPUs, Chiplets (UCIe), and network interconnects cover the remaining 25\%. This distribution empirically confirms that the end of Dennard Scaling pushed toward prioritizing extreme-edge and extreme-accelerator heterogeneity over general-purpose silicon.

\begin{figure}[pos=htbp]
    \centering
    \begin{tikzpicture}
        \begin{axis}[
            ybar,
            width=\columnwidth,
            height=4cm,
            enlarge x limits=0.3,
            ylabel={\% of Surveyed Papers},
            xtick={1,2,3},
            xticklabels={Accelerators, Edge \& IoT, CPUs \& Networks},
            nodes near coords,
            nodes near coords align={vertical},
            bar width=20pt,
            ymin=0, ymax=55,
        ]
            \addplot[fill=blue!60] coordinates {(1,42) (2,33) (3,25)};
        \end{axis}
    \end{tikzpicture}
    \caption{Distribution of recent hardware-related energy-aware studies across the computing continuum.}
    \label{fig:trend_hardware}
\end{figure}
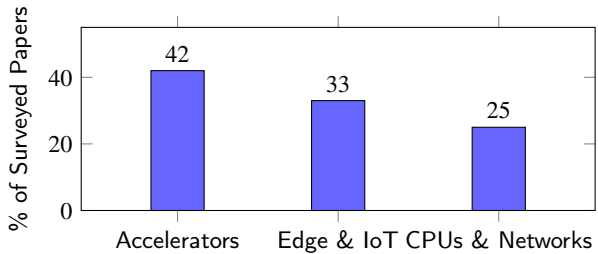

\section{Energy and Carbon Profiling Tools}
\label{sec:tools}

Dynamic energy optimization strategies across the Cloud-Edge-HPC continuum require real-time, fine-grained telemetry. As hardware diversity expands, the abstraction gap between software workflow and hardware profile is widening. The profiling ecosystem can be categorized into three paradigms: \textit{silicon-level middleware}, \textit{institutional facility monitoring}, and \textit{cloud-level carbon estimators}.
\subsection{Silicon Sensors and Abstraction Middleware}
Modern processors incorporate native power sensors to manage power-related aspects. On bleeding-edge 3nm FinFET nodes, On-Die Power Sensors (ODPS) use machine learning calibration to basically limit sub-50$\mu$s over-current excursions before passing abstractions to the OS \cite{Lu2025}. To make these raw registers available to developers, middleware tools consider unified interfaces for machine-specific MSRs. Furthermore, execution frameworks that take parallelization into account (e.g., PARMA) model the hidden parallelism of dynamic workloads on the fly, thus adjusting core voltage domains to improve Energy-Delay-Product (EDP) by nearly 50\% without invasive code instrumentation \cite{Al2020}.

\textbf{Hardware-Accelerator Telemetry:} The NVIDIA Management Library (NVML) is a low-level API that provides instantaneous access to power-related values and programmatic power capping. However, recent profiling work reveals critical flaws: tools like \texttt{nvidia-smi} sample only $\sim$25\% of actual runtime on heavily burdened A100/H100 GPUs, systematically underestimating total job energy due to clock frequency aliasing \cite{Yang2024}.

\textbf{CPU Telemetry (RAPL):} Intel’s RAPL and AMD’s APM stand as the reference of in-band profiling \cite{RAPL2022}. Beyond structural omissions, generic hardware limiters are "application-blind". When a mixed-tenancy node reaches an artificial power cap, RAPL indiscriminately throttles the fastest-running cores, thus aggressively stifling latency-critical high-priority applications as well as demanding software-controlled "Per-Application Power Shares" \cite{Guliani2019, Zhang2024}.

\textbf{Unified Middleware:} LLNL’s Variorum \cite{Variorum} provides a vendor-agnostic C-API to dynamically cap power consumption across heterogeneous nodes. Similarly, the ALUMET energy-measurement framework \cite{Raffin2025, Raffin2025b} provides an extensible, asynchronous Rust-based pipeline designed to minimize the intrusive overhead of high-frequency RAPL polling. Built upon LIKWID \cite{likwid2010}, the EE-HPC energy-related framework \cite{Terboven2024} provides hybrid HPC centers with job-specific monitoring, pairing phase detection with MPI/ OpenMP tuning.

\textbf{Data-Dependent Power Fluctuations:} Macroscopic static modeling tends to be chronically inaccurate due to a strong dependency on static inputs. Benchmarking proves that feeding functionally dissimilar vectors (e.g., vision inputs vs. sparse tensors) diverts the GPU instruction branching, thus generating wild internal silicon footprints even when raw FLOPS remain identical \cite{Gregersen2025, ZhaoZhengji2024}. Similarly, comparative database profiling (MongoDB vs. PostgreSQL) illustrates that identical operational schemes can exhibit severe energy divergence based on the underlying query logic \cite{Lella2024}.

\textbf{The Observer Effect:} The middleware query relies on the execution of monitoring threads by the CPU. Querying MSRs above 100 Hz interrupts the application threads, pollutes the memory caches, and consumes some power, thus affecting the workload's efficiency. To cope with this, state-of-the-art predictive techniques leverage NLP-adapted Deep Learning approaches. Frameworks like \textit{DeepPM} auto-vectorize compiler-generated binaries into Transformer models in order to forecast power consumption, thus bypassing hardware execution \cite{Shim2022}. Similarly, \textit{EffiCast} utilizes sequence-based LSTM/Transformers to resolve dynamic L3-cache contention and project microsecond-latency V/f scaling without static inputs \cite{Sikal2025}. 

\textbf{Standardized ML Telemetry \& Edge AI:} To deeply figure out measurement inconsistencies, the \textit{MLPerf Power} methodology has been standardized \cite{mlcommons_2025}. This was provided through documenting over 1,800 reproducible profiles across 60 systems \cite{Tschand2025}. The \textit{CARAML} suite similarly unifies assessments across different Exascale architectures \cite{John2025, CASTANO2026101347}. Scaling estimation to multi-node domains introduces some profiling bottlenecks; Exascale systems now deploy \textit{Real-Time AI-Powered Monitoring} via neural classification models on peripheral nodes to extrapolate and forecast power deviations, thus preventing thermal faults autonomously \cite{RAHMANI2026108428, Rajput2024, SK2024329}.

\subsection{Institutional-level Facility Monitoring}
To avoid software overloads, national power grid initiatives use out-of-band physical architectures. Platforms like D.A.V.I.D.E. integrate hardware sensors directly into compute backplanes for high-frequency telemetry without context-switching overheads \cite{Abu2017}. Facility monitoring is mostly based on the \textit{Prometheus/Grafana} stack and Time-Series Databases (TSDBs). Deployment examples include CEEMS (Jean Zay supercomputer), which scrapes metrics at 10-second intervals so as to cross-reference jobs with the French carbon factors \cite{Paipuri2025}; and Grid'5000 (\textit{Kwollect}), which synchronizes out-of-band wattmeters with in-band RAPL data. High Definition Energy Efficiency Monitoring (HDEEM) uses FPGA probes and resistive PCIe shunts to capture transient micro-spikes that cannot be caught up by RAPL \cite{Ilsche2015}. However, running 100,000 GPUs generates terabytes of TSDB metrics daily, thus creating a "Big Data" I/O bottleneck.

\subsection{Virtualization Wall and Carbon Estimation}
Public cloud providers restrict access to hardware telemetry because of side-channel threats. Power-consumption alerts can be exploited to infer sensitive system behavior \cite{Dipta2023,Kim2025}. The \textit{GateBleed} attack exposed the fact that {\em hardware power-gating recovery delays} allow malicious intruders to reconstruct hidden Transformer routing paths and proprietary datasets, thus proving that telemetry breaks privacy \cite{Kalyanapu2025}. 

In such a constrained context, direct measurement is replaced by analytical estimation via Performance Monitoring Counters (PMCs) \cite{DACOSTA2025101106}. Lightweight Python libraries (e.g, CodeCarbon, Eco2AI) fall back to analytical estimation when hardware registers are unavailable. Recent methodologies utilize LLMs to predict power variations and drive HPC scheduling directly \cite{Menear2025}. To handle operational uncertainty of the electrical grid, frameworks like \textit{U-DUCT} introduce information-gap theory to sweep server uncertainty bounds without assuming precise power distributions \cite{Guan2024}. For the sake of consistency, container-aware protocols like \textit{Phoeni6} ensure transparent and reproducible neural network benchmarking across different clusters \cite{Zhang2025, OLIVEIRAFILHO2025101172}. Going beyond passive measurement, \textit{Zeus} \cite{You2023} dynamically navigates the Pareto frontier between DNN training time and energy, thus actively adjusting GPU batch sizes at runtime.

\subsection{Trends on Measurement \& Profiling}
The evaluation of profiling methodologies across our corpus of 33 articles reveals the considerable impact of the "virtualization wall". While previous research studies heavily prioritized physical out-of-band monitoring, only 21\% of post-2023 papers focus on hardware shunts or TSDB facility monitors. In addition, 45\% of the literature now strictly evaluates Software-Level Carbon Estimators and Machine Learning predictive telemetry (e.g., LLM-based power prediction and CodeCarbon). In-band silicon sensors and abstraction middleware (e.g., RAPL, eBPF, ALUMET) belong to the remaining 34\%. This proves a pivotal software-engineering observation: because cloud hyperscalers prohibit hardware telemetry for security, researchers are increasingly pushed towards purely virtual and predictive carbon estimation.

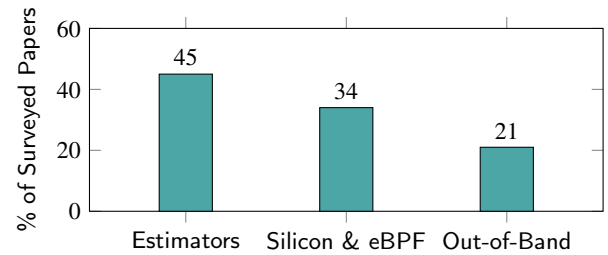
\begin{figure}[pos=htbp]
    \centering
    \begin{tikzpicture}
        \begin{axis}[
            ybar,
            width=\columnwidth,
            height=4cm,
            enlarge x limits=0.3,
            ylabel={\% of Surveyed Papers},
            xtick={1,2,3},
            xticklabels={Estimators, Silicon \& eBPF, Out-of-Band},
            nodes near coords,
            nodes near coords align={vertical},
            bar width=20pt,
            ymin=0, ymax=60,
            ]
            \addplot[fill=teal!70] coordinates {(1,45) (2,34) (3,21)};
        \end{axis}
    \end{tikzpicture}
    \caption{Overview of energy measurement methodologies.}
    \label{fig:trend_tools}
\end{figure}

\section{Software-Level Energy Optimization}
\label{sec:optimization}
While high-performance chips define minimum energy consumption, a system's carbon footprint is largely determined by its software. Thus, this carbon footprint can only be mitigated through advanced and comprehensive software optimization: from code compilation to operating system power management, including distributed coarse-grain scheduling.

\subsection{Algorithmic and Code-Level Optimizations}
Modern design mandates "energy clarity", thus promoting formal energy interfaces that provide maximum power consumption estimates before the integrated circuit is put into operation \cite{Chung2025}. Advanced software engineering incorporates model-driven engineering (e.g., RADIANCE+) to prototype abstract algorithmic configurations and extract power consumption estimates before compilation \cite{ANDRESLARRACOECHEA2026130046, Desai2025}.

\textbf{Data-Oriented ETL and DataFrames:} Preprocessing represents a significant and continuous source of energy consumption. Evaluation of data management frameworks such as Pandas, Vaex, and Dask reveals that multithreaded memory access graphs (Dask) optimize parallelizable filtering, while lazy evaluation processors (Vaex) minimize active RAM consumption during transpositions. In distributed NoSQL ledgers (e.g., Cassandra), adjusting "Consistency Levels" fundamentally acts as a physical power multiplier, mitigating redundant network transmissions \cite{NEJATISHARIFALDIN2026108210}. Bottom-up profiling successfully parameterizes complex relational joins across distributed databases to predict computationally heavy query pathways \cite{Lella2024}.

\textbf{Tiling and Precision Scaling:} Source-code loop transformations such as {\em tiling} and {\em fusion} aim to reduce cache misses and redundant memory accesses. \textit{Energy-Aware tiling} structurally forces auto-tuning compilers to prioritize power grids that constrain GPU register-leakage rather than raw execution latency \cite{Jayaweera2024}. Precision scaling (FP32 to INT8/FP4) decreases memory bandwidth requirements and ALU power activation.

\textbf{Frameworks and JVMs:} Empirical studies show different DL frameworks exhibit measurable efficiency discrepancy \cite{Georgiou2022,Sikand2024,Ournani2021}. At runtime, models compiled with TensorFlow frequently exhibit lower memory-related costs than those obtained with PyTorch, while JVM configurations aggressively shift static consumption from the host \cite{Stoico2025,ZONG201727}.

\textbf{Message Passing Power-Overhead:} MPI processes consume dynamic power during wait times related to communication operations. The (\textit{COUNTDOWN}) tool switches active CPUs into ultra-low ACPI states during these wait times, thus recovering up to 37\% of power at the node level \cite{Cesarini2021}. Furthermore, Green Linear Algebra (GreenLA) restructures the core semantic of the BLAS to minimize VRAM accesses, thus reducing power consumption by $\frac{1}{3}$ \cite{Chen2016}. Automatic kernel optimization through auto-tuning identifies optimal frequency and memory constraints before execution \cite{Schoonhoven2022}.

The lack of parallelization of the processor's data loaders could lead to underloading of the energy-intensive active accelerators, thus increasing the number of joules per epoch.

\textbf{Energy Leaks in AI-Based Systems:} The analysis of code repositories reveals pervasive power leaks (suboptimal API configurations). The lack of (efficient) parallelization of the processor's data loaders could lead to underloaded power-hungry (1000MW) accelerators, thus increasing Joules-per-epoch. \cite{Roque2025, Alizadeh2025}. Furthermore, evaluations of LLM-assisted coding (with GitHub Copilot, for instance) conclude that there is a frequent failure to optimize transient memory thresholds, thus producing solutions that draw significantly higher power consumption than manual refactoring \cite{Swaraj2026, Tunzina2026}.

\textbf{UI/UX and Thread Governance:} The Software Engineering community agrees on the significant power burden of executing heavy interfaces at the system-level; "Persona-Based UI/UX" avoids continuous screen-refresh operations in order to respect the power limits of the device \cite{Zanfardino2025, Desai2025, Tundo2024}. On MPSoCs with heterogeneous edges, interfaces like \textit{HARP} distribute asynchronous threads following a specific scheduling algorithm \cite{Smejkal2025}.

Energy optimization on large-scale computing clusters requires platform-specific co-tuning; JiT (Just in Time) compiled frameworks such as JAX have been shown to exhibit deep operational instability when forced to the lowest dynamic frequencies, which does not appear to be the case with TensorFlow \cite{Tchakoute2025}.

\subsection{Energy-Aware Compilation}
Traditional compilers (e.g., GCC, LLVM) target performance, but minimizing execution time does not linearly translate into energy minimization \cite{Jayaweera2024}. Energy-efficient compilers use instruction scheduling to minimize the Hamming distance between operands (their values), thereby reducing transistor switching. The MLIR framework optimizes tensor graphs based on the thermal profiles of the hardware, by inserting power preventive cut-off instructions. Datasets like \textit{Greenlight} allow profiling the hardware footprint of TensorFlow API calls, thus noticing unpredictable energy spikes \cite{Rajput2024}. For a unify consoderation of extreme heterogeneity, SYCL C++ toolchains (e.g., \textit{SYnergy}) incorporate explicit power limits at compile time, assigning fine-grained core/memory scaling limits on a per-kernel basis \cite{Fan2023, Sikand2024, DESENSI2018136}.

\subsection{Dynamic Power Management}
At runtime, the operating system manages electrical physics in real-time using three main mechanisms: \textit{Dynamic Power Capping}, \textit{Workload Consolidation}, and \textit{Carbon-Aware Scheduling}.

\textbf{DVFS and Q-Learning:} Dynamic Voltage and Frequency Scaling (DVFS) exploits the available computational headroom. Advanced schedulers handle dynamic frequency management following the model of the NP-hard {\em variable-size bin packing puzzle} \cite{Wang2017}. At runtime, MPI "Application-Oblivious" trigger DVFS step-downs automatically \cite{Venkatesh2015}. In edge environments running on Linux, Deep Reinforcement Learning (DRL) adaptively orchestrates DVFS states across asynchronous workloads  \cite{LiXinyi2025}. Q-Learning schedulers continuously build state-action matrices to track the arrival of tasks based on power availability, thus avoiding blockages related to greedy heuristics \cite{DING2020361, Chen2022}.

\textbf{Phase-Aware Core/Uncore Scaling:} Modern architectures allocate substantial power to "uncore" components (caches, interconnects). Phase-aware approaches explicitly coordinate core/uncore adjustments \cite{Carpentieri2025,Shah2026,Zheng2025}.

\textbf{Job-Level Power Allocation (SLURM/PBS):} Facility operators enforce strict power limits \cite{Fraternali2018}. Adaptive strategies combine {\em power capping} and {\em runtime profiling} \cite{Dutot2017}. In congested environments, coordinated control mitigates transient peaks \cite{Malla2020}. Middleware such as  \textit{BAR/BARMAN} intercepts OpenMP tasks to dynamically redistribute unused micro-power among coexisting applications \cite{COSTERO2023100865}.

Imposing strict limits on dense nodes causes cascades of flow limitation.

\textbf{Multi-CPU/GPU Reinforcement Power-Capping:} Imposing strict limits on dense nodes causes a series of limitations. RL frameworks such as \textit{PowerCoord} actively re-allocate power budgets between host CPUs and PCIe accelerators based on {\em phase-prediction} \cite{AZIMI2020100412}. Recent benchmark evaluations of the FRONTIER supercomputer demonstrate that Frequency Capping systematically outperforms raw power-capping on memory-bound applications \cite{Costa2025,KRZYWANIAK2023396}.

\subsection{Cloud/Edge Workload Management}
Since static power leakage pushes {\em hyperscalers} to follow a {\em dense stacking} strategy for the management of containers, an accurate mathematical calculation of the multi-objective ({\em performance and energy}) mapping requires computationally expensive back-end heuristics such as {\em hyperheuristics} for Cloud Scheduling Problems (HHCSP) \cite{Carvalho2025}.

\textbf{Virtualization and Learning-Based Orchestration:} Energy-efficient cloud operation relies on dynamic workload consolidation \cite{Vahabi2025,TOOR20191112,RUAN2019380}. Federated learning techniques enable coordinated scheduling across geo-distributed infrastructures \cite{Sahoo2026,GONZALOSANJOSE2025107593}. To address latency and energy overheads \cite{YAO2025101220}, DRL optimizes migration decisions \cite{Zhao2025,Hidayat2025}, with renewable energy-aware geo-distributed job assignment \cite{Xu2020}. Microarchitectural innovations enable deeper low-power states with reduced wake-up latency \cite{Yahya2022,Antoniou2022}. Orchestrators that use {\em Flower Pollination Optimization} have demonstrably reduced carbon emissions by up to 48.60\% during Virtual Machine Placements (VMP) \cite{Singh2023,CESARIO2026108245}.

\textbf{Green Bin Packing \& Quality-Aware Consolidation:} The {\em online bin packing} approach treats volatile VMs through \textit{Green Bin Packing} heuristics \cite{Bibbens2026}. \textit{EcoCore} and \textit{AgilePkgC} intricately co-adjust CPU voltages under the awareness of tail-latency \cite{Antoniou2022, Park2025}. Topology-Aware approaches seamlessly estimate CPU thermal limits and network switch hop-distances \cite{BANERJEE2024376, QURESHI2019453, HOSSEINISHIRVANI2023100856}. To solve the NP-hard {\em bin packing problem}, hypervisors use evolutionary algorithms that mathematically combine QoS constraints and power consumption limits \cite{LI2020789, YAO2023222}.

\textbf{Dynamic Cloud/Edge Heuristics:} Modern schedulers run complex heuristics such as {\em modified artificial bee colony algorithms}, {\em binary search} \cite{SHE2025107458}, {\em differential evolution} \cite{WANG2026129008}, and {\em genetic algorithms} \cite{LUO2020119, MAGOULA2026108371}. Container-based methodologies enforce hypervisor live-migrations to strictly match dynamic network limits \cite{SINGH2021100463}. The {\em task scheduling} community relies heavily on \textit{Particle Swarm Optimization (PSO)} \cite{AMINOROAYA2026101327, SINGHAL2024100985}. PSO-based multi-objective frameworks explicitly maintain swarm diversity through discrete/quantum variants \cite{BAYDOUN2025101258}. By continuously adjusting particle speed, orchestrators quickly resolve VMP matrices (in real-time), performing hot migrations to optimize host storage while preventing SLA violations. Hybrid intelligent swarms (e.g., Cost-Energy Aware Spider Monkey Optimization) perform an exact routing of payloads to facilities experiencing maximum localized underutilization \cite{VERMA2025101115}.

\textbf{Accelerators Consolidation \& Serverless Cold-Starts:} Dynamic thread block migration redistributes workloads across GPUs \cite{Kempen2025}. Spatial partitioning (MIG slicing) enables multi-tenant execution, but at the cost of noticeable power and thermal interference across logically isolated workloads \cite{Kamatar2025,Antepara2025}. Serverless computing, considering {\em Function as a service} (FaaS), introduces significant cold-start overheads \cite{Stojkovic2024}. Asymmetric multicore architectures assign latency-critical tasks to P-cores and keep-alive mechanisms to E-cores \cite{Basu2025}.

\textbf{Antagonism Between SLAs and Sustainability:} {\em Flow network} algorithms ensure SLA compliance alongside efficient energy management \cite{Sun2022,ZHOU2018836}.  {\em Cuckoo Search} algorithms maximize resource packing, thus preventing violations during DVFS throttling \cite{Wang2024}. However, several analyses confirm a deep uncorrelation: almost all ML engineers fail to implement green metrics into production code due to a lack of standardized guidelines \cite{DeMartino2025}.

\textbf{Carbon-Aware Scheduling:} Optimizing for carbon requires considering spatio-temporal variability \cite{Wajid2016}. {\em Temporal shifting} postpones flexible workloads to low-carbon periods \cite{Beena2025}, while {\em spatial routing} directs computation where the availability of renewable energy is higher \cite{Nan2017, KHODAYARSERESHT2023100888}. Multi-agent RL techniques incorporate {\em dynamic signals} to optimize trade-offs \cite{Jayanetti2024,Hogade2025,CHEN2025107760}. Simulation platforms validate scalable abstractions \cite{Song2024}, while edge devices dynamically offload workloads based on carbon network conditions \cite{Son2025}. Deliberately pausing/resuming scientific processing based on the network's {\em marginal carbon intensity} (MCI) reduces the carbon footprint by more than 80\% \cite{WEST2026108453}. However, optimization exclusively based on MCI has flaws; indeed, missing the evaluation of absolute emission curves leads data centers to unexpected congestion \cite{Wiesner2025}. Hypervisors such as \textit{Ecovisor} and \textit{GreenFlow} transmit physical carbon data directly via the API, ensuring that AI containers self-regulate \cite{Souza2023, Stojkovic2025, Gu2025}.

\textbf{Hypervisor Benchmarking and Cloud Service Brokerage (CSB):} The design of the basic hypervisor (KVM vs. Xen) is inherently subject to power consumption fluctuations \cite{JIANG2019311}. Empirical studies explicitly show the sharp decrease in yields associated with carbon-aware load shifting; frequent temporal shifting leads to tasks generating a significant carbon penalty from the use of intermediate network equipment \cite{Sukprasert2024}. For the purpose of standardization, CSBs deploy mathematical pricing frameworks to incentivize tenants to minimize computing peaks \cite{Qiu2019}.

\textbf{Carbon-Aware DevOps:} 
Carbon-aware computing extends to continuous integration (CI) \& continuous delivery/deployment (CD) pipelines and GitOps frameworks \cite{Dhawan2026}. By combining execution tracing and energy consumption monitoring, CI/CD systems directly associate variations in energy consumption with specific code changes, thus structurally preventing the deployment of suboptimal code in a production environment \cite{Kempen2025,Rajput2024b,Maquoi2025}. Industrial case studies on developers' awareness of green software design and strategies to promote it within the company are a way to encourage green software design \cite{Ournani2020}.

\subsection{Time-to-Solution vs. Energy-to-Solution}
Modern scheduling techniques mainly seek efficiency. Optimizing for energy leads to instant execution of workloads in order to achieve maximum performance efficiency. Conversely, optimizing for carbon leads to delaying workloads following the availability of sustainable grids \cite{Hanafy2024}. 

\textbf{Race-to-Halt Paradigm:} If a task is CPU-bound, the most energy-efficient strategy is to run at maximum clock frequency so as to finish it quickly and switch to a sleep mode. Stochastic queuing-network frameworks mathematically model cloud VM arrival rates, thereby consistently proving that {\em Race-to-Halt} minimizes footprints \cite{Xia2015}.

\textbf{The Software Configurable Paradox:} Reliable investigations show that optimizing the {\em configuration matrix} for the absolute fastest binary runtime frequently spikes maximum baseline voltage states, thus generating highly asymmetric electricity consumption overheads \cite{Weber2023}. 

\textbf{Time vs. Energy:} The basic time/energy correlation seems to be broken during memory-bound workloads, where boosting the CPU frequency has no effect on time reduction, but increases power consumption. Linear Programming algorithms bound the Pareto front for heterogeneous tasks, empowering precise trade-offs \cite{Wang2022, Tarplee2016}. When it comes to managing time/energy trade-offs between independent workloads, energy-efficient schedulers route algorithmic queues to cores experiencing minimal physical thermal stress \cite{KASSAB2021100590}.

\textbf{Distributed Scaling Imbalance (Straggler Effect):} The workload imbalance causes the fastest processors to go into standby mode at the synchronization barriers (i.e. {\em power wastage}) \cite{Chung2024}. Coordinated control strategies (selective {\em frequency boost} or {\em controlled downscaling}) align execution phases.

\textbf{Reliability-Performance Degradation Loop:} Aggressive DVFS increases the vulnerability to soft errors. State-of-the-art dispatchers seek a balance between execution time, power, and reliability \cite{Peng2022,HASSAN2020431,LI2018887}. Due to task dependencies and increased execution time, an energy-efficient scheduling of precedence-constrained applications within the context of SLAs on DVFS-enable clusters has been proposed by Hu et al. \cite{HU2017119}. Failure-Aware VM Consolidation and Asynchronous checkpoints leverage exponential smoothing to predict impending hazards, thereby lowering overall operational energy by 34\% without jeopardizing task recovery \cite{Marahatta2021, SHARMA2019620, RODRIGUEZ2026103752}. 

Following what has been said, energy optimization requires explicitly weighing the cost of static leakage against the exponential penalty of dynamic frequency scaling.

\subsection{Trends on Software Optimization}
An analysis of the selected papers within the literature on software optimization ($N=52$ papers) reveals a sharp systemic shift toward {\em distributed orchestration}. Nearly 46\% of the contributions focus on {\em virtual machine consolidation} and {\em carbon-aware scheduling}, mostly by leveraging meta-heuristics like Particle Swarm Optimization (PSO) and Deep Reinforcement Learning (DRL). Dynamic Power Management (DVFS) and node-level Power Capping cover 31\% of the corpus, while pure algorithmic and code-level optimizations (e.g., compiler loops, dataframes, "Energy Smells") account for 23\%. This uneven distribution illustrates the fact that, while code efficiency remains fundamental, modern academic consensus favors mitigating static leaks from macro-infrastructures through intelligent cloud orchestrators.

\begin{figure}[pos=htbp]
    \centering
    \begin{tikzpicture}
        \begin{axis}[
            ybar,
            width=1.05\columnwidth,
            height=4cm,
            enlarge x limits=0.3,
            ylabel={\% of Surveyed Papers},
            xtick={1,2,3},
            xticklabels={Consolidation, DVFS \& Cap., Code \& Compilers},
            x tick label style={font=\small, align=center, text width=2.5cm},
            nodes near coords,
            nodes near coords align={vertical},
            bar width=20pt,
            ymin=0, ymax=60,
            ]
            \addplot[fill=purple!70] coordinates {(1,46) (2,31) (3,23)};
        \end{axis}
    \end{tikzpicture}
    \caption{Prevalence of software-level energy optimization strategies within recent literature.}
    \label{fig:trend_software}
\end{figure}
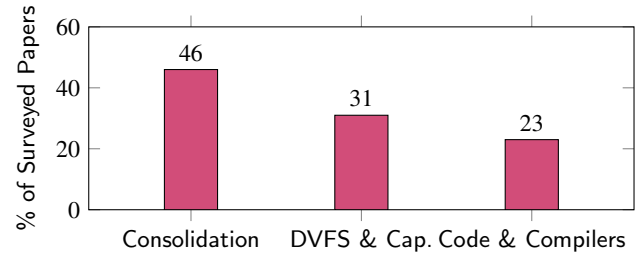
\section{The Energy Wall of Generative AI}
\label{sec:ai}

The impressive emergence of Generative Artificial Intelligence (GenAI) has significantly magnified the level of resources needed in data centers. Since 2012, the computing resources needed to train advanced AI systems is doubling approximately every 3.4 months \cite{openai_compute}. As of early 2026, due to the transition toward trillion-parameter models together with the pervasive deployment of autonomous {\em agentic workflows}, the energy consumption of large-scale AI has moved from an operational challenge to a macro-environmental crisis. While researchers insist on foundational manifestos that chart the course towards "green AI", the rapid growth of needed capabilities requires an immediate architectural revolution that affects {\em IT budgets}, {\em data storage} and global community incentive structures to ensure the {\em sustainability of the AI ecosystem} \cite{Zhao2022}. 

To analyze this landscape, the AI energy footprint must be strictly divided into two categories using a formalized terminology: \textit{Red AI} versus \textit{Green AI}. The term "red AI" refers to any pipeline whose structure requires the use of very large computing resources for the sole purpose of improving performance accuracy by a tiny percentage, thus prioritizing performance over thermodynamic consequences \cite{Barbierato2024}. Conversely, "green AI" systematically imposes efficiency requirements alongside accuracy criteria. To structure this quest for sustainability, the software engineering community has listed robust "green architectural tactics". However, empirical analysis of reference frameworks warns of extreme friction during commercial adaptation; Assessments of more than 160 active open-source ML ecosystems consistently reveal a significant gap between the existence of these ecological tactical protocols and the willingness of AI engineers to integrate them beyond basic performance metrics \cite{Heli2024, Martino2025}.

\subsection{The Global Carbon Cost of Models Training}
Training a dense state-of-the-art Large Language Model (LLM) requires processing tens of trillions of tokens across highly synchronized GPU clusters. A 100,000-GPU cluster consumes more than 100 to 150 Megawatts (MW). Training these dense models structurally requires fragmenting the model state across multiple devices using Pipeline Parallelism (PP) and Tensor Parallelism (TP) (e.g., via Megatron-LM). Recent assessments have proven that arbitrary sizing of these parallel strategies strongly influences the overall efficiency of the cluster. Not optimized tensor parallelism often traps clusters in intensive NVLink synchronization phases, thus consuming a huge amount of inter-GPU static power. Future energy-aware compilers must integrate hyperparameter auto-tuners that calculate exact thermal and synchronization impedance before allocating LLM partition chunks, effectively slashing the hidden network energy cost associated with misaligned topologies \cite{megatron_lm_2019}.

A key operational strategy for green AI training is to mitigate peak fossil-fuel consumption in the electricity grid. Traditionally, schedulers deploy \textit{Suspend/Resume} architectures (completely pausing the cluster and creating checkpoints when green energy drops). However, this severely penalizes the return on investment in equipment. Current state-of-the-art platforms bypass crude suspension entirely in favor of \textit{Carbon Elasticity}. By adapting in real time and transparently the actual dimensions of distributed ML training (by adjusting batch size and number of active workers according to the carbon intensity of the network), elastic AI environments enable additional savings of 37\% to 51\% on the total carbon footprint without suffering crippling synchronization delays. \cite{Hanafy2023}. 

Rather than training AI models with petabytes of highly redundant token distributions, significant power savings are increasingly achieved before VRAM allocation through data-centric AI. Modern evolution-based and data-centric training topologies use "elite sampling" engines to mathematically isolate only hypervariant feature data within a training corpus. By dynamically compressing to only 10\% the footprint of the data used,  data centers completely avoid redundant epoch execution, thus significantly reducing the continuous uptime of the servers by 98\% without substantially penalizing the accuracy of multi-class AI inference \cite{Alswaitti2025, Salehi2024}. 

While large language models is a dominant topic in public discourse, industrial hyperscalers devote the vast majority of their continuous inference energy to running Deep Learning techniques in Recommender Systems (RecSys). Systematic evaluation pipelines allow for rigorous tracing of the Pareto frontier between the accuracy of RecSys clicks and carbon emissions. The evaluation of dozens of distinct algorithmic architectures demonstrates that achieving the last 1\% of recommendation accuracy frequently results in a penalty of roughly 40\% in carbon emissions. By natively mapping these boundaries, engineers can mathematically select slightly smaller matrices that drastically reduce the company's operational carbon emissions \cite{SPILLO2026101286}. Finally, a major aggravating factor in GenAI's carbon crisis is the low sharing of academic and corporate artifacts. When training large language models specifically designed for software engineering, the initial energy expenditure is noticeable. Large-scale literature reviews reveal that nearly 40\% of research institutions and companies do not share their compiled artifacts or model weights, forcing researchers to rerun redundant and extremely computationally intensive training paradigms \cite{Hort2023}.

\subsection{What About Inference ?}
While pre-training dominates headlines, hyperscaler telemetry from 2024–2026 indicates that \textbf{Inference} now accounts for the vast majority of the overall AI energy footprint. It is very important to characterize and understand the differences between {\em training} and {\em inference} power consumption patterns \cite{Patel2024}. Given that Service Level Objectives (SLOs) \cite{Shen2023}  for public LLM endpoints require extremely low latency targets, commercial clusters often keep a large number of GPUs at high-voltage during idle states (because wake-up overhead could impact performance), just waiting for the next avalanche of prompts \cite{Yu2023}. To resolve this ongoing energy drain, automated environment orchestrators (e.g., DynamoLLM) use massive search space formulations to continuously and asynchronously reconfigure dynamic LLM instances. By performing automatic and real-time adjustments of {\em cluster dimension}, {\em clock frequency}, and {\em tensor model parallelism} in response to fluctuations in inference queues, modern streaming platforms manage to save up to 50\% of their total operational carbon footprint without ever violating underlying web latency {\em service level objectives} (SLOs). \cite{Stojkovic2025}.

Furthermore, recent sophisticated energy profilers unequivocally demonstrate that the power consumption profile of inference cannot be considered as uniform. Processing an LLM typically follows two phases: the \textit{prefill phase} and the \textit{decode phase}. The \textit{prefill phase} (which processes the user prompt) mainly involves compute-bound matrix-matrix multiplications, which therefore instantly puts the GPU cores at their highest voltage \cite{Kakolyris2025}. Conversely, the \textit{Decode phase} (which generates tokens one-by-one) translates into strictly memory-bandwidth bound operations, leaving the ALU unsolicited while wasting a huge amount of static high-bandwidth memory (HBM) energy. New-generation inference schedulers proactively apply decoupled DVFS policies based on tokenization state, maximizing clock frequencies during the {\em prefill phase} and significantly reducing logic voltages during the {\em decode phase} to reduce inference carbon by more than 25\% \cite{Tian2026, Stojkovic2025b}.

To prevent hallucinations and access real-time corporate data, enterprises primarily deploy LLMs via Retrieval-Augmented Generation (RAG) pipelines. From an energy perspective, a RAG pipeline creates a severe infrastructure strain: before the LLM generates a response, the system must execute an {\em approximate nearest neighbor} (ANN) search across a massive vector database, which involves a large pool of compute nodes that continuously execute memory-intensive tasks. Integrating these retrieved documents into the LLM significantly expands the context window. Maintaining the Key-Value (KV) Cache in the GPU's VRAM prevents the model from recomputing tokens, but consumes massive dynamic power. A massive paradigm shift revealed in early 2026 directly links this caching physics to external grids through \textit{Carbon-Aware Prompt Caching}. Because the network is heavily powered by fossil fuels, instead of blindly discarding cached requests via generic LRU algorithms, sophisticated broadcast platforms artificially extend the retention lifetime of massive request matrices directly in VRAM,  thus ensuring that subsequent identical requests retrieve pre-computed tokens to physically avoid massive, carbon-emitting redundant recalculations \cite{Kakolyris2025, Tian2026}.

The most advanced AI paradigms rely on {\em autonomous agents} and {\em multi-step reasoning models}. Rather than directly answering a prompt, an agent breaks it into sub-tokens, perform searching (e.g., over the web), generates and runs intermediate codes, performs self-evaluation, and so on (iterates). This introduces a hidden \textbf{Recursive Energy Multiplier}. A single user interaction might trigger 50 background inference calls before a final answer is produced. Advanced 2026 architectures (such as the \textit{PEARL} framework) explicitly perform Performance/Energy-Aware Routing for LLMs. By dynamically intercepting prompts and distributing them asymmetrically between massive monolithic models and lightweight edge models — evaluating semantic complexity in real-time along with active clusters power — orchestrators maintain optimal reasoning accuracy while efficiently reducing the recursive energy multiplier \cite{Antepara2025, CHADLI2026108218}.

\subsection{Evolution of Compression in Green AI}
Historically, the primary constraints in Deep Learning were {\em execution time} (latency) and the {\em amount of GPU VRAM}. To deploy increasingly large models, algorithmic compression techniques have been developed. Initially designed to minimize \textit{time-to-solution} through the corresponding hardware speed-up, these techniques were subsequently redesigned to address \textit{energy-to-solution}. The fundamental principle underlying this adaptation is the physics of silicon: reading data from HBM memory consumes much more joules than the actual ALU calculation. 

However, each new generation of techniques introduces serious technical limitations and energy trade-offs.

\textbf{Pruning and Network Compression:} Earliest compression techniques consisted of algorithmically identifying and removing redundant weight matrices. By removing certain parameters, the total number of multiplication-accumulation (MAC) operations decreases. However, early "unstructured" pruning randomly removed weights, thus creating a \textit{Sparsity Trap}. The systolic arrays of modern GPUs are highly optimized for dense matrix operations. An unstructured sparse matrix forces the GPU to load zeros into memory and multiply by zero anyway, which does not allow for any energy savings on modern accelerators.

\textbf{Knowledge Distillation:} Distillation trains a lightweight "Student" model to mimic a large "Teacher" model. Replacing, for example, a 1000W cloud inference query with a 5W Edge NPU query on a model with 8 billion parameters reduces operational energy consumption per token by nearly 200 times. However, distillation only shifts the energy load backwards, requiring a huge initial carbon investment to execute the {\em Teacher}'s forward passes, which often requires millions of Edge inferences to compensate.

\textbf{Neural Architecture Search (NAS):} Hardware-aware Neural Architecture Search (HW-NAS) injects physical constraints directly into the reward function to discover ideal silicon topologies. However, the search process evaluates thousands of permutations, emitting carbon equivalent to that of five standard cars \cite{strubell2019energy}. To overcome this problem, modern frameworks rely on \textit{training performance estimation} (TPE), using early stop variants to abruptly interrupt training as soon as energy thresholds diverge from accuracy slopes, thus recovering up to 90\% of the search phase energy \cite{DONG2023100926}. Beyond simply rejecting parameters, NAS pipelines actively replace fundamental mathematical operations. Advanced architectures assign custom hardware-friendly inference operators (such as deep-shift or multiplication-free kernels) directly into the DNN structure layer by layer, thus outperforming comparable mixed-precision algorithms by prioritizing the simplification of operations  \cite{Nasrin2022}.

\textbf{Precision Scaling and Quantization:} Instead of removing parameters, quantization reduces their bit-width. The energy cost of arithmetic (resp. memory)  operations scales quadratically (resp. linearly) with bit-width. Recent hardware advances have introduced precision-tailored architectures (e.g., 5.3-bit Block Floating Point formats) that definitively conquer the historic sub-8-bit dynamic range limitations \cite{Sadi2025}. However, as the community attempts to push quantization to extreme limits, a structural limitation in the representation arises. Disruptive 2025 frameworks such as \textit{LUT-DLA} break this limit by substituting scalar computation with vector-quantized Look-Up Tables (LUT), resulting in an impressive area efficiency gain of $\approx 146\times$ \cite{LiGuoyu2025}.

\textbf{Mixture of Experts (MoE) and Speculative Decoding:} Recognizing the limits of extreme quantization has led to considering the alternative of {\em dynamic sparsity} at runtime. MoE uses a router to activate only a sparse subset of "Expert" sub-networks per token, thereby decoupling the total storage capacity from the active compute footprint. However, this creates a \textit{VRAM Leakage Paradox}: Although MoE reduces the power consumption of the dynamic ALU, it requires the entire massive model to reside in the HBM, resulting in significant static power loss, thus greatly limiting its effectiveness on power-constrained deployments.

\textbf{Approximate Accelerators for Transformers:} For spatial mapping engines such as Vision Transformers (ViTs), general approximate computing has shown a negative impact on deep-learning convergence. Advanced methodologies (e.g., the \textit{TransAxx} framework) now securely insert approximate arithmetic multipliers directly into ViT execution paths. Using {\em Monte Carlo Tree Search} (MCTS) algorithms allow to efficiently search the space of possible configurations to perform Vision Transformer (ViT) with structurally defective mathematical chips, thereby enabling an noteworthy reduction in hardware power consumption while preserving visual quality  \cite{Danopoulos2025}.

\subsection{Decentralization: Fog and Mobile Edge}
The ultimate optimization of the infrastructure is to offload heavy computing loads from {\em centralized hyperscalers} to intermediate {\em Mobile Edge Computing} (MEC) and {\em Fog} platforms. Handling the extreme mobility of nodes and decentralized energy ceilings requires replacing hierarchical managers with complex algorithmic coordination. The use of rigorous game theory approaches to offload computation to the mobile cloud allows heterogeneous devices with limited autonomy (battery-powered) to act as autonomous economic agents \cite{LAMBERT2026107968, BELGHACHI2026108452, FUSCO2026108194}. By mathematically modeling the significant penalty of transmission bandwidth delays compared to the relief provided by offloading localized computations \cite{Xu2021}, MEC architectures routinely calculate localized Nash equilibria \cite{Zaw2021}. This structurally prevents decentralized systems from saturating intermediate cloud proxies with computations, thus ensuring optimal and uniform battery survival margin across thousands of highly asynchronous edge appliances \cite{Gai2021}. 

Small language models (SLMs), typically less than 10 billion parameters and optimized by quantization, allow efficient on-device inference on consumer platforms. Empirical lifecycle case studies validate this transition, demonstrating that migrating generative AI inference from centralized cloud infrastructures directly to edge platforms can generate energy savings of over 90\%, while significantly reducing water consumption and carbon footprint through the elimination of network and cloud cooling costs \cite{Tang2025, Li2025, LiPengfei2025}. Expanding Edge capabilities requires that local mobile clusters not only run inferences, but continuously customize models via {\em parameter-efficient fine-tuning} (PEFT) mechanisms, such as {\em low-rank adapting} (LoRA), physically bypassing full gradient weight updates to safely retrain models without inducing thermal throttling.

Federated Learning (FL) prevents centralized data concentration at the expense of a detrimental "wooden barrel effect" regarding carbon emissions. Bidirectional transmission of gigabytes of gradient updates over high-frequency RF networks typically consumes much more radio power than localized computations \cite{Zhang2025, LUO2026108351, Xia2024}. To address this problem in a targeted way, orchestrators perform continuous monitoring based on game theory (e.g., via MARL controllers) to adjust the inclusion of edge-client parameters. Frameworks like CLOVER replace random client inclusion with rigorous participant selection based on their carbon footprint. By mathematically evaluating the instantaneous carbon states of the local power grid alongside the computing speeds of the devices, the orchestrators selectively collect data that forces global convergence while artificially reducing FL carbon emissions by 25\% \cite{Cho2024, YANG2023178}. 

Historically, carbon-aware spatial shifting was strictly limited to the Cloud. Innovative research conducted in mid-2025 establishes the effectiveness \textit{Mesoscale Spatial Shifting}. Architectures such as the \textit{CarbonEdge} framework dynamically track localized variations in "mesoscale" carbon-intensity within interconnected geographic regions. By intelligently micro-migrating CDN workloads between geographically adjacent peripheral hubs with low transient carbon differences, modern orchestrators are able to achieve emission reductions of more than 78\% without triggering latencies of more than 5 milliseconds \cite{WuLi2025}.

\subsection{Main Trends in Green AI}
By analyzing the rapidly expanding field of green AI ($45$ papers), the bibliometric trend reflects the commercial pivot from {\em model creation} to {\em model deployment}. While pre-training of massive reference models initially dominated the discourse, currently only 27\% of our selected papers target optimizations of the training phase (e.g., {\em Carbon Elasticity}, {\em Data-centric pruning}, ...). Conversely, 42\% of recent studies explicitly target inference optimizations, encompassing {\em RAG KV-Cache management}, {\em Agentic routing}, and {\em algorithmic compression} (MoE, Quantization). The remaining 31\% addresses Decentralization, heavily exploring Federated Learning limits and Small Language Models (SLMs) on Edge devices. This confirms that the main challenge of current research (2025 and 2026) is to minimize the continuous and ubiquitous energy consumption of GenAI services.

\begin{figure}[pos=htbp]
    \centering
    \begin{tikzpicture}
        \begin{axis}[
            ybar,
            width=\columnwidth,
            height=4cm,
            enlarge x limits=0.3,
            ylabel={\% of Surveyed Papers},
            xtick={1,2,3},
            xticklabels={Inference \& RAG, Edge \& FL, AI Training},
            nodes near coords,
            nodes near coords align={vertical},
            bar width=20pt,
            ymin=0, ymax=55,
            ]
            \addplot[fill=orange!70] coordinates {(1,42) (2,31) (3,27)};
        \end{axis}
    \end{tikzpicture}
    \caption{Quantitative overview of Green-AI literature.}
    \label{fig:trend_ai}
\end{figure}
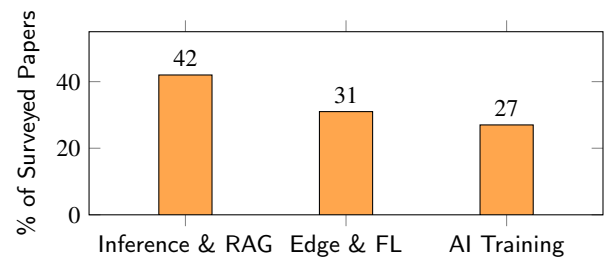

\section{Advanced Cooling Systems}
\label{sec:cooling}

As time goes on, the abstractions of software scheduling and algorithmic compression eventually run up against the unavoidable laws of thermodynamics. All the electrical energy consumed by a computer system is eventually dissipated as heat. Historically, data center cooling was considered a secondary issue at the facility level. Today, the thermal density of Exascale and GenAI workloads has transformed cooling into the primary bottleneck, determining whether a processor can keep running at its peak FLOPs.

\subsection{Air Cooling and Rack Density}
The thermal capacity of air is fundamentally insufficient for modern high-density architectures. Traditionally, the power consumed by {\em Computer Room Air Conditioning} (CRAC) units represents up to 40\% of a facility's total energy. To combat this via software, conventional "Thermal-Aware Schedulers" proactively balanced CPU loads spatially across different racks to eliminate physical hot spots, thus allowing CRAC compressors to safely step down their power levels. However, the shift from provisioning for a single server to that of large AI infrastructures has shattered these parameters at the rack-scale. 

A single NVIDIA GB300 (Blackwell Ultra) NVL72 rack integrates 72 GPUs and 36 CPUs, drawing more than 120 kW in a single physical cabinet. Trying to cool a 120 kW rack by pure ventilation would require localized winds of exceptional magnitude, which would create acoustic damage and consume more energy for the fans than for the processors. Consequently, the industry has unanimously recognized that for any rack exceeding roughly 35 kW, air cooling is no longer a viable solution.

\subsection{Advanced Liquid Cooling Paradigms}
To manage the massive heat flows associated with AI processing, facilities have turned to liquid cooling, which, compared to air, has thermal capacities up to 4 times greater and thermal conductivities $20$ to $100$ times higher.

\textbf{Liquid Cold Plate and the Temperature Paradox:} D2C (or Cold Plate) technology circulates a cold liquid directly over micro-convective blocks mounted on CPU, GPU, and memory chips. To maximize facility-level PUE, data centers increasingly supply "hot water" ($18^{\circ}C$ to $25^{\circ}C$) rather than chilled water, thereby significantly reducing chiller plant power. However, this introduces a serious systemic paradox: the use of warmer inlet water reduces the energy consumption of cooling but inadvertently causes temperature spikes in the silicon chips. This physically exacerbates the {\em leakage current} of the transistors and suddenly triggers a localized DVFS limitation, thus introducing serious execution delays in synchronized MPI workloads that often wipe out the initial cooling energy savings \cite{Conficoni2016}. 

\textbf{Intelligent Cooling Controllers and Joint Thermal-Workload Orchestration:} Traditionally, computer room air conditioning units (CRAC units) operated from static thresholds unrelated to server states. Nowadays, temperature is recognized as an active variable that has a direct impact on hardware transistor leakage. Systematic assessment of \textit{temperature-aware power models} demonstrate that failing to account for the increase in ambient temperature of the incoming air significantly distorts theoretical CPU power consumption predictions. This has led to the deployment of "Joint Data Center Cooling and Workload Management". By merging these logical flows using deep reinforcement learning (DRL), schedulers algorithmically assign high-intensity thermal threads to nodes that exhibit optimal cooling airflow, thereby reducing the overall cooling power of the installation independently of the hardware's P-states \cite{Yang2019, KAHIL2025125734, MIRHOSEININEJAD2020174}.

\textbf{Single-Phase/Two-Phase Immersion:} The most extreme cooling method involves fully immersing the server's motherboard in a dielectric (non-conductive) fluid. In {\em two-phase immersion}, the fluid boils directly upon contact with the silicon chip, exploiting the latent heat of vaporization to achieve unprecedented thermal densities with near-zero pumping power, thus enabling PUEs close to 1.01.

\textbf{Critical Limitation (The PFAS Regulation Crisis):} While two-phase immersion was considered the future of green data centers, it has collided with a recent regulatory hurdle. The dielectric fluids initially required for boiling (e.g., 3M's Novec) are primarily \textit{perfluoroalkyl} and \textit{polyfluoroalkyl substances} (PFAS) — "forever chemicals" exhibiting very high environmental and biological toxicity. Due to severe restrictions and bans from the European Union and the US Environmental Protection Agency (EPA) on PFAS manufacturing, the viability of two-phase immersion has plummeted, forcing hyperscalers to revert to highly complex D2C water configurations. \cite{epa_pfas}.

\subsection{Water and Power Usage Effectiveness}
Optimizing for {\em power usage efficiency} (PUE) raises a major environmental paradox regarding water consumption. Modern data centers often achieve exceptional PUE not through refrigeration compressors, but rather through evaporative cooling towers. The development of very high-density AI clusters requires enormous quantities of chilled water, which constantly evaporates into the atmosphere to release the heat dissipated from the data center.

Merely estimating the raw volume of Water Usage Effectiveness (WUE) does not take into account the geographic hydrology. To address this critical gap, modern supercomputing assessments (such as the ThirstyFLOPS framework \cite{Jiang2025}) introduce the \textit{Adjusted Water Impact (AWI)} indicator. The AWI takes into account the high spatial and temporal volatility of local water stress, demonstrating that a supercomputer consuming 1 million liters of water in the rainy Pacific Northwest region is ecologically harmless, while the same consumption in an arid environment would cripple local public infrastructure. Thus, modern green planning advocates actively anticipating periods of inactivity to offset intensive processing only during droughts \cite{Jiang2025}. In water-stressed regions, municipalities are increasingly refusing building permits for liquid-cooled data centers, despite their low carbon footprint. Therefore, the use of closed-loop "dry coolers" is now mandatory, which inherently leads to higher consumption of electrical energy (thus degrading the PUE) but saves water.

\subsection{Heat Recovery and Energy Reuse}
As ICT-related electricity needs approach 4\% or 5\% of global demand, governments are no longer viewing data centers as mere consumers, but rather as potential localized thermal power plants. The focus has shifted from minimizing cooling overhead to maximizing the \textbf{Energy Reuse Factor (ERF)}. A high heat output (e.g., a $60^{\circ}-80^{\circ}C$ liquid from the cooling of an Exascale system) is potentially very valuable. Thanks to heat exchangers, this wasted energy can be used for external infrastructures:

\textbf{District Heating:} Supercomputers such as LUMI (Finland) integrate their liquid cooling loops directly into Kajaani's district heating network, thus taking advantage of fossil fuel combustion to heat thousands of local residential homes during the winter. 

\textbf{Distributed Edge Heating (e.g., Qarnot Computing):} Innovative European architectures completely decentralize data centers by installing cloud-based computing nodes (viewed here, from a thermal perspective, as electric radiators) directly within residential buildings. This "computing-based heating" approach achieves near 100\% energy efficiency, thus offsetting the cooling costs of the facilities and that of associated infrastructure.

\subsection{Oversizing and Simulation-Driven Design}
In the exascale ecosystem, the evaluation of millions of active sensor parameters reveals a rigid limit of data centers: facilities tend to have more installed computing devices than the electrical grid can run simultaneously. The operation of these legally "oversized" systems requires the direct integration of neural networks into the infrastructure of the facilities. Frameworks such as the \textit{Adaptive Power Management System (DPS)} explicitly map workloads in real time via ML analysis, securely transferring power credits from idle peripheral logic cores exclusively to critical processing queues in order to keep peak Exaflops below regulatory power thresholds \cite{Ding2023, Patki2025}. 

An embarrassing reality arises from the mismatch between the rapid evolution of equipment and the long life cycles of infrastructure. A physical data center is amortized over 15 to 20 years, while new generations of AI accelerators are produced every 12 to 18 months. Upgrading a 2015 air-cooled facility to accommodate a 120 kW liquid-cooled Blackwell rack (planned for 2026) is structurally and financially prohibitive. Future next-generation AI workloads demand bespoke "AI factories." Designing these extreme thermal infrastructures precludes any empirical approach in favor of formal methods. Global hyperscalers are increasingly relying on holistic operational simulators, a major one being the \textit{Carbon Explorer} framework. This topology integrates the interaction between local geography, PUE degradation, and daytime renewable energy to mathematically advise on structural deployment trade-offs between battery CapEx and direct utility energy consumption, thus enabling early exploration of sustainability-focused design space even before concrete realizations  \cite{Acun2023}.

\section{Power-aware Computing Perspectives}
\label{sec:future}

The shift from a paradigm focused on absolute performance to one that considers strict energy constraints has highlighted corresponding conceptual gaps in modern computing. On the cusp of the exascale era and with the expansion of generative AI, addressing the energy crisis requires moving beyond iterative software optimization alone. Based on critical limitations identified across the Cloud-Edge-HPC continuum, we have outlined five major and disruptive future directions encompassing {\em hardware, software telemetry, supply chain auditing}, and {\em global legislation}. These aspects will shape the future of computing through 2030.

\subsection{Advanced Measurement: OS and On-chip}
As highlighted in Section \ref{sec:tools}, current high-frequency monitoring relies heavily on user-space polling or kernel-level interrupts. This induces the "observer effect": continuously polling hardware registers, such as MSRs, frequently triggers context switches, thus {\em saturating CPU caches}, {\em interrupting AI workloads}, and {\em artificially amplifying node-level power consumption}, simply to measure it. For large-scale telemetry with no overhead, the community is rapidly turning to the following two fundamental paradigms.

\textbf{Kernel-Bypass via eBPF:} Extended Berkeley Packet Filter (eBPF) technology (e.g., CNCF Kepler) allows researchers to securely integrate event-driven polling programs directly into the Linux kernel, without requiring custom kernel modules. Currently, frameworks using eBPF represent the core of the state-of-the-art for containerized cloud deployments. By intercepting OS-level events, eBPF architectures accurately map the energy footprint of highly ephemeral microservices and serverless FaaS workloads that cannot be detected by traditional MSR polling, thus eliminating blind spots in energy consumption measurements of modern virtualized grids \cite{Qiao2025}.

\textbf{In-Network Telemetry via DPUs:} To eliminate the monitoring task at the host CPU, Data Processing Units (DPUs) and SmartNICs (e.g., NVIDIA BlueField, AMD Pensando) are absorbing the telemetry stack. DPUs aggregate, compress, and stream time-series data directly into facility management pipelines via a dedicated PCIe path, thereby ensuring that host CPUs are exclusively dedicated to computation.

\subsection{In-Memory Computing and Hardware Co-Design}
For decades, von Neumann architectures have separated the arithmetic logic unit (ALU) from main memory. In GenAI inference paradigms, transferring terabytes of LLM-related data via PCIe buses and memory interconnects consumes roughly two orders of magnitude more energy than that of the arithmetic operations. Next-generation hardware must clearly eliminate this "memory wall".

\textbf{Memory Disaggregation via Compute Express Link:} Historically, administrators oversized local server RAM to prevent AI workloads from crashing (due to lack of memory), resulting in terabytes of unused but powered DRAM throughout the facility. Currently, the transition to the CXL interconnect standard enables \textit{memory disaggregation}, grouping petabytes of memory into distributed chassis that are dynamically accessible by any processor via PCIe logic. Decoupling memory allocations from physical hosts safely eliminates current leakage due to "unused memory," thereby reducing the ongoing carbon footprint of RAM (integrated and operational) in a data center by more than 25\% \cite{Shen2025, Ifath2026}.

\textbf{Near-Data Processing (NDP):} Processing-in-Memory (PIM) and in-storage processing (ISP) mitigate data movement by offloading the instructions to the cores integrated with the main memory level. PIM architectures (e.g., UPMEM DIMMs and HBM-PIMs) fundamentally reverse the conventional hardware paradigm by moving the ALU directly inside the DRAM arrays, thereby reducing parasitic capacitance energy from the memory buses \cite{Chen2023}. However, a significant loss of efficiency occurs during the conversion of memory voltages. Representing the elite architectural limits, the systems now utilize Analog-Mixed-Signal (AMS) capacitive coupling directly inside standard 6T and 8T SRAM matrices. The close association of XNOR multipliers with shared in-memory {\em analog-to-digital converters} (ADCs) \cite{Chang2025} enables remarkable advances in the computation of binarized neural networks, achieving macroscopic efficiencies of nearly 490 TOPS/W through the complete elimination of column-dense read paths \cite{Yang2025}. 

\textbf{Chiplets and Heterogeneous Integration:} Instead of manufacturing massive monolithic chips prone to significant current leakage, silicon is fragmented into "chiplets" using the UCIe (Universal Chiplet Interconnect Express) standard. This allows manufacturers to combine high-performance logic components etched at 2 nm with I/O and analog components etched at 6 nm, thus offering more concentrated power and more stable consumption. Architectural carbon footprint modeling frameworks, such as the \textit{Architectural $CO_2$ Footprint Tool (ACT)}, \textit{ECO-CHIP}, and \textit{3D-Carbon}, empirically prove that the migration to {\em heterogeneous integration} (HI) strategies reduces embodied carbon emissions by 30\% compared to traditional monolithic systems \cite{Gupta2023, Sudarshan2024, Zhao2024}.

\textbf{Wafer-Scale Systems and Carbon Dependency:} While the traditional industry is turning to UCIe chiplets, the extreme counter-paradigm involves monolithic \textit{wafer-scale integration (WSI)}. Architectures like the Cerebras Wafer-Scale Engine (e.g., the CS-3) completely abandon chiplets, sculpting an entire supercomputer from a single continuous silicon wafer. Recent life cycle assessments of the {\em total Carbon-Delay Product} (tCDP) have empirically concluded that, despite the huge manufacturing challenges of wafer-scale processors, their complete elimination from off-chip diffusion dies generates a design space with uniquely optimized carbon performance \cite{Golden2025}.

\textbf{Photonic computing:} Replacing copper interconnects with photonic (light-based) data buses significantly reduces energy consumption related to static impedance, thus overcoming memory bandwidth limitations. Next-generation frameworks (such as the open-source tool \textit{EPiCarbon}) demonstrate that photonics far surpasses traditional semiconductors structurally: manufacturing the hardware requires at least $4.1$ times less manufacturing energy (embedded carbon) while offering significantly higher yields than equivalent-sized CMOS chips \cite{Fayza2025}. By considering moderate and aggressive photonic scaling, frameworks such as Albireo and ROCKET show that EDP and Energy can be reduced by at least $229 \times$ when compared to state-of-the-art electronic accelerators and GPUs on AI workloads \cite{Shiflett2021, ZhangHao2025}.

\textbf{Neuromorphic Computing vs Analog Computing:} For edge endpoints, pulsed neural networks mimic the biological brain by performing calculations only during discrete events. This enables the design of true "event-driven silicon", reducing standby power consumption to a few microwatts \cite{Sanni2019}. Furthermore, analog computing natively processes deep neural networks in the continuous load domain. Recent advances in mixed-signal and charge-based multiply-add cores allow matrix algebra to be run at the fundamental thermal noise limit, thue leveraging large onboard capacitor arrays to routinely reduce computing power consumption by more than 37\% compared to equivalent dense digital logic paths.

\subsection{Operational  Carbon vs Embodied  Carbon}
Until now, the green software community has primarily focused on \textit{operational carbon} (emissions mainly related to operational electricity consumption). However, with the rapid decarbonization of power grids, the operational footprint of a cluster ideally tends toward zero. Consequently, the predominant environmental impact of computing is shifting toward \textit{embodied carbon}, which corresponds to the significant greenhouse gas emissions associated with {\em semiconductor manufacturing}, {\em rare earth mining}, {\em server production}, and {\em transportation activities}\cite{Wu2024, Eeckhout2023}. 

Life cycle assessment (LCA) studies unequivocally state that the exponential proliferation of custom AI architectures necessitates a unified Pareto LCA frontier; hyperscale environments must quantitatively balance the operational carbon savings of an ASIC with the much larger carbon footprint required for semiconductor manufacturing. Replacing server arrays every two to three years structurally prevents data centers from achieving carbon neutrality, regardless of their power usage efficiency (PUE) \cite{Eilam2024, Mersy2025}.

\textbf{Lifecycle-Aware Software and System Sustainability:} Software overload and unoptimized libraries inevitably lead to premature hardware obsolescence. We anticipate a paradigm shift toward \textit{software lifecycle engineering}, where software is explicitly optimized to maintain backward compatibility, thereby extending the operational lifespan of a data center's IT equipment from 3 to 7 years \cite{Ayoola2025, Wang2025}. To address this problem structurally, the software engineering community is mandatorily integrating sustainability across four key dimensions: environmental, economic, social, and technical. Integrating global sociological and economic constraints directly into developers' workflows ensures that they prioritize compatibility with moderately innovative architectures by deliberately limiting algorithmic feature extensions that absolutely require next-generation hardware.

\textbf{Analytical Modeling and Rebound Effects:} Establishing rigorous parameterized baselines (such as the First-Order analytical CArbon modeL (FOCAL) \cite{Eeckhout2025}) inherently isolates rebound effects (the Jevons Paradox). In-depth assessments clearly highlight that when cloud instances achieve significant efficiency gains, the absolute capacity of data centers tends to increase primarily to absorb the financial benefit associated with the energy margin. New analytical criteria make it possible to decouple this cycle by mathematically defining the strict limits necessary to transform "weakly" sustainable operational accelerators into closed-loop and strictly \textit{controlled life cycle assessments} (LCAs), thus guaranteeing overall societal reduction margins in relation to the expansion of digital consumption habits \cite{Elgamal2025, Panteleaki2025, Halder2025, Roelandts2025, MENCARONI2025146787}.

\subsection{Compliance with Standards and Regulations}
The development of green IT is driven more by regulatory incentives than technological ones. Sustainable IT will now move from a voluntary initiative, driven by corporate social responsibility (CSR), to a globally audited regulatory compliance framework.

\textbf{Macro-Policy and Granular Reporting:} Under the European Energy Efficiency Directive (EED), data centers with computing power exceeding 500 kW are legally required to publish comprehensive sustainability indicators. In parallel, the {\em Corporate Sustainability Reporting Directive} (CSR) subjects IT companies to rigorous emissions audits related to {\em scopes 1, 2, and 3} \cite{eu_csrd_2022}. Current corporate sustainability reports fundamentally obscure critical architectural realities beneath broad overviews. In-depth analyses of computer hardware reveal a growing and unreported disparity: the absolute carbon footprint of local integrated circuits is increasing independently year after year. To put an end to this "greenwashing", future compliance initiatives must require extremely accurate reporting of hardware components \cite{Sudarshan2024c, Peltonen2025}.

\textbf{The EU AI Act and Ethical Prompts:} Specific to the software field, the final clauses of the {\em European AI Act} require creators of general-purpose AI (GPA) to explicitly document the energy consumption and overall ecological footprint associated with training their fundamental models \cite{eu_ai_act_2024}. According to our vision of the 2030 Sustainable Development Roadmap, governance is undeniably shifting from training centers to the user interface. New paradigms of "ethical prompt engineering" (EPE) integrated into interactions impose auditable compliance controls specific to recovery architectures. EPE structurally limits unrestricted LLM requests by treating user prompts as versioned constraints specifically designed to block the execution of resource-intensive or redundant models before they trigger the hardware logic layer \cite{Migliarini2026, Cruz2025}.

\textbf{Digital Sustainability Labels:} The criteria for acquiring computer equipment are crystallizing around ecolabels (e.g., EPEAT and Energy Star). In order to fill this regulatory gap at the software level, international consortia have published the "Green Software Measurement Model" (GSMM) \cite{GULDNER2024402}. This {\em validated reference model} standardizes how auditors empirically and dynamically extract resource efficiency gains from isolated software products. Furthermore, running models like the GSMM enables the imminent transition to \textit{software sustainability labels}; European and international legislation now requires enterprise software suites to actively calculate and display, in real time, "green labels" similar to energy labels for appliances \cite{Sudarshan2024c}.

\textbf{Economic Catalysts and Carbon Market Exchanges:} To enforce global thermodynamic limits, providers are natively integrating energy-efficient algorithmic structures into their middleware. Emerging cloud brokers are deploying fully automated \textit{carbon trading data markets}. Drawing inspiration from Nash auction systems, geodistributed Kubernetes clusters actively trade and bid for prioritized execution windows, exclusively during periods of high renewable energy production (e.g., local solar power peaks). By monetizing carbon allowances, the hypervisor infrastructure inherently prevents network capacity overflows and incentivizes edge users to dynamically suspend lower-priority tasks in exchange for direct financial dividends, thus realizing "green AI" through a rigid market capitalization \cite{Georgiou2015, Wu2025, LI201763}. Furthermore, large-scale operations conducted onboard the \textit{Fugaku supercomputer} have deployed robust and operational "incentive-based energy efficiency programs"  \cite{Ana2024}. By directly returning to the teams of supercomputers, who voluntarily activated deeper P states in their scientific codes, a digital currency corresponding to their exact use, the system administrators have concretely validated that gamification on an institutional scale effectively redirects collective behaviors in order to optimize planetary impact.

\textbf{Microgrids and Wind Energy Deployments:} The development of zero-carbon integrations requires the dismantling of traditional facilities. To address the drastic restrictions on distribution lines that are hindering the global energy transition to wind power, the macroeconomic frameworks of July 2025 advocate the dynamic deployment of high-performance data center equipment directly within local wind farms \cite{KUMAR2026116664, Kilian2025}. Placing the AI training logic close to the power generation source acts as a highly elastic storage buffer. This proximity configuration systematically absorbs and modulates the raw transient power well before it is transmitted over high-voltage municipal power lines, thus minimizing extreme losses due to voltage conversion and significantly increasing the overall capacity factors of the power grid. Operators are deploying architectures such as \textit{PowerMorph} as part of demand management contracts. By precisely manipulating the electrical footprint of QoS-aware servers to the microsecond, the data center acts as an electrical regulation well, rapidly stabilizing the AC frequency of the regional electrical grid for significant financial gains. \cite{Jahanshahi2022}. Within the framework of pan-European projects (\textit{Green.Dat.AI}), strict structural "data spaces" harmonize semantic datasets well before the activation of deep learning architectures \cite{Chrysakis2025}.

\section{Conclusion}
\label{sec:conclusion}

Over the past decade, energy consumption has shifted from a secondary operating expense to the primary physical, financial, and environmental bottleneck in the computing ecosystem. As this comprehensive survey demonstrates, the challenge of energy-aware computing is ubiquitous across the entire cloud-edge-HPC continuum. From low-power microcontrollers performing milliwatt-scale inferences to exascale supercomputers of hundred-megawatt AI data centers, the laws of thermodynamics impose absolute limits on performance scaling.

The rapid proliferation of generative artificial intelligence, LLM inference pipelines, and agentic AI workflows has accelerated the energy crisis. The traditional response, which relies solely on semiconductor manufacturing to produce low-power transistors, is no longer sufficient. Genuine sustainability requires a comprehensive and cross-cutting overhaul of the entire ecosystem.

We have established that cutting-edge chips rely on extreme hardware heterogeneity, memory disaggregation (CXL), and drastic precision reduction (FP4/LUT) to maintain AI throughput. However, to measure this efficiency, the software community must overcome virtualization-related hurdles by deploying overhead-free eBPF polling tools, predictive tracing models for transformers, and reproducible carbon tracking that avoids the standard overhead of memory jump registers (MSRs). Furthermore, static code optimizations and energy-efficient compilation must be tightly integrated with dynamic schedulers. These orchestrators require advanced particle swarm optimization and deep reinforcement learning techniques to seamlessly ensure workload consolidation, C-state latency balancing, and energy-efficient spatiotemporal shifting. Finally, as air cooling reaches its absolute physical limits, liquid-to-chip infrastructure, intelligent DRL HVAC controllers, and waste heat recovery (energy reuse factor) must be standardized, thus bridging the severe gap between 15-year facility lifecycles and 18-month GPU iterations.

As is widely documented in the field of sustainable digital infrastructure, addressing the energy crisis requires viewing computing limits as a multidisciplinary continuum. The transition to a net-zero carbon architecture necessitates abandoning the historical criterion of raw performance at all costs. By leveraging renewable sources intrinsically linked to dynamic planning, establishing rigorous transparency to combat the rebound effects of embodied carbon, and directly integrating environmental indicators into global legislation (e.g., the European AI law and CSRD mandates), the research and industry communities can successfully transition to the era of exascale AI while respecting the energy constraints of our time \cite{VERDECCHIA2022100767}.




\bibliographystyle{model1-num-names}

\bibliography{cas-refs}

@INPROCEEDINGS{Shin2024,
  author={Shin, Woong and White, James B. and Elwasif, Wael and Da Silva, Rafael Ferreira and Zimmer, Christopher and Messer, Bronson and Budiardja, Reuben and Georgiadou, Antigoni and Vergara, Verónica Melesse and Lange, Jack and Maiterth, Matthias and Osborne, Tim and Huk, Leah and Holmen, John and Hagerty, Nick and Karimi, Ahmad Maroof and Naughton, Thomas and Adamson, Ryan and Prout, Ryan and Wang, Feiyi and Atchley, Scott and Thach, Kevin G. and Beck, Thomas and Oral, Sarp},
  booktitle={Proc. SC '24 Workshops}, 
  title={Towards Sustainable Post-Exascale Leadership Computing}, 
  year={2024},
  volume={},
  number={},
  pages={1790-1794},
  doi={10.1109/SCW63240.2024.00225}}

@INPROCEEDINGS{Shiflett2021,
  author={Shiflett, Kyle and Karanth, Avinash and Bunescu, Razvan and Louri, Ahmed},
  booktitle={ACM/IEEE 48th Annu. Int. Symp. Comput. Archit. (ISCA)}, 
  title={Albireo: Energy-Efficient Acceleration of Convolutional Neural Networks via Silicon Photonics}, 
  year={2021},
  volume={},
  number={},
  pages={860-873},
  doi={10.1109/ISCA52012.2021.00072}}

@ARTICLE{ZhaoYingnan2024,
  author={Zhao, Yingnan and Wang, Ke and Louri, Ahmed},
  journal={IEEE Trans. Computer-Aided Design of Integr. Circuits and Systems}, 
  title={OPT-GCN: A Unified and Scalable Chiplet-Based Accelerator for High-Performance and Energy-Efficient GCN Computation}, 
  year={2024},
  volume={43},
  number={12},
  pages={4827-4840},
  doi={10.1109/TCAD.2024.3401543}}

@INPROCEEDINGS{Chen2023,
  author={Chen, Jiaxian and Zhong, Zhaoyu and Sun, Kaoyi and Ma, Chenlin and Mao, Rui and Wang, Yi},
  booktitle={2023 60th ACM/IEEE Design Autom. Conf. (DAC)}, 
  title={Lift: Exploiting Hybrid Stacked Memory for Energy-Efficient Processing of Graph Convolutional Networks}, 
  year={2023},
  volume={},
  number={},
  pages={1-6},
  doi={10.1109/DAC56929.2023.10247871}}

@INPROCEEDINGS{Chang2025,
  author={Chang, Wen-Tse and Wu, Chun-Feng and Lo, Yun-Chen},
  booktitle={2025 62nd ACM/IEEE Design Automation Conference (DAC)}, 
  title={P-DAC: Power-Efficient Photonic Accelerators for LLM Inference}, 
  year={2025},
  volume={},
  number={},
  pages={1-7},
  doi={10.1109/DAC63849.2025.11132618}}

@inproceedings{ZhangHao2025,
author = {Zhang, Hao and Zhang, Haibo and Xia, Chengpeng and Huang, Zhiyi and Chen, Yawen and Barnard, Amanda},
title = {ROCKET: An RNS-based Photonic Accelerator for High-Precision and Energy-Efficient DNN Training},
year = {2025},
isbn = {9798400715372},
publisher = {ACM},
address = {New York, NY, USA},
doi = {10.1145/3721145.3734529},
booktitle = {Proc. of the ACM Int. Conf. on Supercomputing},
pages = {1020–1033},
numpages = {14},
location = {
},
series = {ICS '25}
}

@INPROCEEDINGS{Terboven2024,
  author={Terboven, C. and Liem, R. and Gracia, J. and Haldar, K. and Engels, J.F. and Giesselmann, P. and Brayford, D. and Wilde, T. and Simmendinger, C. and Marquardt, M. and Eitzinger, J. and Gruber, T.},
  booktitle={Proc. SC ’24 Workshops}, 
  title={EE-HPC a Framework for Energy Efficient HPC System Management}, 
  year={2024},
  volume={},
  number={},
  pages={1878-1882},
  doi={10.1109/SCW63240.2024.00236}}

@ARTICLE{Fraternali2018,
  author={Fraternali, Francesco and Bartolini, Andrea and Cavazzoni, Carlo and Benini, Luca},
  journal={IEEE Trans. Parallel. Distrib. Syst.}, 
  title={Quantifying the Impact of Variability and Heterogeneity on the Energy Efficiency for a Next-Generation Ultra-Green Supercomputer}, 
  year={2018},
  volume={29},
  number={7},
  pages={1575-1588},
  doi={10.1109/TPDS.2017.2766151}}

@ARTICLE{Raffin2025,
  author={Raffin, Guillaume and Trystram, Denis},
  journal={IEEE Trans. Parallel. Distrib. Syst.},  
  title={Dissecting the Software-Based Measurement of CPU Energy Consumption: A Comparative Analysis}, 
  year={2025},
  volume={36},
  number={1},
  pages={96-107},
  doi={10.1109/TPDS.2024.3492336}}

@inproceedings{Raffin2025b,
  TITLE = {{Alumet: a Modular Framework to Standardize the Measurement of Energy Consumption}},
  AUTHOR = {Raffin, Guillaume and Trystram, Denis and Richard, Olivier},
  URL = {https://hal.science/hal-05246933},
  BOOKTITLE = {{PECS 2025 - Workshop on Performance and Energy Efficiency in Concurrent and Distributed Systems}},
  ADDRESS = {Dresden, Germany},
  ORGANIZATION = {{T{\"U}D Dresden University of Technology}},
  PUBLISHER = {{Springer}},
  PAGES = {1-12},
  YEAR = {2025},
  MONTH = Aug,
  HAL_ID = {hal-05246933},
  HAL_VERSION = {v1},
}

@ARTICLE{Sun2022,
  author={Sun, Joohyung and Cho, Hyeonjoong},
  journal={IEEE Access}, 
  title={A Lightweight Optimal Scheduling Algorithm for Energy-Efficient and Real-Time Cloud Services}, 
  year={2022},
  volume={10},
  number={},
  pages={5697-5714},
  doi={10.1109/ACCESS.2022.3141086}}

@ARTICLE{Wang2024,
  author={Wang, Zhehui and Luo, Tao and Goh, Rick Siow Mong and Zhou, Joey Tianyi},
  journal={IEEE Trans. Neural Net. Learn. Syst.}, 
  title={EDCompress: Energy-Aware Model Compression for Dataflows}, 
  year={2024},
  volume={35},
  number={1},
  pages={208-220},
  doi={10.1109/TNNLS.2022.3172941}}

@ARTICLE{Wajid2016,
  author={Wajid, Usman and Cappiello, Cinzia and Plebani, Pierluigi and Pernici, Barbara and Mehandjiev, Nikolay and Vitali, Monica and Gienger, Michael and Kavoussanakis, Kostas and Margery, David and Perez, David Garcia and Sampaio, Pedro},
  journal={IEEE Trans. Cloud Comput.}, 
  title={On Achieving Energy Efficiency and Reducing CO2 Footprint in Cloud Computing}, 
  year={2016},
  volume={4},
  number={2},
  pages={138-151},
  doi={10.1109/TCC.2015.2453988}}

@ARTICLE{Conficoni2016,
  author={Conficoni, Christian and Bartolini, Andrea and Tilli, Andrea and Cavazzoni, Carlo and Benini, Luca},
  journal={IEEE Trans. Indus. Inform.}, 
  title={Integrated Energy-Aware Management of Supercomputer Hybrid Cooling Systems}, 
  year={2016},
  volume={12},
  number={4},
  pages={1299-1311},
  doi={10.1109/TII.2016.2569399}}

@INPROCEEDINGS{Abu2017,
  author={Abu Ahmad, Wissam and Bartolini, Andrea and Beneventi, Francesco and Benini, Luca and Borghesi, Andrea and Cicala, Marco and Forestieri, Privato and Gianfreda, Cosimo and Gregori, Daniele and Libri, Antonio and Spiga, Filippo and Tinti, Simone},
  booktitle={IEEE IPDPS' 17 Workshops}, 
  title={Design of an Energy Aware Petaflops Class High Performance Cluster Based on Power Architecture}, 
  year={2017},
  volume={},
  number={},
  pages={964-973},
  doi={10.1109/IPDPSW.2017.22}}

@ARTICLE{Tchakoute2025,
  author={Tchakoute, Roblex Nana and Tadonki, Claude and Dokladal, Petr and Mesri, Youssef},
  journal={IEEE Access}, 
  title={Benchmark-Based Study of CPU/GPU Power-Related Features Through JAX and TensorFlow}, 
  year={2025},
  volume={13},
  number={},
  pages={184543-184560},
  doi={10.1109/ACCESS.2025.3625414}}

@ARTICLE{Vahabi2025,
  author={Vahabi, Shahrokh and Righetti, Francesca and Vallati, Carlo and Tonellotto, Nicola},
  journal={IEEE Access}, 
  title={The Impact of Prediction Models on Energy-Aware Resource Management in FaaS Platforms}, 
  year={2025},
  volume={13},
  number={},
  pages={85711-85727},
  doi={10.1109/ACCESS.2025.3569068}}

@ARTICLE{Tarplee2016,
  author={Tarplee, Kyle M. and Friese, Ryan and Maciejewski, Anthony A. and Siegel, Howard Jay and Chong, Edwin K. P.},
  journal={IEEE Trans. Parallel. Distrib. Syst.}, 
  title={Energy and Makespan Tradeoffs in Heterogeneous Computing Systems using Efficient Linear Programming Techniques}, 
  year={2016},
  volume={27},
  number={6},
  pages={1633-1646},
  doi={10.1109/TPDS.2015.2456020}}

@ARTICLE{Nan2017,
  author={Nan, Yucen and Li, Wei and Bao, Wei and Delicato, Flavia C. and Pires, Paulo F. and Dou, Yong and Zomaya, Albert Y.},
  journal={IEEE Access}, 
  title={Adaptive Energy-Aware Computation Offloading for Cloud of Things Systems}, 
  year={2017},
  volume={5},
  number={},
  pages={23947-23957},
  doi={10.1109/ACCESS.2017.2766165}}

@ARTICLE{Gai2021,
  author={Gai, Keke and Qin, Xiao and Zhu, Liehuang},
  journal={IEEE Trans. Comput.}, 
  title={An Energy-Aware High Performance Task Allocation Strategy in Heterogeneous Fog Computing Environments}, 
  year={2021},
  volume={70},
  number={4},
  pages={626-639},
  doi={10.1109/TC.2020.2993561}}

@INPROCEEDINGS{Menear2025,
  author={Menear, Kevin and Wilkinson, Alex and Dykes, Tim and Haus, Utz-Uwe and Duplyakin, Dmitry},
  booktitle={Proc. SC ’25 Workshops}, 
  title={Energy-Aware HPC Scheduling with LLM-Based Power Prediction}, 
  year={2025},
  volume={},
  number={},
  pages={1997-2006},
  doi={10.1145/3731599.3767563}}

@INPROCEEDINGS{Alan2015,
  author={Alan, Ismail and Arslan, Engin and Kosar, Tevfik},
  booktitle={SC '15: Proc. Int. Conf. High Perform. Comput., Netw., Storage Anal}, 
  title={Energy-aware data transfer algorithms}, 
  year={2015},
  volume={},
  number={},
  pages={1-12},
  doi={10.1145/2807591.2807628}}

@INPROCEEDINGS{Georgiou2015,
  author={Georgiou, Yiannis and Glesser, David and Rzadca, Krzysztof and Trystram, Denis},
  booktitle={15th IEEE/ACM Int. Symp. Cluster, Cloud Grid Comput. (CCGRID)}, 
  title={A Scheduler-Level Incentive Mechanism for Energy Efficiency in HPC}, 
  year={2015},
  volume={},
  number={},
  pages={617-626},
  doi={10.1109/CCGrid.2015.101}}

@ARTICLE{Sanni2019,
  author={Sanni, Kayode A. and Andreou, Andreas G.},
  journal={IEEE J. Emerg. Sel Top. Circ. Syst.}, 
  title={A Historical Perspective on Hardware AI Inference, Charge-Based Computational Circuits and an 8 bit Charge-Based Multiply-Add Core in 16 nm FinFET CMOS}, 
  year={2019},
  volume={9},
  number={3},
  pages={532-543},
  doi={10.1109/JETCAS.2019.2933795}}

@ARTICLE{Wang2017,
  author={Wang, Songyun and Qian, Zhuzhong and Yuan, Jiabin and You, Ilsun},
  journal={IEEE Access}, 
  title={A DVFS Based Energy-Efficient Tasks Scheduling in a Data Center}, 
  year={2017},
  volume={5},
  number={},
  pages={13090-13102},
  doi={10.1109/ACCESS.2017.2724598}}

@INPROCEEDINGS{Venkatesh2015,
  author={Venkatesh, Akshay and Vishnu, Abhinav and Hamidouche, Khaled and Tallent, Nathan and Panda, Dhabaleswar and Kerbyson, Darren and Hoisie, Adolfy},
  booktitle={SC '15: Proc. Int. Conf. High Perform. Comput., Netw., Storage Anal}, 
  title={A case for application-oblivious energy-efficient MPI runtime}, 
  year={2015},
  volume={},
  number={},
  pages={1-12},
  doi={10.1145/2807591.2807658}}

@INPROCEEDINGS{Dutot2017,
  author={Dutot, Pierre-François and Georgiou, Yiannis and Glesser, David and Lefevre, Laurent and Poquet, Millian and Rais, Issam},
  booktitle={17th IEEE/ACM Int. Symp. Cluster, Cloud Grid Comput. (CCGRID)}, 
  title={Towards Energy Budget Control in HPC}, 
  year={2017},
  volume={},
  number={},
  pages={381-390},
  doi={10.1109/CCGRID.2017.16}}

@ARTICLE{Xia2015,
  author={Xia, YunNi and Zhou, MengChu and Luo, Xin and Pang, ShanChen and Zhu, QingSheng},
  journal={IEEE Trans. Syst., Man, Cybern.: Syst.}, 
  title={A Stochastic Approach to Analysis of Energy-Aware DVS-Enabled Cloud Datacenters}, 
  year={2015},
  volume={45},
  number={1},
  pages={73-83},
  doi={10.1109/TSMC.2014.2331022}}

@ARTICLE{Wang2022,
  author={Wang, Qiang and Mei, Xinxin and Liu, Hai and Leung, Yiu-Wing and Li, Zongpeng and Chu, Xiaowen},
  journal={IEEE Trans. Parallel. Distrib. Syst.}, 
  title={Energy-Aware Non-Preemptive Task Scheduling With Deadline Constraint in DVFS-Enabled Heterogeneous Clusters}, 
  year={2022},
  volume={33},
  number={12},
  pages={4083-4099},
  doi={10.1109/TPDS.2022.3181096}}

@ARTICLE{Ali2018,
  author={Ali, Haider and Tariq, Umair Ullah and Zheng, Yongjun and Zhai, Xiaojun and Liu, Lu},
  journal={IEEE Access}, 
  title={Contention \& Energy-Aware Real-Time Task Mapping on NoC Based Heterogeneous MPSoCs}, 
  year={2018},
  volume={6},
  number={},
  pages={75110-75123},
  doi={10.1109/ACCESS.2018.2882941}}

@INPROCEEDINGS{Yang2024,
  author={Yang, Zeyu and Adamek, Karel and Armour, Wesley},
  booktitle={SC24: Proc. Int. Conf. High Perform. Comput., Netw., Storage Anal}, 
  title={Accurate and Convenient Energy Measurements for GPUs: A Detailed Study of NVIDIA GPU’s Built-In Power Sensor}, 
  year={2024},
  volume={},
  number={},
  pages={1-17},
  doi={10.1109/SC41406.2024.00028}}

@ARTICLE{Al2020,
  author={Al-hayanni, Mohammed A. Noaman and Rafiev, Ashur and Xia, Fei and Shafik, Rishad and Romanovsky, Alexander and Yakovlev, Alex},
  journal={IEEE Trans. Comput.}, 
  title={PARMA: Parallelization-Aware Run-Time Management for Energy-Efficient Many-Core Systems}, 
  year={2020},
  volume={69},
  number={10},
  pages={1507-1518},
  doi={10.1109/TC.2020.2975787}}

@INPROCEEDINGS{Rajput2024,
  author={Rajput, Saurabhsingh and Kechagia, Maria and Sarro, Federica and Sharma, Tushar},
  booktitle={Proc. IEEE/ACM 21st Int. Conf. Min. Softw. Repos. (MSR)}, 
  title={Greenlight: Highlighting TensorFlow APIs Energy Footprint}, 
  year={2024},
  volume={},
  number={},
  pages={304-308},
  doi={}}

@INPROCEEDINGS{Yu2023,
  author={Yu, Junyeol and Kim, Jongseok and Seo, Euiseong},
  booktitle={IEEE Int. Symp. High-Perform. Comput. Archit. (HPCA)}, 
  title={Know Your Enemy To Save Cloud Energy: Energy-Performance Characterization of Machine Learning Serving}, 
  year={2023},
  volume={},
  number={},
  pages={842-854},
  doi={10.1109/HPCA56546.2023.10070943}}

@INPROCEEDINGS{Stojkovic2025,
  author={Stojkovic, Jovan and Zhang, Chaojie and Goiri, Inigo and Torrellas, Josep and Choukse, Esha},
  booktitle={IEEE Int. Symp. High-Perform. Comput. Archit. (HPCA)}, 
  title={DynamoLLM: Designing LLM Inference Clusters for Performance and Energy Efficiency}, 
  year={2025},
  volume={},
  number={},
  pages={1348-1362},
  doi={10.1109/HPCA61900.2025.00102}}

@INPROCEEDINGS{Li2024,
  author={Li, Qing and Wang, Shangguang and Xu, Chenren and Ma, Xiao and Xu, Mengwei and Zhou, Ao and Xing, Ruolin and Yang, Boyuan and Zhu, Zuo and Zhang, Ying and Liu, Xuanzhe},
  booktitle={2024 IEEE Real-Time Systems Symposium (RTSS)}, 
  title={Exploring Real-Time Satellite Computing: From Energy and Thermal Perspectives}, 
  year={2024},
  volume={},
  number={},
  pages={161-173},
  doi={10.1109/RTSS62706.2024.00023}}

@ARTICLE{Chen2025,
  author={Chen, Yijie and Zhang, Qiyang and Xing, Ruolin and Li, Yuanzhe and Ma, Xiao and Zhang, Yiran and Zhou, Ao and Wang, Shangguang},
  journal={IEEE Trans. Serv. Comput.}, 
  title={SLICE: Energy-Efficient Satellite-Ground Co-Inference via Layer-Wise Scheduling Optimization}, 
  year={2025},
  volume={18},
  number={4},
  pages={2388-2402},
  doi={10.1109/TSC.2025.3577451}}

@ARTICLE{Xu2021,
  author={Xu, Zichuan and Zhao, Liqian and Liang, Weifa and Rana, Omer F. and Zhou, Pan and Xia, Qiufen and Xu, Wenzheng and Wu, Guowei},
  journal={IEEE Trans. Parallel. Distrib. Syst.}, 
  title={Energy-Aware Inference Offloading for DNN-Driven Applications in Mobile Edge Clouds}, 
  year={2021},
  volume={32},
  number={4},
  pages={799-814},
  doi={10.1109/TPDS.2020.3032443}}

@ARTICLE{Zaw2021,
  author={Zaw, Chit Wutyee and Pandey, Shashi Raj and Kim, Kitae and Hong, Choong Seon},
  journal={IEEE Access}, 
  title={Energy-Aware Resource Management for Federated Learning in Multi-Access Edge Computing Systems}, 
  year={2021},
  volume={9},
  number={},
  pages={34938-34950},
  doi={10.1109/ACCESS.2021.3055523}}

@ARTICLE{Sahoo2026,
  author={Sahoo, Subham K. and Dash, Abdhisuta and Mishra, Sambit K. and Puthal, Deepak},
  journal={IEEE Access}, 
  title={Energy-Efficient Federated Learning Scheduler for VM Allocation in Edge-Cloud Continuum}, 
  year={2026},
  volume={14},
  number={},
  pages={4274-4291},
  doi={10.1109/ACCESS.2026.3650945}}

@ARTICLE{Zhao2025,
  author={Zhao, Daming and Zhou, Jian-tao and Li, Keqin},
  journal={IEEE Trans. Sustain. Comput.}, 
  title={CFWS: DRL-Based Framework for Energy Cost and Carbon Footprint Optimization in Cloud Data Centers}, 
  year={2025},
  volume={10},
  number={1},
  pages={95-107},
  doi={10.1109/TSUSC.2024.3391791}}

@ARTICLE{Hidayat2025,
  author={Hidayat, Taufik and Ramli, Kalamullah and Harwahyu, Ruki and Salman, Muhammad and Surya Gunawan, Teddy},
  journal={IEEE Access}, 
  title={Reinforcement Learning-Driven Hybrid Precopy/Postcopy VM Migration for Energy-Efficient Data Centers}, 
  year={2025},
  volume={13},
  number={},
  pages={169521-169533},
  doi={10.1109/ACCESS.2025.3613235}}

@ARTICLE{Jayanetti2024,
  author={Jayanetti, Amanda and Halgamuge, Saman and Buyya, Rajkumar},
  journal={IEEE Trans. Parallel. Distrib. Syst.}, 
  title={Multi-Agent Deep Reinforcement Learning Framework for Renewable Energy-Aware Workflow Scheduling on Distributed Cloud Data Centers}, 
  year={2024},
  volume={35},
  number={4},
  pages={604-615},
  doi={10.1109/TPDS.2024.3360448}}

@ARTICLE{Xu2020,
  author={Xu, Chenhan and Wang, Kun and Li, Peng and Xia, Rui and Guo, Song and Guo, Minyi},
  journal={IEEE Trans. Net. Sci Eng.}, 
  title={Renewable Energy-Aware Big Data Analytics in Geo-Distributed Data Centers with Reinforcement Learning}, 
  year={2020},
  volume={7},
  number={1},
  pages={205-215},
  doi={10.1109/TNSE.2018.2813333}}

@ARTICLE{Hogade2025,
  author={Hogade, Ninad and Pasricha, Sudeep},
  journal={IEEE Trans. Sustain. Comput.}, 
  title={Game-Theoretic Deep Reinforcement Learning to Minimize Carbon Emissions and Energy Costs for AI Inference Workloads in Geo-Distributed Data Centers}, 
  year={2025},
  volume={10},
  number={4},
  pages={628-641},
  doi={10.1109/TSUSC.2024.3520969}}

@ARTICLE{Carvalho2025,
  author={De Carvalho, Vinicius Renan and Sichman, Jaime Simão},
  journal={IEEE Access}, 
  title={Toward Cost-Efficiency and Reduced Carbon Footprint: A Multi-Armed Bandit Hyper-Heuristic for Cloud Scheduling Problems}, 
  year={2025},
  volume={13},
  number={},
  pages={204656-204675},
  doi={10.1109/ACCESS.2025.3639273}}

@ARTICLE{Beena2025,
  author={Beena, B. M. and Ranga, Prashanth Cheluvasai and Manideep, Thotapalli Sri Surya and Saragadam, Sneha and Karthik, Garikipati},
  journal={IEEE Access}, 
  title={A Green Cloud-Based Framework for Energy-Efficient Task Scheduling Using Carbon Intensity Data for Heterogeneous Cloud Servers}, 
  year={2025},
  volume={13},
  number={},
  pages={73916-73938},
  doi={10.1109/ACCESS.2025.3562882}}

@ARTICLE{Zhong2026,
  author={Zhong, Ailing and Wu, Dapeng and Zhu, Zhiyuan and Yang, Boran and Wang, Ruyan},
  journal={IEEE Trans. Cloud Comput.}, 
  title={Dual-Timescale Nonlinear Energy Optimization for Cloud Resource Provisioning}, 
  year={2026},
  volume={14},
  number={1},
  pages={177-191},
  doi={10.1109/TCC.2025.3641292}}

@ARTICLE{Tang2025,
  author={Tang, Xuhao and Liu, Fagui and Xu, Dishi and Jiang, Jun and Tang, Quan and Wang, Bin and Wu, Qingbo and Philip Chen, C. L.},
  journal={IEEE Trans. Cons. Electr.}, 
  title={LLM-Assisted Reinforcement Learning: Leveraging Lightweight Large Language Model Capabilities for Efficient Task Scheduling in Multi-Cloud Environment}, 
  year={2025},
  volume={71},
  number={2},
  pages={5631-5644},
  doi={10.1109/TCE.2024.3524612}}

@ARTICLE{Li2025,
  author={Li, Yuanzhuang and He, Yifan and Lin, Jian and Xu, Zhan and Zhang, Shiyu},
  journal={IEEE Trans. Serv. Comput.}, 
  title={A Reinforcement Learning-Based Population Hyper-Heuristic for Energy-Efficient Cloud Workflow Scheduling Problem}, 
  year={2025},
  volume={18},
  number={5},
  pages={2545-2558},
  doi={10.1109/TSC.2025.3589126}}

@ARTICLE{Huang2025,
  author={Huang, Huikang and Lin, Weiwei and Lin, Jianpeng and Li, Keqin},
  journal={IEEE Trans. Sustain. Comput.}, 
  title={Power Management Optimization for Data Centers: A Power Supply Perspective}, 
  year={2025},
  volume={10},
  number={4},
  pages={784-803},
  doi={10.1109/TSUSC.2025.3542779}}

@ARTICLE{Shen2023,
  author={Shen, Haiying and Wang, Haoyu and Gao, Jiechao and Buyya, Rajkumar},
  journal={IEEE Trans. Parallel. Distrib. Syst.}, 
  title={An Instability-Resilient Renewable Energy Allocation System for a Cloud Datacenter}, 
  year={2023},
  volume={34},
  number={3},
  pages={1020-1034},
  doi={10.1109/TPDS.2023.3235957}}

@ARTICLE{Song2024,
  author={Song, Jie and Zhu, Peimeng and Zhang, Yanfeng and Yu, Ge},
  journal={IEEE Trans. Parallel. Distrib. Syst.}, 
  title={CloudSimPer: Simulating Geo-Distributed Datacenters Powered by Renewable Energy Mix}, 
  year={2024},
  volume={35},
  number={4},
  pages={531-547},
  doi={10.1109/TPDS.2024.3357532}}

@INPROCEEDINGS{Shim2022,
  author={Shim, Jun S. and Han, Bogyeong and Kim, Yeseong and Kim, Jihong},
  booktitle={Des. Autom. Test Eur. Conf. Exhib. (DATE)}, 
  title={DeepPM: Transformer-based Power and Performance Prediction for Energy-Aware Software}, 
  year={2022},
  volume={},
  number={},
  pages={1491-1496},
  doi={10.23919/DATE54114.2022.9774589}}

@INPROCEEDINGS{Sikal2025,
  author={Sikal, Mohammed Bakr and González-Gómez, Jeferson and Khdr, Heba and Henkel, Jörg},
  booktitle={2025 62nd ACM/IEEE Design Autom. Conf. (DAC)}, 
  title={Contention-Aware Forecasting of Energy Efficiency through Sequence-Based Models in Modern Heterogeneous Processors}, 
  year={2025},
  volume={},
  number={},
  pages={1-7},
  doi={10.1109/DAC63849.2025.11132825}}

@INPROCEEDINGS{Wang2019,
  author={Wang, Ke and Louri, Ahmed and Karanth, Avinash and Bunescu, Razvan},
  booktitle={ACM/IEEE 46th Annu. Int. Symp. Comput. Archit. (ISCA)}, 
  title={IntelliNoC: A Holistic Design Framework for Energy-Efficient and Reliable On-Chip Communication for Manycores}, 
  year={2019},
  volume={},
  number={},
  pages={1-12},
  doi={}}

@ARTICLE{WangKe2022,
  author={Wang, Ke and Zheng, Hao and Li, Yuan and Louri, Ahmed},
  journal={IEEE Trans. Sustain. Comput.}, 
  title={SecureNoC: A Learning-Enabled, High-Performance, Energy-Efficient, and Secure On-Chip Communication Framework Design}, 
  year={2022},
  volume={7},
  number={3},
  pages={709-723},
  doi={10.1109/TSUSC.2021.3138279}}

@ARTICLE{Fettes2019,
  author={Fettes, Quintin and Clark, Mark and Bunescu, Razvan and Karanth, Avinash and Louri, Ahmed},
  journal={IEEE Trans. Comput.}, 
  title={Dynamic Voltage and Frequency Scaling in NoCs with Supervised and Reinforcement Learning Techniques}, 
  year={2019},
  volume={68},
  number={3},
  pages={375-389},
  doi={10.1109/TC.2018.2875476}}

@INPROCEEDINGS{Xia2024,
  author={Xia, Jun and Zhang, Yi and Shi, Yiyu},
  booktitle={ACM/IEEE Int. Conf. Computer Aided Des. (ICCAD)}, 
  title={Towards Energy-Aware Federated Learning via MARL: A Dual-Selection Approach for Model and Client}, 
  year={2024},
  volume={},
  number={},
  pages={1-9},
  doi={10.1145/3676536.3676815}}

@ARTICLE{LiXinyi2025,
  author={Li, Xinyi and Zhou, Ti and Wang, Haoyu and Lin, Man},
  journal={IEEE Trans. Sustain. Comput.}, 
  title={Energy-Efficient Computation With DVFS Using Deep Reinforcement Learning for Multi-Task Systems in Edge Computing}, 
  year={2025},
  volume={10},
  number={6},
  pages={1116-1127},
  doi={10.1109/TSUSC.2025.3593971}}

@INPROCEEDINGS{Zhao2022,
  author={Zhao, Dan and Frey, Nathan C. and McDonald, Joseph and Hubbell, Matthew and Bestor, David and Jones, Michael and Prout, Andrew and Gadepally, Vijay and Samsi, Siddharth},
  booktitle={IEEE IPDPS '22 Workshops}, 
  title={A Green(er) World for A.I.}, 
  year={2022},
  volume={},
  number={},
  pages={742-750},
  doi={10.1109/IPDPSW55747.2022.00126}}

@INPROCEEDINGS{Carpentieri2025,
  author={Carpentieri, Lorenzo and De Caro, Antonio and Beni, Majid Salimi and Fan, Kaijie and Cosenza, Biagio},
  booktitle={IEEE Int. Par Distr. Process. Symp. (IPDPS)}, 
  title={Phase-Based Frequency Scaling for Energy-Efficient Heterogeneous Computing}, 
  year={2025},
  volume={},
  number={},
  pages={824-836},
  doi={10.1109/IPDPS64566.2025.00078}}

@INPROCEEDINGS{Chen2016,
  author={Chen, Jieyang and Tan, Li and Wu, Panruo and Tao, Dingwen and Li, Hongbo and Liang, Xin and Li, Sihuan and Ge, Rong and Bhuyan, Laxmi and Chen, Zizhong},
  booktitle={SC '16: Proc. Int. Conf. High Perform. Comput., Netw., Storage Anal}, 
  title={GreenLA: Green Linear Algebra Software for GPU-accelerated Heterogeneous Computing}, 
  year={2016},
  volume={},
  number={},
  pages={667-677},
  doi={10.1109/SC.2016.56}}

@INPROCEEDINGS{Schoonhoven2022,
  author={Schoonhoven, Richard and Veenboer, Bram and Van Werkhoven, Ben and Batenburg, K. Joost},
  booktitle={PMBS' 22 Workshop}, 
  title={Going green: optimizing GPUs for energy efficiency through model-steered auto-tuning}, 
  year={2022},
  volume={},
  number={},
  pages={48-59},
  doi={10.1109/PMBS56514.2022.00010}}

@ARTICLE{Nasrin2022,
  author={Nasrin, Shamma and Shylendra, Ahish and Darabi, Nastaran and Tulabandhula, Theja and Gomes, Wilfred and Chakrabarty, Ankush and Trivedi, Amit Ranjan},
  journal={IEEE Access}, 
  title={ENOS: Energy-Aware Network Operator Search in Deep Neural Networks}, 
  year={2022},
  volume={10},
  number={},
  pages={81447-81457},
  doi={10.1109/ACCESS.2022.3192515}}

@INPROCEEDINGS{Ilsche2015,
  author={Ilsche, Thomas and Hackenberg, Daniel and Graul, Stefan and Schöne, Robert and Schuchart, Joseph},
  booktitle={IEEE IGSC'15}, 
  title={Power measurements for compute nodes: Improving sampling rates, granularity and accuracy}, 
  year={2015},
  volume={},
  number={},
  pages={1-8},
  doi={10.1109/IGCC.2015.7393710}}

@ARTICLE{Peng2022,
  author={Peng, Jiwu and Li, Kenli and Chen, Jianguo and Li, Keqin},
  journal={IEEE Trans. Sustain. Comput.}, 
  title={Reliability/Performance-Aware Scheduling for Parallel Applications With Energy Constraints on Heterogeneous Computing Systems}, 
  year={2022},
  volume={7},
  number={3},
  pages={681-695},
  doi={10.1109/TSUSC.2022.3146138}}

@ARTICLE{Swaraj2026,
  author={Swaraj, Aman and Kumar, Sandeep},
  journal={IEEE Software}, 
  title={Is ChatGPT-Generated Code Really Green?: Evaluating AI-Generated Solutions for Energy-Efficient Coding Practices}, 
  year={2026},
  volume={43},
  number={2},
  pages={42-50},
  doi={10.1109/MS.2025.3644903}}

@ARTICLE{Tunzina2026,
  author={Tunzina, Tayrin and Mashira, Mysun and Islam, Md. Motaharul},
  journal={IEEE Access}, 
  title={GreenAI: A Comparative Analysis of Environmental Efficiency in LLM-Generated Code}, 
  year={2026},
  volume={14},
  number={},
  pages={18654-18672},
  doi={10.1109/ACCESS.2026.3658813}}

@INPROCEEDINGS{Heli2024,
  author={Järvenpää, Heli and Lago, Patricia and Bogner, Justus and Lewis, Grace and Muccini, Henry and Ozkaya, Ipek},
  booktitle={IEEE/ACM 46th Int. Conf. Soft. Eng. (ICSE-SEIS)}, 
  title={A Synthesis of Green Architectural Tactics for ML-Enabled Systems}, 
  year={2024},
  volume={},
  number={},
  pages={130-141},
  doi={}}

@INPROCEEDINGS{Martino2025,
  author={De Martino, Vincenzo and Martínez-Fernández, Silverio and Palomba, Fabio},
  booktitle={IEEE/ACM 47th Int. Conf. Soft. Eng. (ICSE-SEIS)}, 
  title={Do Developers Adopt Green Architectural Tactics for ML-Enabled Systems? A Mining Software Repository Study}, 
  year={2025},
  volume={},
  number={},
  pages={135-139},
  doi={10.1109/ICSE-SEIS66351.2025.00019}}

@ARTICLE{Alswaitti2025,
  author={Alswaitti, Mohammed and Verdecchia, Roberto and Danoy, Grégoire and Bouvry, Pascal and Pecero, Johnatan E.},
  journal={IEEE Trans. Sustain. Comput.}, 
  title={Training Green AI Models Using Elite Samples}, 
  year={2025},
  volume={10},
  number={5},
  pages={858-872},
  doi={10.1109/TSUSC.2025.3544430}}

@INPROCEEDINGS{Ana2024,
  author={Solórzano, Ana Luisa Veroneze and Sato, Kento and Yamamoto, Keiji and Shoji, Fumiyoshi and Brandt, Jim M. and Schwaller, Benjamin and Walton, Sara Petra and Green, Jennifer and Tiwari, Devesh},
  booktitle={SC24: Proc. Int. Conf. High Perform. Comput., Netw., Storage Anal}, 
  title={Toward Sustainable HPC: In-Production Deployment of Incentive-Based Power Efficiency Mechanism on the Fugaku Supercomputer}, 
  year={2024},
  volume={},
  number={},
  pages={1-16},
  doi={10.1109/SC41406.2024.00030}}

@INPROCEEDINGS{Chrysakis2025,
  author={Chrysakis, Ioannis and Agorogiannis, Evangelos and Tsampanaki, Nikoleta and Vourtzoumis, Michalis and Chondrodima, Eva and Theodoridis, Yannis and Mongus, Domen and Capper, Ben and Wagner, Martin and Sotiropoulos, Aris and Coelho, Fábio André and Brito, Cláudia Vanessa and Protopapas, Panos and Brasinika, Despina and Fergadiotou, Ioanna and Doulkeridis, Christos},
  booktitle={Des. Autom. Test Eur. Conf. Exhib. (DATE)}, 
  title={Multi-Partner Project: Green.Dat.AI: A Data Spaces Architecture for Enhancing Green AI Services}, 
  year={2025},
  volume={},
  number={},
  pages={1-7},
  doi={10.23919/DATE64628.2025.10992729}}

@ARTICLE{Ebrahimi2024,
  author={Ebrahimi, Zahra and Kumar, Akash},
  journal={IEEE Trans. Comp.-Aid. Des. Integr. Circ. Syst.}, 
  title={GREEN: An Approximate SIMD/MIMD CGRA for Energy-Efficient Processing at the Edge}, 
  year={2024},
  volume={43},
  number={10},
  pages={2874-2887},
  doi={10.1109/TCAD.2024.3383349}}

@ARTICLE{Philipo2025,
  author={Philipo, Adamu Gaston and Ning, Huansheng and Sarwatt, Doreen Sebastian and Mohamed, Jumanne Ally and Yusufu, Afidhu Swaibu and Shi, Feifei and Urenje, Shepherd and Ding, Jianguo},
  journal={IEEE Trans. Sustain. Comput.}, 
  title={Sustainable AI: Emerging Trends, Impacts, and Future Challenges}, 
  year={2025},
  volume={10},
  number={6},
  pages={1278-1291},
  doi={10.1109/TSUSC.2025.3611272}}

@INPROCEEDINGS{Kim2025,
  author={Kim, Hyosang and Kang, Ki-Dong and Park, Gyeongseo and Lee, Seungkyu and Kim, Daehoon},
  booktitle={IEEE Int. Symp. High-Perform. Comput. Archit. (HPCA)}, 
  title={BrokenSleep: Remote Power Timing Attack Exploiting Processor Idle States}, 
  year={2025},
  volume={},
  number={},
  pages={409-422},
  doi={10.1109/HPCA61900.2025.00040}}

@INPROCEEDINGS{Fayza2025,
  author={Fayza, Farbin and Demirkiran, Cansu and Rao, Satyavolu Papa and Bunandar, Darius and Gupta, Udit and Joshi, Ajay},
  booktitle={2025 IEEE/ACM Int. Conf. Computer Aided Des. (ICCAD)}, 
  title={EPiCarbon: A Carbon Modeling Tool for Electro-Photonic Accelerators}, 
  year={2025},
  volume={},
  number={},
  pages={1-9},
  doi={10.1109/ICCAD66269.2025.11240809}}

@INPROCEEDINGS{Cochet2024,
  author={Cochet, Martin and Swaminathan, Karthik and Loscalzo, Erik and Zuckerman, Joseph and Santos, Maico Cassel Dos and Giri, Davide and Buyuktosunoglu, Alper and Jia, Tianyu and Brooks, David and Wei, Gu-Yeon and Shepard, Kenneth and Carloni, Luca P. and Bose, Pradip},
  booktitle={ACM/IEEE 51st Annu. Int. Symp. Comput. Archit. (ISCA)}, 
  title={BlitzCoin: Fully Decentralized Hardware Power Management for Accelerator-Rich SoCs}, 
  year={2024},
  volume={},
  number={},
  pages={801-817},
  doi={10.1109/ISCA59077.2024.00063}}

@INPROCEEDINGS{Lu2025,
  author={Lu, Chien-Yu and Huang, Bo-Jr and Chen, Min-Chieh and Tsai, Alfred and Fang, Eric Jia-Wei and Cho, Yuju and Liu, Rex Che-Yuan and Wang, Ericbill and Tsao, You-Ming and Mair, Hugh and Hwang, Shih-Arn},
  booktitle={2025 IEEE International Solid-State Circuits Conference (ISSCC)}, 
  title={8.2: Run-Time Power Management System by on-Die Power Sensor with Silicon Machine Learning-Based Calibration in a 3nm Octa-Core CPU}, 
  year={2025},
  volume={68},
  number={},
  pages={160-162},
  doi={10.1109/ISSCC49661.2025.10904564}}

@INPROCEEDINGS{Yahya2022,
  author={Yahya, Jawad Haj and Volos, Haris and Bartolini, Davide B. and Antoniou, Georgia and Kim, Jeremie S. and Wang, Zhe and Kalaitzidis, Kleovoulos and Rollet, Tom and Chen, Zhirui and Geng, Ye and Mutlu, Onur and Sazeides, Yiannakis},
  booktitle={2022 55th IEEE/ACM International Symposium on Microarchitecture (MICRO)}, 
  title={AgileWatts: An Energy-Efficient CPU Core Idle-State Architecture for Latency-Sensitive Server Applications}, 
  year={2022},
  volume={},
  number={},
  pages={835-850},
  doi={10.1109/MICRO56248.2022.00063}}

@INPROCEEDINGS{Antoniou2022,
  author={Antoniou, Georgia and Volos, Haris and Bartolini, Davide B. and Rollet, Tom and Sazeides, Yiannakis and Yahya, Jawad Haj},
  booktitle={2022 55th IEEE/ACM International Symposium on Microarchitecture (MICRO)}, 
  title={AgilePkgC: An Agile System Idle State Architecture for Energy Proportional Datacenter Servers}, 
  year={2022},
  volume={},
  number={},
  pages={851-867},
  doi={10.1109/MICRO56248.2022.00065}}

@ARTICLE{Eilam2024,
  author={Eilam, Tamar and Bose, Pradip and Carloni, Luca P. and Cidon, Asaf and Franke, Hubertus and Kim, Martha A. and Lee, Eun K. and Naghshineh, Mahmoud and Parida, Pritish and Stein, Clifford S. and Tantawi, Asser N.},
  journal={IEEE Trans. Semic. Manuf.}, 
  title={Reducing Datacenter Compute Carbon Footprint by Harnessing the Power of Specialization: Principles, Metrics, Challenges and Opportunities}, 
  year={2024},
  volume={37},
  number={4},
  pages={481-488},
  doi={10.1109/TSM.2024.3434331}}

@INPROCEEDINGS{Costa2025,
  author={Costa, Mariana Toledo and Georgiadou, Antigoni and White, Tames B. and Alvarez, Bruno Villasenor and Polo, Jordà and Shin, Woong and Olivier, Philippe and Messer, Bronson and Lorenzon, Arthur Francisco},
  booktitle={Proc. SC ’25 Workshops}, 
  title={Characterizing the Impact of GPU Power Management on an Exascale System}, 
  year={2025},
  volume={},
  number={},
  pages={1524-1533},
  doi={}}

@ARTICLE{Barbierato2024,
  author={Barbierato, Enrico and Gatti, Alice},
  journal={IEEE Access}, 
  title={Toward Green AI: A Methodological Survey of the Scientific Literature}, 
  year={2024},
  volume={12},
  number={},
  pages={23989-24013},
  doi={10.1109/ACCESS.2024.3360705}}

@ARTICLE{Cesarini2021,
  author={Cesarini, Daniele and Bartolini, Andrea and Bonfà, Pietro and Cavazzoni, Carlo and Benini, Luca},
  journal={IEEE Trans. Comput.}, 
  title={COUNTDOWN: A Run-Time Library for Performance-Neutral Energy Saving in MPI Applications}, 
  year={2021},
  volume={70},
  number={5},
  pages={682-695},
  doi={10.1109/TC.2020.2995269}}

@ARTICLE{Zhou2022,
  author={Zhou, Zhou and Shojafar, Mohammad and Abawajy, Jemal and Yin, Hui and Lu, Hongming},
  journal={IEEE Trans. Green Comm. Network.}, 
  title={ECMS: An Edge Intelligent Energy Efficient Model in Mobile Edge Computing}, 
  year={2022},
  volume={6},
  number={1},
  pages={238-247},
  doi={10.1109/TGCN.2021.3121961}}

@ARTICLE{Yang2019,
  author={Yang, Jun and Xiao, Wenjing and Jiang, Chun and Hossain, M. Shamim and Muhammad, Ghulam and Amin, Syed Umar},
  journal={IEEE Access}, 
  title={AI-Powered Green Cloud and Data Center}, 
  year={2019},
  volume={7},
  number={},
  pages={4195-4203},
  doi={10.1109/ACCESS.2018.2888976}}

@ARTICLE{Eeckhout2025,
  author={Eeckhout, Lieven},
  journal={IEEE Micro}, 
  title={Assessing Processor Sustainability Using the First-Order FOCAL Carbon Model}, 
  year={2025},
  volume={45},
  number={4},
  pages={11-18},
  doi={10.1109/MM.2025.3576714}}

@ARTICLE{Roelandts2025,
  author={Roelandts, Jaime and Naithani, Ajeya and Eeckhout, Lieven},
  journal={IEEE Comp. Arch. Lett.}, 
  title={The Architectural Sustainability Indicator}, 
  year={2025},
  volume={24},
  number={2},
  pages={205-208},
  doi={10.1109/LCA.2025.3576891}}

@ARTICLE{Sadi2025,
  author={Sadi, Mohammad Hassani and Sudarshan, Chirag and Nassif, Sani R. and Wehn, Norbert},
  journal={IEEE Trans. Circ Syst. A.I.}, 
  title={Efficient Deep Neural Network Training With a Novel 5.3-Bit Block Floating Point Data Format}, 
  year={2025},
  volume={2},
  number={2},
  pages={150-161},
  doi={10.1109/TCASAI.2025.3532254}}

@ARTICLE{Son2025,
  author={Son, Yonglak and Gupta, Udit and McCrabb, Andrew and Kim, Young Geun and Bertacco, Valeria and Brooks, David and Wu, Carole-Jean},
  journal={IEEE Internet of Things J.}, 
  title={GreenScale: Carbon Optimization for Edge Computing}, 
  year={2025},
  volume={12},
  number={16},
  pages={32379-32393},
  doi={10.1109/JIOT.2025.3555153}}

@INPROCEEDINGS{Elgamal2025,
  author={Elgamal, Mariam and Carmean, Doug and Ansari, Elnaz and Zed, Okay and Peri, Ramesh and Manne, Srilatha and Gupta, Udit and Wei, Gu-Yeon and Brooks, David and Hills, Gage and Wu, Carole-Jean},
  booktitle={IEEE Int. Symp. High-Perform. Comput. Archit. (HPCA)}, 
  title={CORDOBA: Carbon-Efficient Optimization Framework for Computing Systems}, 
  year={2025},
  volume={},
  number={},
  pages={1289-1303},
  doi={10.1109/HPCA61900.2025.00098}}

@INPROCEEDINGS{Panteleaki2025,
  author={Panteleaki, Aikaterini Maria and Balaskas, Konstantinos and Zervakis, Georgios and Amrouch, Hussam and Anagnostopoulos, Iraklis},
  booktitle={Des. Autom. Test Eur. Conf. Exhib. (DATE)}, 
  title={Late Breaking Results: Leveraging Approximate Computing for Carbon-Aware DNN Accelerators}, 
  year={2025},
  volume={},
  number={},
  pages={1-2},
  doi={10.23919/DATE64628.2025.10993191}}

@INPROCEEDINGS{Stojkovic2024,
  author={Stojkovic, Jovan and Iliakopoulou, Nikoleta and Xu, Tianyin and Franke, Hubertus and Torrellas, Josep},
  booktitle={ACM/IEEE 51st Annu. Int. Symp. Comput. Archit. (ISCA)}, 
  title={EcoFaaS: Rethinking the Design of Serverless Environments for Energy Efficiency}, 
  year={2024},
  volume={},
  number={},
  pages={471-486},
  doi={10.1109/ISCA59077.2024.00042}}

@ARTICLE{Wu2024,
  author={Wu, Carole-Jean and Acun, Bilge and Raghavendra, Ramya and Hazelwood, Kim},
  journal={IEEE Micro}, 
  title={Beyond Efficiency: Scaling AI Sustainably}, 
  year={2024},
  volume={44},
  number={5},
  pages={37-46},
  doi={10.1109/MM.2024.3409275}}

@ARTICLE{Eeckhout2023,
  author={Eeckhout, Lieven},
  journal={IEEE Micro}, 
  title={Kaya for Computer Architects: Toward Sustainable Computer Systems}, 
  year={2023},
  volume={43},
  number={1},
  pages={9-18},
  doi={10.1109/MM.2022.3218034}}

@inproceedings{Gupta2023,
author = {Gupta, Udit and Elgamal, Mariam and Hills, Gage and Wei, Gu-Yeon and Lee, Hsien-Hsin S. and Brooks, David and Wu, Carole-Jean},
title = {ACT: designing sustainable computer systems with an architectural carbon modeling tool},
year = {2022},
isbn = {9781450386104},
publisher = {ACM},
doi = {10.1145/3470496.3527408},
booktitle = {ACM/IEEE 49th Annu. Int. Symp. Comput. Archit. (ISCA)},
pages = {784–799},
numpages = {16},
location = {New York, New York},
series = {ISCA '22}
}

@INPROCEEDINGS{Sudarshan2024,
  author={Sudarshan, Chetan Choppali and Matkar, Nikhil and Vrudhula, Sarma and Sapatnekar, Sachin S. and Chhabria, Vidya A.},
  booktitle={IEEE Int. Symp. High-Perform. Comput. Archit. (HPCA)}, 
  title={ECO-CHIP: Estimation of Carbon Footprint of Chiplet-based Architectures for Sustainable VLSI}, 
  year={2024},
  volume={},
  number={},
  pages={671-685},
  doi={10.1109/HPCA57654.2024.00058}}

@INPROCEEDINGS{Zhao2024,
  author={Zhao, Yujie and Zhao, Yang Katie and Wan, Cheng and Lin, Yingyan Celine},
  booktitle={2024 61st ACM/IEEE Design Autom. Conf. (DAC)}, 
  title={3D-Carbon: An Analytical Carbon Modeling Tool for 3D and 2.5D Integrated Circuits}, 
  year={2024},
  volume={},
  number={},
  pages={1-6},
  doi={}}

@INPROCEEDINGS{Tschand2025,
  author={Tschand, Arya and Rajan, Arun Tejusve Raghunath and Idgunji, Sachin and Ghosh, Anirban and Holleman, Jeremy and Kiraly, Csaba and Ambalkar, Pawan and Borkar, Ritika and Chukka, Ramesh and Cockrell, Trevor and Curtis, Oliver and Fursin, Grigori and Hodak, Miro and Kassa, Hiwot and Lokhmotov, Anton and Miskovic, Dejan and Pan, Yuechao and Manmathan, Manu Prasad and Raymond, Liz and John, Tom St. and Suresh, Arjun and Taubitz, Rowan and Zhan, Sean and Wasson, Scott and Kanter, David and Reddi, Vijay Janapa},
  booktitle={IEEE Int. Symp. High-Perform. Comput. Archit. (HPCA)}, 
  title={MLPerf Power: Benchmarking the Energy Efficiency of Machine Learning Systems from µWatts to MWatts for Sustainable AI}, 
  year={2025},
  volume={},
  number={},
  pages={1201-1216},
  doi={10.1109/HPCA61900.2025.00092}}

@INPROCEEDINGS{Guan2024,
  author={Guan, Wenkai and Zhao, Yang Katie and Ababei, Cristinel},
  booktitle={IEEE IGSC'24}, 
  title={U-DUCT: Uncertainty-aware Dynamic Unified Carbon Modeling Tool for Datacenter Scheduling}, 
  year={2024},
  volume={},
  number={},
  pages={29-34},
  doi={10.1109/IGSC64514.2024.00014}}

@INPROCEEDINGS{Sudarshan2024b,
  author={Sudarshan, Chetan Choppali and Arora, Aman and Chhabria, Vidya A.},
  booktitle={2024 61st ACM/IEEE Design Autom. Conf. (DAC)}, 
  title={GreenFPGA: Evaluating FPGAs as Environmentally Sustainable Computing Solutions}, 
  year={2024},
  volume={},
  number={},
  pages={1-6},
  doi={}}

@INPROCEEDINGS{Spieck2025,
  author={Spieck, Jan and Walter, Dominik and Waschkeit, Jan and Teich, Jürgen},
  booktitle={Des. Autom. Test Eur. Conf. Exhib. (DATE)}, 
  title={Co-Design of Sustainable Embedded Systems-on-Chip}, 
  year={2025},
  volume={},
  number={},
  pages={1-2},
  doi={10.23919/DATE64628.2025.10992914}}

@ARTICLE{Peltonen2025,
  author={Peltonen, Ella and Bayhan, Suzan and Bermbach, David and Buschjäger, Sebastian and Degeler, Victoria and Ding, Aaron Yi and Incel, Ozlem Durmaz and Katare, Dewant and Kjærgaard, Mikkel Baun and Leroux, Sam and Mahmoodi, Toktam and Mann, Zoltn Adam and Meratnia, Nirvana and Pimentel, Andy D. and Rellermeyer, Jan S. and Rivière, Etienne and Sapra, Dolly and Solmaz, Gürkan and van der Waaij, Bram},
  journal={IEEE Internet Computing}, 
  title={Rethinking Computing Systems in the Era of Climate Crisis: A Call for a Sustainable Computing Continuum}, 
  year={2025},
  volume={29},
  number={2},
  pages={8-18},
  doi={10.1109/MIC.2025.3566642}}

@INPROCEEDINGS{Sudarshan2024c,
  author={Sudarshan, Chetan Choppali and Arora, Aman and Chhabria, Vidya A.},
  booktitle={IGSC'24}, 
  title={Beyond the Surface: The Necessity for Detailed Metrics in Corporate Sustainability Reports}, 
  year={2024},
  volume={},
  number={},
  pages={145-150},
  doi={10.1109/IGSC64514.2024.00035}}

@ARTICLE{Salehi2024,
  author={Salehi, Shirin and Schmeink, Anke},
  journal={IEEE Trans. on Art. Intell.}, 
  title={Data-Centric Green Artificial Intelligence: A Survey}, 
  year={2024},
  volume={5},
  number={5},
  pages={1973-1989},
  doi={10.1109/TAI.2023.3315272}}

@ARTICLE{Singh2023,
  author={Singh, Ashutosh Kumar and Swain, Smruti Rekha and Saxena, Deepika and Lee, Chung-Nan},
  journal={IEEE Systems Journal}, 
  title={A Bio-Inspired Virtual Machine Placement Toward Sustainable Cloud Resource Management}, 
  year={2023},
  volume={17},
  number={3},
  pages={3894-3905},
  doi={10.1109/JSYST.2023.3248118}}

@INPROCEEDINGS{Malla2020,
  author={Malla, Sulav and Deng, Qingyuan and Ebrahimzadeh, Zoh and Gasperetti, Joe and Jain, Sajal and Kondety, Parimala and Ortiz, Thiara and Vieira, Debra},
  booktitle={2020 53rd Annual IEEE/ACM International Symposium on Microarchitecture (MICRO)}, 
  title={Coordinated Priority-aware Charging of Distributed Batteries in Oversubscribed Data Centers}, 
  year={2020},
  volume={},
  number={},
  pages={839-851},
  doi={10.1109/MICRO50266.2020.00073}}

@INPROCEEDINGS{Shah2026,
  author={Shah, Nilesh Rajendra and Kumar, M V V S Manoj and Baxi, Dhairya and Upadrasta, Ramakrishna},
  booktitle={2026 IEEE/ACM International Symposium on Code Generation and Optimization (CGO)}, 
  title={PolyUFC: Polyhedral Compilation Meets Roofline Analysis for Uncore Frequency Capping}, 
  year={2026},
  volume={},
  number={},
  pages={563-576},
  doi={10.1109/CGO68049.2026.11395211}}

@ARTICLE{Pashaeifar2019,
  author={Pashaeifar, Masoud and Kamal, Mehdi and Afzali-Kusha, Ali and Pedram, Massoud},
  journal={IEEE Trans. Circ Syst.}, 
  title={A Theoretical Framework for Quality Estimation and Optimization of DSP Applications Using Low-Power Approximate Adders}, 
  year={2019},
  volume={66},
  number={1},
  pages={327-340},
  doi={10.1109/TCSI.2018.2856757}}

@ARTICLE{Qiu2019,
  author={Qiu, Chenxi and Shen, Haiying and Chen, Liuhua},
  journal={IEEE Trans. Big Data}, 
  title={Towards Green Cloud Computing: Demand Allocation and Pricing Policies for Cloud Service Brokerage}, 
  year={2019},
  volume={5},
  number={2},
  pages={238-251},
  doi={10.1109/TBDATA.2018.2823330}}

@INPROCEEDINGS{Kakolyris2025,
  author={Kakolyris, Andreas Kosmas and Masouros, Dimosthenis and Vavaroutsos, Petros and Xydis, Sotirios and Soudris, Dimitrios},
  booktitle={IEEE Int. Symp. High-Perform. Comput. Archit. (HPCA)}, 
  title={throttLL’eM: Predictive GPU Throttling for Energy Efficient LLM Inference Serving}, 
  year={2025},
  volume={},
  number={},
  pages={1363-1378},
  doi={10.1109/HPCA61900.2025.00103}}

@ARTICLE{Dhawan2026,
  author={Dhawan, Rohit and Suresh Patel, Minav},
  journal={IEEE Access}, 
  title={Carbon-Aware AI Control Plane for DevOps Automation: A Reference Architecture and Next-Generation Sustainability Framework}, 
  year={2026},
  volume={14},
  number={},
  pages={11163-11184},
  doi={10.1109/ACCESS.2026.3656467}}

@ARTICLE{Dipta2023,
  author={Dipta, Debopriya Roy and Gulmezoglu, Berk},
  journal={IEEE Trans. Inform. Forens. Secur.}, 
  title={MAD-EN: Microarchitectural Attack Detection Through System-Wide Energy Consumption}, 
  year={2023},
  volume={18},
  number={},
  pages={3006-3017},
  doi={10.1109/TIFS.2023.3272748}}

@ARTICLE{Danopoulos2025,
  author={Danopoulos, Dimitrios and Zervakis, Georgios and Soudris, Dimitrios and Henkel, Jörg},
  journal={IEEE Trans. Circ Syst. Art. Intell.}, 
  title={TransAxx: Efficient Transformers With Approximate Computing}, 
  year={2025},
  volume={2},
  number={4},
  pages={288-301},
  doi={10.1109/TCASAI.2025.3565685}}

@ARTICLE{Hasan2025,
  author={Hasan, Syed Mhamudul and Islam, Taminul and Saifuzzaman, Munshi and Ahmed, Khaled R. and Huang, Chun-Hsi and Shahid, Abdur R.},
  journal={IEEE Trans. Sustain. Comput.}, 
  title={Carbon Emission Quantification of Machine Learning: A Review}, 
  year={2025},
  volume={10},
  number={6},
  pages={1085-1102},
  doi={10.1109/TSUSC.2025.3578834}}

@INPROCEEDINGS{McGuire2023,
  author={McGuire, Sean and Schultz, Erin and Ayoola, Bimpe and Ralph, Paul},
  booktitle={IEEE/ACM 45th Int. Conf. Soft. Eng. (ICSE)}, 
  title={Sustainability is Stratified: Toward a Better Theory of Sustainable Software Engineering}, 
  year={2023},
  volume={},
  number={},
  pages={1996-2008},
  doi={10.1109/ICSE48619.2023.00169}}

@INPROCEEDINGS{LiGuoyu2025,
  author={Li, Guoyu and Ye, Shengyu and Chen, Chunyun and Wang, Yang and Yang, Fan and Cao, Ting and Liu, Cheng and Aly, Mohamed M. Sabry and Yang, Mao},
  booktitle={IEEE Int. Symp. High-Perform. Comput. Archit. (HPCA)}, 
  title={LUT-DLA: Lookup Table as Efficient Extreme Low-Bit Deep Learning Accelerator}, 
  year={2025},
  volume={},
  number={},
  pages={671-684},
  doi={10.1109/HPCA61900.2025.00057}}

@INPROCEEDINGS{LiJiajun2021,
  author={Li, Jiajun and Louri, Ahmed and Karanth, Avinash and Bunescu, Razvan},
  booktitle={IEEE Int. Symp. High-Perform. Comput. Archit. (HPCA)},  
  title={GCNAX: A Flexible and Energy-efficient Accelerator for Graph Convolutional Neural Networks}, 
  year={2021},
  volume={},
  number={},
  pages={775-788},
  doi={10.1109/HPCA51647.2021.00070}}

@INPROCEEDINGS{Hort2023,
  author={Hort, Max and Grishina, Anastasiia and Moonen, Leon},
  booktitle={ACM/IEEE Int. Symp. Emp. Soft. Eng Meas. (ESEM)}, 
  title={An Exploratory Literature Study on Sharing and Energy Use of Language Models for Source Code}, 
  year={2023},
  volume={},
  number={},
  pages={1-12},
  doi={10.1109/ESEM56168.2023.10304803}}

@INPROCEEDINGS{Ayoola2025,
  author={Ayoola, Bimpe and Kuutila, Miikka and Wehbe, Rina R. and Ralph, Paul},
  booktitle={IEEE/ACM 47th In. Conf. Soft. Eng. (ICSE)}, 
  title={User Personas Improve Social Sustainability by Encouraging Software Developers to Deprioritize Antisocial Features}, 
  year={2025},
  volume={},
  number={},
  pages={833-845},
  doi={10.1109/ICSE55347.2025.00135}}

@INPROCEEDINGS{Kempen2025,
  author={Kempen, Nicolas van and Kwon, Hyuk-Je and Nguyen, Dung Tuan and Berger, Emery D.},
  booktitle={40th IEEE/ACM Int. Conf. Aut. Soft. Eng. (ASE)}, 
  title={It’s Not Easy Being Green: On the Energy Efficiency of Programming Languages}, 
  year={2025},
  volume={},
  number={},
  pages={1553-1565},
  doi={10.1109/ASE63991.2025.00131}}

@ARTICLE{Gu2025,
  author={Gu, Diandian and Zhao, Yihao and Sun, Peng and Jin, Xin and Liu, Xuanzhe},
  journal={IEEE Trans. Parallel. Distrib. Syst.}, 
  title={GreenFlow: A Carbon-Efficient Scheduler for Deep Learning Workloads}, 
  year={2025},
  volume={36},
  number={2},
  pages={168-184},
  doi={10.1109/TPDS.2024.3470074}}

@article{Golden2025,
author = {Golden, Alicia and Elgamal, Mariam and Mahmoud, Abdulrahman and Hills, Gage and Wu, Carole-Jean and Wei, Gu-Yeon and Brooks, David},
title = {Wafer-Scale Systems: A Carbon Perspective},
year = {2025},
issue_date = {July 2025},
publisher = {ACM},
volume = {5},
number = {2},
doi = {10.1145/3757892.3757909},
journal = {SIGENERGY Energy Inform. Rev.},
month = aug,
pages = {118–124},
numpages = {7}
}

@inproceedings{Jiang2025,
author = {Jiang, Yankai and Kanakagiri, Raghavendra and Basu Roy, Rohan and Tiwari, Devesh},
title = {ThirstyFLOPS: Water Footprint Modeling and Analysis Toward Sustainable HPC Systems},
year = {2025},
isbn = {9798400714665},
publisher = {ACM},
doi = {10.1145/3712285.3759804},
booktitle = {Proc. Int. Conf. High Perform. Comput., Netw., Storage Anal},
pages = {855–869},
numpages = {15},
location = {},
series = {SC '25}
}

@article{LiPengfei2025,
author = {Li, Pengfei and Islam, Mohammad J. and Ren, Shaolei},
title = {A Case Study of Environmental Footprints for Generative AI Inference: Cloud versus Edge},
year = {2025},
issue_date = {September 2025},
publisher = {ACM},
volume = {53},
number = {2},
issn = {0163-5999},
doi = {10.1145/3764944.3764950},
journal = {SIGMETRICS Perform. Eval. Rev.},
month = aug,
pages = {21–26},
numpages = {6}
}

@article{Migliarini2026,
author = {Migliarini, Patrizio and Autili, Marco and Inverardi, Paola and Pelliccione, Patrizio},
title = {Ethical Prompt Engineering for AI-driven SE: Evidence-informed Interaction-time Governance Roadmap to 2030},
year = {2026},
publisher = {ACM},
issn = {1049-331X},
doi = {10.1145/3801980},
note = {Just Accepted},
journal = {ACM Trans. Softw. Eng. Methodol.},
month = mar
}

@article{Cruz2025,
author = {Cruz, Lu\'{\i}s and Franch, Xavier and Mart\'{\i}nez-Fern\'{a}ndez, Silverio},
title = {Innovating for Tomorrow: The Convergence of Software Engineering and Green AI},
year = {2025},
issue_date = {June 2025},
publisher = {ACM},
volume = {34},
number = {5},
issn = {1049-331X},
doi = {10.1145/3712007},
journal = {ACM Trans. Softw. Eng. Methodol.},
month = may,
articleno = {138},
numpages = {13}
}

@article{Mersy2025,
author = {Mersy, Gabriel and Krishnan, Sanjay},
title = {Toward a Life Cycle Assessment for the Carbon Footprint of Data},
year = {2025},
issue_date = {December 2024},
publisher = {ACM},
volume = {4},
number = {5},
doi = {10.1145/3727200.3727206},
journal = {SIGENERGY Energy Inform. Rev.},
month = apr,
pages = {25–33},
numpages = {9}
}

@article{DeMartino2025,
author = {De Martino, Vincenzo and Lambiase, Stefano and Pecorelli, Fabiano and van den Heuvel, Willem-Jan and Ferrucci, Filomena and Palomba, Fabio},
title = {Sustainability of Machine Learning-Enabled Systems: The Machine Learning Practitioner’s Perspective},
year = {2025},
publisher = {ACM},
issn = {1049-331X},
doi = {10.1145/3777553},
note = {Just Accepted},
journal = {ACM Trans. Softw. Eng. Methodol.},
month = nov
}

@inbook{Roque2025,
author = {Roque, Enrique Barba and Cruz, Luis and Durieux, Thomas},
title = {Unveiling the Energy Vampires: A Methodology for Debugging Software Energy Consumption},
year = {2025},
isbn = {9798331505691},
publisher = {IEEE Press},
booktitle = {Proceedings of the IEEE/ACM 47th International Conference on Software Engineering},
pages = {2406–2418},
numpages = {13}
}

@inproceedings{Alizadeh2025,
author = {Alizadeh, Negar and Belchev, Boris and Saurabh, Nishant and Kelbert, Patricia and Castor, Fernando},
title = {Language Models in Software Development Tasks: An Experimental Analysis of Energy and Accuracy},
year = {2025},
publisher = {IEEE Press},
doi = {10.1109/MSR66628.2025.00109},
booktitle = {Proc. IEEE/ACM 22nd Int. Conf. Min. Softw. Repos. (MSR)},
pages = {725–736},
numpages = {12},
location = {Ottawa, ON, Canada}
}

@inproceedings{Kamatar2025,
author = {Kamatar, Alok and Gonthier, Maxime and Hayot-Sasson, Valerie and Bauer, Andr\'{e} and Copik, Marcin and Castro Fernandez, Raul and Hoefler, Torsten and Chard, Kyle and Foster, Ian},
title = {Core Hours and Carbon Credits: Incentivizing Sustainability in HPC},
year = {2025},
isbn = {9798400714665},
publisher = {ACM},
doi = {10.1145/3712285.3759858},
booktitle = {Proc. Int. Conf. High Perform. Comput., Netw., Storage Anal},
pages = {870–887},
numpages = {18},
location = {
},
series = {SC '25}
}

@inproceedings{Antepara2025,
author = {Antepara, Oscar and Zhao, Zhengji and Austin, Brian and Ding, Nan and Oliker, Leonid and Wright, Nicholas J. and Williams, Samuel},
title = {Benchmark-driven Models for Energy Analysis and Attribution of GPU-Accelerated Supercomputing},
year = {2025},
isbn = {9798400714665},
publisher = {ACM},
doi = {10.1145/3712285.3759815},
booktitle = {Proc. Int. Conf. High Perform. Comput., Netw., Storage Anal},
pages = {888–904},
numpages = {17},
location = {
},
series = {SC '25}
}

@inproceedings{Shen2025,
author = {Shen, Michael and Umar, Muhammad and Maeng, Kiwan and Suh, G. Edward and Gupta, Udit},
title = {Hermes: Algorithm-System Co-design for Efficient Retrieval-Augmented Generation At-Scale},
year = {2025},
isbn = {9798400712616},
publisher = {ACM},
doi = {10.1145/3695053.3731076},
booktitle = {ACM/IEEE 52nd Annu. Int. Symp. Comput. Archit. (ISCA)},
pages = {958–973},
numpages = {16},
location = {
},
series = {ISCA '25}
}

@article{Ifath2026,
author = {Ifath, Md. Monzurul Amin and Haque, Israat},
title = {Characterizing Performance–Energy Trade-offs of Large Language Models in Multi-Request Workflows},
year = {2026},
issue_date = {March 2026},
publisher = {ACM},
volume = {10},
number = {1},
doi = {10.1145/3788089},
journal = {Proc. ACM Meas. Anal. Comput. Syst.},
month = mar,
articleno = {7},
numpages = {26}
}

@article{Wu2025,
author = {Wu, Yanran and Hua, Inez and Ding, Yi},
title = {Not All Water Consumption Is Equal: A Water Stress Weighted Metric for Sustainable Computing},
year = {2025},
issue_date = {July 2025},
publisher = {ACM},
volume = {5},
number = {2},
doi = {10.1145/3757892.3757904},
journal = {SIGENERGY Energy Inform. Rev.},
month = aug,
pages = {84–90},
numpages = {7}
}

@article{Rajput2024b,
author = {Rajput, Saurabhsingh and Widmayer, Tim and Shang, Ziyuan and Kechagia, Maria and Sarro, Federica and Sharma, Tushar},
title = {Enhancing Energy-Awareness in Deep Learning through Fine-Grained Energy Measurement},
year = {2024},
issue_date = {November 2024},
publisher = {ACM},
volume = {33},
number = {8},
issn = {1049-331X},
doi = {10.1145/3680470},
journal = {ACM Trans. Softw. Eng. Methodol.},
month = dec,
articleno = {211},
numpages = {34}
}

@inproceedings{Patel2024,
author = {Patel, Pratyush and Choukse, Esha and Zhang, Chaojie and Goiri, \'{I}\~{n}igo and Warrier, Brijesh and Mahalingam, Nithish and Bianchini, Ricardo},
title = {Characterizing Power Management Opportunities for LLMs in the Cloud},
year = {2024},
isbn = {9798400703867},
publisher = {ACM},
doi = {10.1145/3620666.3651329},
booktitle = {Proc. 29th ACM Int. Conf. Archit. Support Program. Lang. Oper. Syst. (ASPLOS), Vol. 3},
pages = {207–222},
numpages = {16},
location = {La Jolla, CA, USA},
series = {ASPLOS '24}
}

@inproceedings{Stojkovic2025b,
author = {Stojkovic, Jovan and Zhang, Chaojie and Goiri, \'{I}\~{n}igo and Choukse, Esha and Qiu, Haoran and Fonseca, Rodrigo and Torrellas, Josep and Bianchini, Ricardo},
title = {TAPAS: Thermal- and Power-Aware Scheduling for LLM Inference in Cloud Platforms},
year = {2025},
isbn = {9798400710797},
publisher = {ACM},
doi = {10.1145/3676641.3716025},
booktitle = {Proc. 30th ACM Int. Conf. Archit. Support Program. Lang. Oper. Syst. (ASPLOS), Vol. 2},
pages = {1266–1281},
numpages = {16},
location = {Rotterdam, Netherlands},
series = {ASPLOS '25}
}

@inproceedings{Chung2024,
author = {Chung, Jae-Won and Gu, Yile and Jang, Insu and Meng, Luoxi and Bansal, Nikhil and Chowdhury, Mosharaf},
title = {Reducing Energy Bloat in Large Model Training},
year = {2024},
isbn = {9798400712517},
publisher = {ACM},
doi = {10.1145/3694715.3695970},
booktitle = {Proceedings of the ACM SIGOPS 30th Symp. Oper. Syst. Princ.},
pages = {144–159},
numpages = {16},
location = {Austin, TX, USA},
series = {SOSP '24}
}

@inproceedings{Xue2025,
author = {Xue, Yuqi and Huang, Jian},
title = {ReGate: Enabling Power Gating in Neural Processing Units},
year = {2025},
isbn = {9798400715730},
publisher = {ACM},
doi = {10.1145/3725843.3756038},
booktitle = {Proc. 58th IEEE/ACM Int. Symp. Microarch.},
pages = {1160–1177},
numpages = {18},
location = {
},
series = {MICRO '25}
}

@inproceedings{Laskar2025,
author = {Laskar, Sabuj and Majhi, Pranati and Muzahid, Abdullah and Kim, Eun Jung},
title = {SuperMesh: Energy-Efficient Collective Communications for Accelerators},
year = {2025},
isbn = {9798400715730},
publisher = {ACM},
doi = {10.1145/3725843.3756085},
booktitle = {Proceedings of the 58th IEEE/ACM International Symposium on Microarchitecture},
pages = {1640–1655},
numpages = {16},
location = {
},
series = {MICRO '25}
}

@inbook{Liu2025,
author = {Liu, Yiqi and Xue, Yuqi and Crawford, Noelle and Xue, Jilong and Huang, Jian},
title = {Elk: Exploring the Efficiency of Inter-core Connected AI Chips with Deep Learning Compiler Techniques},
year = {2025},
isbn = {9798400715730},
publisher = {ACM},
booktitle = {Proc. 58th IEEE/ACM Int. Symp. Microarch.},
pages = {1284–1299},
numpages = {16}
}

@article{Wiesner2025,
author = {Wiesner, Philipp and Kao, Odej},
title = {Moving Beyond Marginal Carbon IntensityA Poor Metric for Both Carbon Accounting and Grid Flexibility},
year = {2025},
issue_date = {September 2025},
publisher = {ACM},
volume = {53},
number = {2},
issn = {0163-5999},
doi = {10.1145/3764944.3764969},
journal = {SIGMETRICS Perform. Eval. Rev.},
month = aug,
pages = {108–111},
numpages = {4}
}

@inproceedings{Ding2023,
author = {Ding, Jianru and Hoffmann, Henry},
title = {DPS: Adaptive Power Management for Overprovisioned Systems},
year = {2023},
isbn = {9798400701092},
publisher = {ACM},
doi = {10.1145/3581784.3607091},
booktitle = {Proc. Int. Conf. High Perform. Comput., Netw., Storage Anal},
articleno = {27},
numpages = {14},
location = {Denver, CO, USA},
series = {SC '23}
}

@inproceedings{Patki2025,
author = {Patki, Tapasya and Rountree, Barry and Wilde, Torsten and Bartolini, Andrea and Brink, Stephanie and Heiskanen, Esa and Idgunji, Sachin and Maiterth, Matthias and Rogers, James and Rrapaj, Ermal and Schneider, Ralf and Shin, Woong and Shoga, Kathleen and Simmendinger, Christian and Wright, Nicholas J. and Zhao, Zhengji},
title = {A Global Perspective on Supercomputer Power Provisioning: Case Studies from United States and Europe},
year = {2025},
isbn = {9798400715372},
publisher = {ACM},
doi = {10.1145/3721145.3734532},
booktitle = {Proceedings of the 39th ACM International Conference on Supercomputing},
pages = {1034–1051},
numpages = {18},
location = {
},
series = {ICS '25}
}

@inproceedings{Gregersen2025,
author = {Gregersen, Theo and Patel, Pratyush and Choukse, Esha},
title = {Input-Dependent Power Usage in GPUs},
year = {2025},
isbn = {9798350355543},
publisher = {IEEE Press},
doi = {10.1109/SCW63240.2024.00235},
booktitle = {Proc. SC ’24 Workshops},
pages = {1872–1877},
numpages = {6},
location = {Atlanta, GA, USA},
series = {SC-W '24}
}

@inproceedings{ZhaoZhengji2024,
author = {Zhao, Zhengji and Austin, Brian and Rrapaj, Ermal and Wright, Nicholas J.},
title = {Understanding VASP Power Profiles on NVIDIA A100 GPUs},
year = {2025},
isbn = {9798350355543},
publisher = {IEEE Press},
doi = {10.1109/SCW63240.2024.00189},
booktitle = {Proc. SC ’24 Workshops},
pages = {1496–1505},
numpages = {10},
location = {Atlanta, GA, USA},
series = {SC-W '24}
}

@inproceedings{Lella2024,
author = {Lella, Hemasri Sai and Chattaraj, Rajrupa and Chimalakonda, Sridhar and Kurra, Manasa},
title = {Towards Comprehending Energy Consumption of Database Management Systems - A Tool and Empirical Study},
year = {2024},
isbn = {9798400717017},
publisher = {ACM},
doi = {10.1145/3661167.3661174},
booktitle = {Proc. 28th Int. Conf. Eval. Assess. Softw. Eng. (EASE)},
pages = {272–281},
numpages = {10},
location = {Salerno, Italy},
series = {EASE '24}
}

@inproceedings{Sukprasert2024,
author = {Sukprasert, Thanathorn and Souza, Abel and Bashir, Noman and Irwin, David and Shenoy, Prashant},
title = {On the Limitations of Carbon-Aware Temporal and Spatial Workload Shifting in the Cloud},
year = {2024},
isbn = {9798400704376},
publisher = {ACM},
doi = {10.1145/3627703.3650079},
booktitle = {Proc. 19th Eur. Conf. Comput. Syst. (EuroSys)},
pages = {924–941},
numpages = {18},
location = {Athens, Greece},
series = {EuroSys '24}
}

@article{Hanafy2023,
author = {Hanafy, Walid A. and Liang, Qianlin and Bashir, Noman and Irwin, David and Shenoy, Prashant},
title = {CarbonScaler: Leveraging Cloud Workload Elasticity for Optimizing Carbon-Efficiency},
year = {2023},
issue_date = {December 2023},
publisher = {ACM},
volume = {7},
number = {3},
doi = {10.1145/3626788},
journal = {Proc. ACM Meas. Anal. Comput. Syst.},
month = dec,
articleno = {57},
numpages = {28}
}

@inproceedings{WuLi2025,
author = {Wu, Li and Hanafy, Walid A. and Souza, Abel and Nguyen, Khai and Harkes, Jan and Irwin, David and Satyanarayanan, Mahadev and Shenoy, Prashant},
title = {CarbonEdge: Leveraging Mesoscale Spatial Carbon-Intensity Variations for Low Carbon Edge Computing},
year = {2025},
isbn = {9798400718694},
publisher = {ACM},
doi = {10.1145/3731545.3731576},
booktitle = {Proc. 34th Int. Symp. High-Perf. Par Distr.Comput.},
articleno = {12},
numpages = {13},
location = {University of Notre Dame Conference Facilities, Notre Dame, IN, USA},
series = {HPDC '25}
}

@article{Jahanshahi2022,
author = {Jahanshahi, Ali and Yu, Nanpeng and Wong, Daniel},
title = {PowerMorph: QoS-Aware Server Power Reshaping for Data Center Regulation Service},
year = {2022},
issue_date = {September 2022},
publisher = {ACM},
volume = {19},
number = {3},
issn = {1544-3566},
doi = {10.1145/3524129},
journal = {ACM Trans. Archit. Code Optim.},
month = aug,
articleno = {36},
numpages = {27}
}

@article{Bibbens2026,
author = {Bibbens, Jackson and Sigrist, Cooper and Sun, Bo and Kamali, Shahin and Hajiesmaili, Mohammad},
title = {Green Bin Packing},
year = {2026},
issue_date = {March 2026},
publisher = {ACM},
volume = {10},
number = {1},
doi = {10.1145/3788093},
journal = {Proc. ACM Meas. Anal. Comput. Syst.},
month = mar,
articleno = {11},
numpages = {51}
}

@inproceedings{Park2025,
author = {Park, Gyeongseo and Kim, Minho and Kang, Ki-Dong and Jeon, Yunhyeong and Kim, Seulki and Kim, Daehoon},
title = {EcoCore: Dynamic Core Management for Improving Energy Efficiency in Latency-Critical Applications},
year = {2025},
isbn = {9798400715730},
publisher = {ACM},
doi = {10.1145/3725843.3756021},
booktitle = {Proceedings of the 58th IEEE/ACM International Symposium on Microarchitecture},
pages = {1132–1146},
numpages = {15},
location = {
},
series = {MICRO '25}
}

@inproceedings{Zheng2025,
author = {Zheng, Zhong and Sultanov, Seyfal and Papka, Michael E. and Lan, Zhiling},
title = {Minimizing Power Waste in Heterogenous Computing via Adaptive Uncore Scaling},
year = {2025},
isbn = {9798400714665},
publisher = {ACM},
doi = {10.1145/3712285.3759879},
booktitle = {Proc. Int. Conf. High Perform. Comput., Netw., Storage Anal},
pages = {505–518},
numpages = {14},
location = {
},
series = {SC '25}
}

@article{Hanafy2024,
author = {Hanafy, Walid A. and Bostandoost, Roozbeh and Bashir, Noman and Irwin, David and Hajiesmaili, Mohammad and Shenoy, Prashant},
title = {The War of the Efficiencies: Understanding the Tension between Carbon and Energy Optimization},
year = {2024},
issue_date = {July 2024},
publisher = {ACM},
volume = {4},
number = {3},
doi = {10.1145/3698365.3698379},
journal = {SIGENERGY Energy Inform. Rev.},
month = sep,
pages = {87–93},
numpages = {7}
}

@inproceedings{Maquoi2025,
author = {Maquoi, J\'{e}r\^{o}me and Cauz, Maxime and Vanderose, Benoit and Devroey, Xavier},
title = {Energy Codesumption, Leveraging Test Execution for Source Code Energy Consumption Analysis},
year = {2025},
isbn = {9798400712760},
publisher = {ACM},
doi = {10.1145/3696630.3728707},
booktitle = {Proc.33rd ACM Int. Conf. Found. Soft. Eng.},
pages = {1432–1436},
numpages = {5},
location = {Clarion Hotel Trondheim, Trondheim, Norway},
series = {FSE Companion '25}
}

@inproceedings{Ournani2020,
author = {Ournani, Zakaria and Rouvoy, Romain and Rust, Pierre and Penhoat, Joel},
title = {On Reducing the Energy Consumption of Software: From Hurdles to Requirements},
year = {2020},
isbn = {9781450375801},
publisher = {ACM},
doi = {10.1145/3382494.3410678},
booktitle = {Proc. 14th ACM/IEEE Int. Symp. Emp. Soft. Eng Meas. (ESEM)},
articleno = {14},
numpages = {12},
location = {Bari, Italy},
series = {ESEM '20}
}

@article{Moro_Ragazzi_Valgimigli_2023, 
title={Carburacy: Summarization Models Tuning and Comparison in Eco-Sustainable Regimes with a Novel Carbon-Aware Accuracy}, 
volume={37}, 
DOI={10.1609/aaai.v37i12.26686}, 
number={12}, 
journal={Proceedings of the AAAI Conference on Artificial Intelligence}, 
author={Moro, Gianluca and Ragazzi, Luca and Valgimigli, Lorenzo}, 
year={2023}, 
month={Jun.}, 
pages={14417-14425} 
}

@article{Halder2025,
author = {Halder, Debajyoti and Banerjee, Deboparna and Mani, Akash and Gandhi, Anshul and Zadok, Erez},
title = {How Carbon Metrics Impact Device Selection},
year = {2025},
issue_date = {September 2025},
publisher = {ACM},
volume = {53},
number = {2},
issn = {0163-5999},
doi = {10.1145/3764944.3764968},
journal = {SIGMETRICS Perform. Eval. Rev.},
month = aug,
pages = {104–107},
numpages = {4}
}

@inproceedings{Hampau2022,
author = {Hampau, Raluca Maria and Kaptein, Maurits and van Emden, Robin and Rost, Thomas and Malavolta, Ivano},
title = {An empirical study on the Performance and Energy Consumption of AI Containerization Strategies for Computer-Vision Tasks on the Edge},
year = {2022},
isbn = {9781450396134},
publisher = {ACM},
doi = {10.1145/3530019.3530025},
booktitle = {Proc. 26th Int. Conf. Eval. Assess. Softw. Eng. (EASE)},
pages = {50–59},
numpages = {10},
location = {Gothenburg, Sweden},
series = {EASE '22}
}

@inproceedings{Georgiou2022,
author = {Georgiou, Stefanos and Kechagia, Maria and Sharma, Tushar and Sarro, Federica and Zou, Ying},
title = {Green AI: do deep learning frameworks have different costs?},
year = {2022},
isbn = {9781450392211},
publisher = {ACM},
doi = {10.1145/3510003.3510221},
booktitle = {Proc. 44th Int. Conf. Soft. Eng.},
pages = {1082–1094},
numpages = {13},
location = {Pittsburgh, Pennsylvania},
series = {ICSE '22}
}

@inproceedings{Sikand2024,
author = {Sikand, Samarth and Sharma, Vibhu Saujanya and Kaulgud, Vikrant and Podder, Sanjay},
title = {Green AI Quotient: Assessing Greenness of AI-Based Software and the Way Forward},
year = {2024},
isbn = {9798350329964},
publisher = {IEEE Press},
doi = {10.1109/ASE56229.2023.00115},
booktitle = {Proceedings of the 38th IEEE/ACM International Conference on Automated Software Engineering},
pages = {1828–1833},
numpages = {6},
location = {Echternach, Luxembourg},
series = {ASE '23}
}

@inproceedings{Ournani2021,
author = {Ournani, Zakaria and Belgaid, Mohammed Chakib and Rouvoy, Romain and Rust, Pierre and Penhoat, Jo\"{e}l},
title = {Evaluating the Impact of Java Virtual Machines on Energy Consumption},
year = {2021},
isbn = {9781450386654},
publisher = {ACM},
doi = {10.1145/3475716.3475774},
booktitle = {Proc. 15th ACM/IEEE Int. Symp. Emp. Soft. Eng Meas. (ESEM)},
articleno = {15},
numpages = {11},
location = {Bari, Italy},
series = {ESEM '21}
}

@article{Chen2022,
author = {Chen, Jing and Manivannan, Madhavan and Abduljabbar, Mustafa and Peric\`{a}s, Miquel},
title = {ERASE: Energy Efficient Task Mapping and Resource Management for Work Stealing Runtimes},
year = {2022},
issue_date = {June 2022},
publisher = {ACM},
volume = {19},
number = {2},
issn = {1544-3566},
doi = {10.1145/3510422},
journal = {ACM Trans. Archit. Code Optim.},
month = mar,
articleno = {27},
numpages = {29}
}

@inbook{Kalyanapu2025,
author = {Kalyanapu, Joshua and Dizani, Farshad and Asher, Darsh and Ghanbari, Azam and Cammarota, Rosario and Aysu, Aydin and Ajorpaz, Samira Mirbagher},
title = {GateBleed: Exploiting On-Core Accelerator Power Gating for High Performance and Stealthy Attacks on AI},
year = {2025},
isbn = {9798400715730},
publisher = {ACM},
booktitle = {Proc. 58th IEEE/ACM Int. Symp. Microarch.},
pages = {308–325},
numpages = {18}
}

@article{Tian2026,
author = {Tian, Yuyang and Sun, Desen and Ding, Yi and Liu, Sihang},
title = {Cache Your Prompt When It's Green — Carbon-Aware Caching for Large Language Model Serving},
year = {2026},
issue_date = {March 2026},
publisher = {ACM},
volume = {10},
number = {1},
doi = {10.1145/3788087},
journal = {Proc. ACM Meas. Anal. Comput. Syst.},
month = mar,
articleno = {5},
numpages = {28}
}

@article{Zhang2025,
author = {Zhang, Jiaru and Wang, Zesong and Wang, Hao and Song, Tao and Su, Huai-an and Chen, Rui and Hua, Yang and Zhou, Xiangwei and Ma, Ruhui and Pan, Miao and Guan, Haibing},
title = {AMPERE: A Generic Energy Estimation Approach for On-Device Training},
year = {2025},
issue_date = {September 2025},
publisher = {ACM},
volume = {53},
number = {2},
issn = {0163-5999},
doi = {10.1145/3764944.3764951},
journal = {SIGMETRICS Perform. Eval. Rev.},
month = aug,
pages = {27–32},
numpages = {6}
}

@inproceedings{Souza2023,
author = {Souza, Abel and Bashir, Noman and Murillo, Jorge and Hanafy, Walid and Liang, Qianlin and Irwin, David and Shenoy, Prashant},
title = {Ecovisor: A Virtual Energy System for Carbon-Efficient Applications},
year = {2023},
isbn = {9781450399166},
publisher = {ACM},
doi = {10.1145/3575693.3575709},
booktitle = {Proc. 28th ACM Int. Conf. Archit. Support Program. Lang. Oper. Syst. (ASPLOS), Vol. 2},
pages = {252–265},
numpages = {14},
location = {Vancouver, BC, Canada},
series = {ASPLOS 2023}
}

@inproceedings{Guliani2019,
author = {Guliani, Akhil and Swift, Michael M.},
title = {Per-Application Power Delivery},
year = {2019},
isbn = {9781450362818},
publisher = {ACM},
doi = {10.1145/3302424.3303981},
booktitle = {Proc. 14th Eur. Conf. Comput. Syst. (EuroSys)},
articleno = {5},
numpages = {16},
location = {Dresden, Germany},
series = {EuroSys '19}
}

@inproceedings{Zhang2024,
author = {Zhang, Yijia and Wang, Qiang and Lin, Zhe and Xu, Pengxiang and Wang, Bingqiang},
title = {Improving GPU Energy Efficiency through an Application-transparent Frequency Scaling Policy with Performance Assurance},
year = {2024},
isbn = {9798400704376},
publisher = {ACM},
doi = {10.1145/3627703.3629584},
booktitle = {Proc. 19th Eur. Conf. Comput. Syst. (EuroSys)},
pages = {769–785},
numpages = {17},
location = {Athens, Greece},
series = {EuroSys '24}
}

@inproceedings{Acun2023,
author = {Acun, Bilge and Lee, Benjamin and Kazhamiaka, Fiodar and Maeng, Kiwan and Gupta, Udit and Chakkaravarthy, Manoj and Brooks, David and Wu, Carole-Jean},
title = {Carbon Explorer: A Holistic Framework for Designing Carbon Aware Datacenters},
year = {2023},
isbn = {9781450399166},
publisher = {ACM},
doi = {10.1145/3575693.3575754},
booktitle = {Proc. 28th ACM Int. Conf. Archit. Support Program. Lang. Oper. Syst. (ASPLOS), Vol. 2},
pages = {118–132},
numpages = {15},
location = {Vancouver, BC, Canada},
series = {ASPLOS 2023}
}

@inproceedings{Wang2025,
author = {Wang, Jaylen and Berger, Daniel S. and Kazhamiaka, Fiodar and Irvene, Celine and Zhang, Chaojie and Choukse, Esha and Frost, Kali and Fonseca, Rodrigo and Warrier, Brijesh and Bansal, Chetan and Stern, Jonathan and Bianchini, Ricardo and Sriraman, Akshitha},
title = {Designing Cloud Servers for Lower Carbon},
year = {2025},
isbn = {9798350326581},
publisher = {IEEE Press},
doi = {10.1109/ISCA59077.2024.00041},
booktitle = {ACM/IEEE 51st Annu. Int. Symp. Comput. Archit. (ISCA)},
pages = {452–470},
numpages = {19},
location = {Buenos Aires, Argentina},
series = {ISCA '24}
}

@inproceedings{Cho2024,
author = {Cho, Chanwoo and Son, Yonglak and Park, Seongbin and Kim, Young Geun},
title = {CLOVER: Carbon Optimization of Federated Learning over Heterogeneous Clients},
year = {2024},
isbn = {9798400706882},
publisher = {ACM},
doi = {10.1145/3665314.3670829},
booktitle = {Proc. 29th ACM/IEEE Int. Symp. Low Power Electr. Des.},
pages = {1–6},
numpages = {6},
location = {Newport Beach, CA, USA},
series = {ISLPED '24}
}

@inproceedings{Weber2023,
author = {Weber, Max and Kaltenecker, Christian and Sattler, Florian and Apel, Sven and Siegmund, Norbert},
title = {Twins or False Friends? A Study on Energy Consumption and Performance of Configurable Software},
year = {2023},
isbn = {9781665457019},
publisher = {IEEE Press},
doi = {10.1109/ICSE48619.2023.00177},
booktitle = {Proc.45th Int. Conf. Soft. Eng.},
pages = {2098–2110},
numpages = {13},
location = {Melbourne, Victoria, Australia},
series = {ICSE '23}
}

@inproceedings{Stoico2025,
author = {Stoico, Vincenzo and Dragomir, Andrei Calin and Lago, Patricia},
title = {An Empirical Study on the Performance and Energy Usage of Compiled Python Code},
year = {2025},
isbn = {9798400713859},
publisher = {ACM},
doi = {10.1145/3756681.3756972},
booktitle = {Proc. 29th Int. Conf. Eval. Assess. Softw. Eng. (EASE)},
pages = {46–56},
numpages = {11},
location = {
},
series = {EASE '25}
}

@inproceedings{Weber2025,
author = {Weber, Max and Dorn, Johannes and Apel, Sven and Siegmund, Norbert},
title = {FAMLEM, the FAst ModuLar Energy Meter at Code Level},
year = {2025},
isbn = {9798400712760},
publisher = {ACM},
doi = {10.1145/3696630.3728595},
booktitle = {Proc. 33rd ACM Int. Conf. Found. Soft. Eng.},
pages = {1129–1133},
numpages = {5},
location = {Clarion Hotel Trondheim, Trondheim, Norway},
series = {FSE Companion '25}
}

@inproceedings{Yang2025,
author = {Yang, Junyi and Dong, Shuai and Fu, Zhengnan and Shang, Hongyang and Basu, Arindam},
title = {High Energy-Efficiency and Low Latency In-Memory Computing Using Analog Accumulator and In-Memory ADC with Shared},
year = {2025},
isbn = {9798331503048},
publisher = {IEEE Press},
doi = {10.1109/DAC63849.2025.11133014},
booktitle = {Proceedings of the 62nd Annual ACM/IEEE Design Autom. Conf.},
articleno = {273},
numpages = {7},
location = {San Francisco, California, United States},
series = {DAC '25}
}

@inproceedings{LiTaixin2025,
author = {Li, Taixin and Zhong, Hongtao and Xu, Yixin and Narayanan, Vijaykrishnan and Ni, Kai and Yang, Huazhong and K\"{a}mpfe, Thomas and Li, Xueqing},
title = {REMNA: Variation-Resilient and Energy-Efficient MLC FeFET Computing-in-Memory Using NAND Flash-Like Read and Adaptive Control},
year = {2025},
isbn = {9798400710773},
publisher = {ACM},
doi = {10.1145/3676536.3676715},
booktitle = {Proc. 43rd IEEE/ACM Int. Conf. Computer Aided Des. (ICCAD)},
articleno = {76},
numpages = {9},
location = {Newark Liberty International Airport Marriott, New York, NY, USA},
series = {ICCAD '24}
}

@inproceedings{Jayaweera2024,
author = {Jayaweera, Malith and Kong, Martin and Wang, Yanzhi and Kaeli, David},
title = {Energy-Aware Tile Size Selection for Affine Programs on GPUs},
year = {2024},
isbn = {9798350395099},
publisher = {IEEE Press},
doi = {10.1109/CGO57630.2024.10444795},
booktitle = {Proceedings of the 2024 IEEE/ACM Int. Symp. Cod. Gen Optim.},
pages = {13–27},
numpages = {15},
location = {Edinburgh, United Kingdom},
series = {CGO '24}
}

@inproceedings{Wang2025b,
author = {Wang, Zibo and Zhang, Yijia and Wei, Fuchun and Wang, Bingqiang and Liu, Yanlin and Hu, Zhiheng and Zhang, Jingyi and Xu, Xiaoxin and He, Jian and Wang, Xiaoliang and Dou, Wanchun and Chen, Guihai and Tian, Chen},
title = {Using Analytical Performance/Power Model and Fine-Grained DVFS to Enhance AI Accelerator Energy Efficiency},
year = {2025},
isbn = {9798400706981},
publisher = {ACM},
doi = {10.1145/3669940.3707231},
booktitle = {Proc. 30th ACM Int. Conf. Archit. Support Program. Lang. Oper. Syst. (ASPLOS), Vol. 1},
pages = {1118–1132},
numpages = {15},
location = {Rotterdam, Netherlands},
series = {ASPLOS '25}
}

@article{Han2025,
author = {Han, Bing-Shiun and Parekh, Kunaal and Lin, Wan-Chu and Paul, Tathagata and Gandhi, Anshul and Liu, Zhenhua},
title = {Energy-efficient GPU SM allocation},
year = {2025},
issue_date = {September 2025},
publisher = {ACM},
volume = {53},
number = {2},
issn = {0163-5999},
doi = {10.1145/3764944.3764952},
journal = {SIGMETRICS Perform. Eval. Rev.},
month = aug,
pages = {33–38},
numpages = {6}
}

@inproceedings{John2025,
author = {John, Chelsea Maria and Nassyr, Stepan and Penke, Carolin and Herten, Andreas},
title = {Performance and Power: Systematic Evaluation of AI Workloads on Accelerators with CARAML},
year = {2025},
isbn = {9798350355543},
publisher = {IEEE Press},
doi = {10.1109/SCW63240.2024.00158},
booktitle = {Proc. SC ’24 Workshops},
pages = {1164–1176},
numpages = {13},
location = {Atlanta, GA, USA},
series = {SC-W '24}
}

@inproceedings{Zanfardino2025,
author = {Zanfardino, Gennaro and Memon, Mashal Afzal and Tucci, Michele},
title = {Enhancing Energy Efficiency with Reusable Software Ecosystems and Persona-Based UI/UX},
year = {2025},
isbn = {9798400712760},
publisher = {ACM},
doi = {10.1145/3696630.3728710},
booktitle = {Proc.33rd ACM Int. Conf. Found. Soft. Eng.},
pages = {1449–1452},
numpages = {4},
location = {Clarion Hotel Trondheim, Trondheim, Norway},
series = {FSE Companion '25}
}

@inproceedings{Desai2025,
author = {Desai, Harsh and Wang, Xinye and Lucia, Brandon},
title = {Energy-aware Scheduling and Input Buffer Overflow Prevention for Energy-harvesting Systems},
year = {2025},
isbn = {9798400710797},
publisher = {ACM},
doi = {10.1145/3676641.3715995},
booktitle = {Proc. 30th ACM Int. Conf. Archit. Support Program. Lang. Oper. Syst. (ASPLOS), Vol. 2},
pages = {339–354},
numpages = {16},
location = {Rotterdam, Netherlands},
series = {ASPLOS '25}
}

@inproceedings{Tundo2024,
author = {Tundo, Alessandro and Mobilio, Marco and Ilager, Shashikant and Brandi\'{c}, Ivona and Bartocci, Ezio and Mariani, Leonardo},
title = {An Energy-Aware Approach to Design Self-Adaptive AI-based Applications on the Edge},
year = {2024},
isbn = {9798350329964},
publisher = {IEEE Press},
doi = {10.1109/ASE56229.2023.00046},
booktitle = {Proceedings of the 38th IEEE/ACM International Conference on Automated Software Engineering},
pages = {281–293},
numpages = {13},
location = {Echternach, Luxembourg},
series = {ASE '23}
}

@inproceedings{Paipuri2025,
author = {Paipuri, Mahendra},
title = {CEEMS: A Resource Manager Agnostic Energy and Emissions Monitoring Stack},
year = {2025},
isbn = {9798350355543},
publisher = {IEEE Press},
doi = {10.1109/SCW63240.2024.00233},
booktitle = {Proc. SC ’24 Workshops},
pages = {1862–1866},
numpages = {5},
location = {Atlanta, GA, USA},
series = {SC-W '24}
}

@inproceedings{Chung2025,
author = {Chung, Fan and Kuo, Henry and Candea, George},
title = {The Case for Energy Clarity},
year = {2025},
isbn = {9798400714757},
publisher = {ACM},
doi = {10.1145/3713082.3730370},
booktitle = {Proceedings of the 2025 Workshop on Hot Topics in Operating Systems},
pages = {202–209},
numpages = {8},
location = {Banff, AB, Canada},
series = {HotOS '25}
}

@inproceedings{Fan2023,
author = {Fan, Kaijie and D'Antonio, Marco and Carpentieri, Lorenzo and Cosenza, Biagio and Ficarelli, Federico and Cesarini, Daniele},
title = {SYnergy: Fine-grained Energy-Efficient Heterogeneous Computing for Scalable Energy Saving},
year = {2023},
isbn = {9798400701092},
publisher = {ACM},
doi = {10.1145/3581784.3607055},
booktitle = {Proc. Int. Conf. High Perform. Comput., Netw., Storage Anal},
articleno = {69},
numpages = {13},
location = {Denver, CO, USA},
series = {SC '23}
}

@inproceedings{Basu2025,
author = {Basu Roy, Rohan and Patel, Tirthak and Li, Baolin and Samsi, Siddharth and Gadepally, Vijay and Tiwari, Devesh},
title = {GreenMix: Energy-Efficient Serverless Computing via Randomized Sketching on Asymmetric Multi-Cores},
year = {2025},
isbn = {9798400714665},
publisher = {ACM},
doi = {10.1145/3712285.3759861},
booktitle = {Proc. Int. Conf. High Perform. Comput., Netw., Storage Anal},
pages = {475–489},
numpages = {15},
location = {
},
series = {SC '25}
}

@article{Kilian2025,
author = {Kilian, Alexander and de Meer, Hermann and Schomaker, Gunnar},
title = {Energy-Optimized Supercomputer Networks Using Wind Energy},
year = {2025},
issue_date = {July 2025},
publisher = {ACM},
volume = {68},
number = {7},
issn = {0001-0782},
doi = {10.1145/3725981},
journal = {Commun. ACM},
month = jun,
pages = {74–79},
numpages = {6}
}

@inproceedings{Smejkal2025,
author = {Smejkal, Till and Khasanov, Robert and Castrillon, Jeronimo and H\"{a}rtig, Hermann},
title = {HARP: Energy-Aware and Adaptive Management of Heterogeneous Processors},
year = {2025},
isbn = {9798400715549},
publisher = {ACM},
doi = {10.1145/3721462.3770774},
booktitle = {Proceedings of the 26th International Middleware Conference},
pages = {270–284},
numpages = {15},
location = {Vanderbilt University, Nashville, TN, USA},
series = {Middleware '25}
}

@article{Qiao2025,
author = {Qiao, Feitong and Fang, Yiming and Cidon, Asaf},
title = {Energy-Aware Process Scheduling in Linux},
year = {2025},
issue_date = {December 2024},
publisher = {ACM},
volume = {4},
number = {5},
doi = {10.1145/3727200.3727214},
journal = {SIGENERGY Energy Inform. Rev.},
month = apr,
pages = {91–97},
numpages = {7}
}

@article{NEJATISHARIFALDIN2026108210,
title = {From watts to writes: The role of frequency, replication strategies and consistency levels in performance and energy-aware cassandra cluster operations},
journal = {Future Gener. Comput. Syst.},
volume = {176},
pages = {108210},
year = {2026},
issn = {0167-739X},
doi = {https://doi.org/10.1016/j.future.2025.108210},
author = {Hesam {Nejati Sharif Aldin} and Farnoush {Nayebi Pour}}
}

@article{WEST2026108453,
title = {A systematic evaluation of the potential of carbon-aware execution for scientific workflows},
journal = {Future Gener. Comput. Syst.},
volume = {182},
pages = {108453},
year = {2026},
issn = {0167-739X},
doi = {https://doi.org/10.1016/j.future.2026.108453},
url = {https://www.sciencedirect.com/science/article/pii/S0167739X26000877},
author = {Kathleen West and Youssef Moawad and Fabian Lehmann and Vasilis Bountris and Ulf Leser and Yehia Elkhatib and Lauritz Thamsen}
}

@article{COSTERO2023100865,
title = {Dynamic power budget redistribution under a power cap on multi-application environments},
journal = { Sustain. Comput. Inf. Syst.},
volume = {38},
pages = {100865},
year = {2023},
issn = {2210-5379},
doi = {https://doi.org/10.1016/j.suscom.2023.100865},
url = {https://www.sciencedirect.com/science/article/pii/S2210537923000203},
author = {Luis Costero and Francisco D. Igual and Katzalin Olcoz}
}

@article{KAHIL2025125734,
title = {Reinforcement learning for data center energy efficiency optimization: A systematic literature review and research roadmap},
journal = {Applied Energy},
volume = {389},
pages = {125734},
year = {2025},
issn = {0306-2619},
doi = {https://doi.org/10.1016/j.apenergy.2025.125734},
url = {https://www.sciencedirect.com/science/article/pii/S0306261925004647},
author = {Hussain Kahil and Shiva Sharma and Petri Välisuo and Mohammed Elmusrati}
}

@article{KUMAR2026116664,
title = {Toward sustainable data center operation: A review on existing infrastructures, integrated smart energy management frameworks, and future perspectives},
journal = {Renewable and Sustainable Energy Reviews},
volume = {230},
pages = {116664},
year = {2026},
issn = {1364-0321},
doi = {https://doi.org/10.1016/j.rser.2025.116664},
url = {https://www.sciencedirect.com/science/article/pii/S1364032125013371},
author = {Amit Kumar and Naran M. Pindoriya}
}

@article{CHADLI2026108218,
title = {PEARL: Performance and energy aware routing for LLMs},
journal = {Future Gener. Comput. Syst.},
volume = {176},
pages = {108218},
year = {2026},
issn = {0167-739X},
doi = {https://doi.org/10.1016/j.future.2025.108218},
url = {https://www.sciencedirect.com/science/article/pii/S0167739X25005126},
author = {Kouider Chadli and Goetz Botterweck and Takfarinas Saber}
}

@article{CHEN2025107760,
title = {GAS-MARL: Green-Aware job Scheduling algorithm for HPC clusters based on Multi-Action Deep Reinforcement Learning},
journal = {Future Gener. Comput. Syst.},
volume = {167},
pages = {107760},
year = {2025},
issn = {0167-739X},
doi = {https://doi.org/10.1016/j.future.2025.107760},
url = {https://www.sciencedirect.com/science/article/pii/S0167739X2500055X},
author = {Rui Chen and Weiwei Lin and Huikang Huang and Xiaoying Ye and Zhiping Peng}
}

@article{CHEN2026108009,
title = {QoS-aware placement of interdependent services in energy-harvesting-enabled multi-access edge computing},
journal = {Future Gener. Comput. Syst.},
volume = {174},
pages = {108009},
year = {2026},
issn = {0167-739X},
doi = {https://doi.org/10.1016/j.future.2025.108009},
url = {https://www.sciencedirect.com/science/article/pii/S0167739X25003048},
author = {Shuyi Chen and Panagiotis Oikonomou and Zhengchang Hua and Nikos Tziritas and Karim Djemame and Nan Zhang and Georgios Theodoropoulos}
}

@article{SK2024329,
title = {PowerTrain: Fast, generalizable time and power prediction models to optimize DNN training on accelerated edges},
journal = {Future Gener. Comput. Syst.},
volume = {161},
pages = {329-344},
year = {2024},
issn = {0167-739X},
doi = {https://doi.org/10.1016/j.future.2024.07.001},
url = {https://www.sciencedirect.com/science/article/pii/S0167739X24003649},
author = {Prashanthi S.K. and Saisamarth Taluri and Beautlin S and Lakshya Karwa and Yogesh Simmhan}
}

@article{YANG2023178,
title = {Joint heterogeneity-aware personalized federated search for energy efficient battery-powered edge computing},
journal = {Future Gener. Comput. Syst.},
volume = {146},
pages = {178-194},
year = {2023},
issn = {0167-739X},
doi = {https://doi.org/10.1016/j.future.2023.04.024},
url = {https://www.sciencedirect.com/science/article/pii/S0167739X23001644},
author = {Zhao Yang and Shengbing Zhang and Chuxi Li and Miao Wang and Jiaying Yang and Meng Zhang},
}

@article{GONZALOSANJOSE2025107593,
title = {NARA: Network-Aware Resource Allocation mechanism for minimizing quality-of-service impact while dealing with energy consumption in volunteer networks},
journal = {Future Gener. Comput. Syst.},
volume = {164},
pages = {107593},
year = {2025},
issn = {0167-739X},
url = {https://www.sciencedirect.com/science/article/pii/S0167739X24005570},
author = {Sergio {Gonzalo San José} and Joan Manuel Marquès and Javier Panadero and Laura Calvet}
}

@article{TOOR20191112,
title = {Energy and performance aware fog computing: A case of DVFS and green renewable energy},
journal = {Future Gener. Comput. Syst.},
volume = {101},
pages = {1112-1121},
year = {2019},
issn = {0167-739X},
url = {https://www.sciencedirect.com/science/article/pii/S0167739X19310234},
author = {Asfa Toor and Saif ul Islam and Nimra Sohail and Adnan Akhunzada and Jalil Boudjadar and Hasan Ali Khattak and Ikram Ud Din and Joel J.P.C. Rodrigues}
}

@article{CESARIO2026108245,
title = {Enhancing energy efficiency in cloud computing through regression models: A data-driven approach with experimental validation},
journal = {Future Gener. Comput. Syst.},
volume = {177},
pages = {108245},
year = {2026},
issn = {0167-739X},
url = {https://www.sciencedirect.com/science/article/pii/S0167739X25005394},
author = {Eugenio Cesario and Paolo Lindia and Federica Lobello and Andrea Vinci and Santina Capalbo}
}

@ARTICLE{Marahatta2021,
  author={Marahatta, Avinab and Xin, Qin and Chi, Ce and Zhang, Fa and Liu, Zhiyong},
  journal={IEEE Trans. Sustain. Comput.}, 
  title={PEFS: AI-Driven Prediction Based Energy-Aware Fault-Tolerant Scheduling Scheme for Cloud Data Center}, 
  year={2021},
  volume={6},
  number={4},
  pages={655-666},
  doi={10.1109/TSUSC.2020.3015559}}

@article{SHARMA2019620,
title = {Failure-aware energy-efficient VM consolidation in cloud computing systems},
journal = {Future Gener. Comput. Syst.},
volume = {94},
pages = {620-633},
year = {2019},
issn = {0167-739X},
doi = {https://doi.org/10.1016/j.future.2018.11.052},
url = {https://www.sciencedirect.com/science/article/pii/S0167739X1831700X},
author = {Yogesh Sharma and Weisheng Si and Daniel Sun and Bahman Javadi}
}

@article{RODRIGUEZ2026103752,
title = {Predictively controlling the computing continuum with distributed energy-aware orchestration},
journal = {J. Syst. Arch.},
volume = {175},
pages = {103752},
year = {2026},
issn = {1383-7621},
doi = {https://doi.org/10.1016/j.sysarc.2026.103752},
url = {https://www.sciencedirect.com/science/article/pii/S1383762126000706},
author = {Pablo Rodríguez and Javier Mateos-Bravo and Sergio Laso and Juan Luis Herrera and Javier Berrocal}
}

@article{ZONG201727,
title = {Marcher: A Heterogeneous System Supporting Energy-Aware High Performance Computing and Big Data Analytics},
journal = {Big Data Research},
volume = {8},
pages = {27-38},
year = {2017},
note = {Tutorials on Tools and Methods using High Performance Computing resources for Big Data},
issn = {2214-5796},
doi = {https://doi.org/10.1016/j.bdr.2017.01.003},
url = {https://www.sciencedirect.com/science/article/pii/S221457961630048X},
author = {Ziliang Zong and Rong Ge and Qijun Gu}
}

@article{KRZYWANIAK2023396,
title = {Dynamic GPU power capping with online performance tracing for energy efficient GPU computing using DEPO tool},
journal = {Future Gener. Comput. Syst.},
volume = {145},
pages = {396-414},
year = {2023},
issn = {0167-739X},
url = {https://www.sciencedirect.com/science/article/pii/S0167739X23001267},
author = {Adam Krzywaniak and Paweł Czarnul and Jerzy Proficz}
}

@article{BANERJEE2024376,
title = {Towards energy and QoS aware dynamic VM consolidation in a multi-resource cloud},
journal = {Future Gener. Comput. Syst.},
volume = {157},
pages = {376-391},
year = {2024},
issn = {0167-739X},
url = {https://www.sciencedirect.com/science/article/pii/S0167739X24001274},
author = {Sounak Banerjee and Sarbani Roy and Sunirmal Khatua}
}

@article{QURESHI2019453,
title = {Profile-based power-aware workflow scheduling framework for energy-efficient data centers},
journal = {Future Gener. Comput. Syst.},
volume = {94},
pages = {453-467},
year = {2019},
issn = {0167-739X},
doi = {https://doi.org/10.1016/j.future.2018.11.010},
url = {https://www.sciencedirect.com/science/article/pii/S0167739X18318491},
author = {Basit Qureshi}
}

@article{DESENSI2018136,
title = {Simplifying self-adaptive and power-aware computing with Nornir},
journal = {Future Gener. Comput. Syst.},
volume = {87},
pages = {136-151},
year = {2018},
issn = {0167-739X},
doi = {https://doi.org/10.1016/j.future.2018.05.012},
url = {https://www.sciencedirect.com/science/article/pii/S0167739X17326699},
author = {Daniele {De Sensi} and Tiziano {De Matteis} and Marco Danelutto}
}

@article{HASSAN2020431,
title = {A smart energy and reliability aware scheduling algorithm for workflow execution in DVFS-enabled cloud environment},
journal = {Future Gener. Comput. Syst.},
volume = {112},
pages = {431-448},
year = {2020},
issn = {0167-739X},
doi = {https://doi.org/10.1016/j.future.2020.05.040},
url = {https://www.sciencedirect.com/science/article/pii/S0167739X19322307},
author = {Hadeer A. Hassan and Sameh A. Salem and Elsayed M. Saad}
}

@article{LI2018887,
title = {Holistic energy and failure aware workload scheduling in Cloud datacenters},
journal = {Future Gener. Comput. Syst.},
volume = {78},
pages = {887-900},
year = {2018},
issn = {0167-739X},
doi = {https://doi.org/10.1016/j.future.2017.07.044},
url = {https://www.sciencedirect.com/science/article/pii/S0167739X17315650},
author = {Xiang Li and Xiaohong Jiang and Peter Garraghan and Zhaohui Wu}
}

@article{ZHOU2018836,
title = {Minimizing SLA violation and power consumption in Cloud data centers using adaptive energy-aware algorithms},
journal = {Future Gener. Comput. Syst.},
volume = {86},
pages = {836-850},
year = {2018},
issn = {0167-739X},
doi = {https://doi.org/10.1016/j.future.2017.07.048},
url = {https://www.sciencedirect.com/science/article/pii/S0167739X17316059},
author = {Zhou Zhou and Jemal Abawajy and Morshed Chowdhury and Zhigang Hu and Keqin Li and Hongbing Cheng and Abdulhameed A. Alelaiwi and Fangmin Li}
}

@article{BANCHELLI2026108125,
title = {Introducing MareNostrum5: A European pre-exascale energy-efficient system designed to serve a broad spectrum of scientific workloads},
journal = {Future Gener. Comput. Syst.},
volume = {176},
pages = {108125},
year = {2026},
issn = {0167-739X},
doi = {https://doi.org/10.1016/j.future.2025.108125},
url = {https://www.sciencedirect.com/science/article/pii/S0167739X25004194},
author = {Fabio Banchelli and Marta Garcia-Gasulla and Filippo Mantovani and Joan Vinyals and Josep Pocurull and David Vicente and Beatriz Eguzkitza and Flavio C.C. Galeazzo and Mario C. Acosta and Sergi Girona}
}

@article{MIRHOSEININEJAD2020174,
title = {Joint data center cooling and workload management: A thermal-aware approach},
journal = {Future Gener. Comput. Syst.},
volume = {104},
pages = {174-186},
year = {2020},
issn = {0167-739X},
url = {https://www.sciencedirect.com/science/article/pii/S0167739X19302547},
author = {SeyedMorteza MirhoseiniNejad and Hosein Moazamigoodarzi and Ghada Badawy and Douglas G. Down}
}

@article{LI2020789,
title = {Energy-efficient and quality-aware VM consolidation method},
journal = {Future Gener. Comput. Syst.},
volume = {102},
pages = {789-809},
year = {2020},
issn = {0167-739X},
doi = {https://doi.org/10.1016/j.future.2019.08.004},
url = {https://www.sciencedirect.com/science/article/pii/S0167739X18324713},
author = {Zhihua Li and Xinrong Yu and Lei Yu and Shujie Guo and Victor Chang}
}

@article{LUO2020119,
title = {Solving the dynamic energy aware job shop scheduling problem with the heterogeneous parallel genetic algorithm},
journal = {Future Gener. Comput. Syst.},
volume = {108},
pages = {119-134},
year = {2020},
issn = {0167-739X},
doi = {https://doi.org/10.1016/j.future.2020.02.019},
url = {https://www.sciencedirect.com/science/article/pii/S0167739X19314189},
author = {Jia Luo and Didier {El Baz} and Rui Xue and Jinglu Hu}
}

@article{RAHMANI2026108428,
title = {Real-time AI-powered monitoring for energy-efficient scheduling in multi-node heterogeneous systems},
journal = {Future Gener. Comput. Syst.},
volume = {181},
pages = {108428},
year = {2026},
issn = {0167-739X},
doi = {https://doi.org/10.1016/j.future.2026.108428},
url = {https://www.sciencedirect.com/science/article/pii/S0167739X26000622},
author = {Taha Abdelazziz Rahmani and Ghalem Belalem and Sidi Ahmed Mahmoudi and Omar Rafik Merad-Boudia}
}

@article{SHE2025107458,
title = {Energy-efficiency optimization for heterogeneous computing-assisted NOMA-MEC edge AI tasks},
journal = {Future Gener. Comput. Syst.},
volume = {162},
pages = {107458},
year = {2025},
issn = {0167-739X},
doi = {https://doi.org/10.1016/j.future.2024.07.036},
url = {https://www.sciencedirect.com/science/article/pii/S0167739X24003984},
author = {Rui She and Yuting Wu and Enfang Cui and Mengyu Sun and Wei Zhao and Deji Fu}
}

@article{RUAN2019380,
title = {Virtual machine allocation and migration based on performance-to-power ratio in energy-efficient clouds},
journal = {Future Gener. Comput. Syst.},
volume = {100},
pages = {380-394},
year = {2019},
issn = {0167-739X},
doi = {https://doi.org/10.1016/j.future.2019.05.036},
url = {https://www.sciencedirect.com/science/article/pii/S0167739X18321629},
author = {Xiaojun Ruan and Haiquan Chen and Yun Tian and Shu Yin}
}

@article{YAO2023222,
title = {An energy-efficient load balance strategy based on virtual machine consolidation in cloud environment},
journal = {Future Gener. Comput. Syst.},
volume = {146},
pages = {222-233},
year = {2023},
issn = {0167-739X},
doi = {https://doi.org/10.1016/j.future.2023.04.014},
url = {https://www.sciencedirect.com/science/article/pii/S0167739X23001498},
author = {Wenbin Yao and Zhuqing Wang and Yingying Hou and Xikang Zhu and Xiaoyong Li and Yamei Xia}
}

@article{HU2017119,
title = {Slack allocation algorithm for energy minimization in cluster systems},
journal = {Future Gener. Comput. Syst.},
volume = {74},
pages = {119-131},
year = {2017},
issn = {0167-739X},
doi = {https://doi.org/10.1016/j.future.2016.08.022},
url = {https://www.sciencedirect.com/science/article/pii/S0167739X16302941},
author = {Yikun Hu and Chubo Liu and Kenli Li and Xuedi Chen and Keqin Li}
}

@article{DING2020361,
title = {Q-learning based dynamic task scheduling for energy-efficient cloud computing},
journal = {Future Gener. Comput. Syst.},
volume = {108},
pages = {361-371},
year = {2020},
issn = {0167-739X},
doi = {https://doi.org/10.1016/j.future.2020.02.018},
url = {https://www.sciencedirect.com/science/article/pii/S0167739X19313858},
author = {Ding Ding and Xiaocong Fan and Yihuan Zhao and Kaixuan Kang and Qian Yin and Jing Zeng}
}

@article{LUO2026108351,
title = {Semi-asynchronous energy-efficient federated prototype learning for end-edge-cloud architectures},
journal = {Future Gener. Comput. Syst.},
volume = {179},
pages = {108351},
year = {2026},
issn = {0167-739X},
doi = {https://doi.org/10.1016/j.future.2025.108351},
url = {https://www.sciencedirect.com/science/article/pii/S0167739X25006454},
author = {Wendian Luo and Tong Yu and Shengxin Dai and Bing Guo and Xuesen Lin and Yanglin Pu}
}

@article{FUSCO2026108194,
title = {On-device training and pruning for energy saving and continuous learning in resource-constrained MCUs},
journal = {Future Gener. Comput. Syst.},
volume = {176},
pages = {108194},
year = {2026},
issn = {0167-739X},
doi = {https://doi.org/10.1016/j.future.2025.108194},
url = {https://www.sciencedirect.com/science/article/pii/S0167739X25004881},
author = {P. Fusco and G.P. Rimoli and A. Guerriero and F. Palmieri and M. Ficco}
}

@article{LAMBERT2026107968,
title = {Consolidation of virtual machines to reduce energy consumption of data centers by using ballooning, sharing and swapping mechanisms},
journal = {Future Gener. Comput. Syst.},
volume = {174},
pages = {107968},
year = {2026},
issn = {0167-739X},
doi = {https://doi.org/10.1016/j.future.2025.107968},
url = {https://www.sciencedirect.com/science/article/pii/S0167739X25002638},
author = {Simon Lambert and Eddy Caron and Laurent Lefevre and Rémi Grivel}
}

@article{GULDNER2024402,
title = {Development and evaluation of a reference measurement model for assessing the resource and energy efficiency of software products and components—Green Software Measurement Model (GSMM)},
journal = {Future Gener. Comput. Syst.},
volume = {155},
pages = {402-418},
year = {2024},
issn = {0167-739X},
doi = {https://doi.org/10.1016/j.future.2024.01.033},
url = {https://www.sciencedirect.com/science/article/pii/S0167739X24000384},
author = {Achim Guldner and Rabea Bender and Coral Calero and Giovanni S. Fernando and Markus Funke and Jens Gröger and Lorenz M. Hilty and Julian Hörnschemeyer and Geerd-Dietger Hoffmann and Dennis Junger and Tom Kennes and Sandro Kreten and Patricia Lago and Franziska Mai and Ivano Malavolta and Julien Murach and Kira Obergöker and Benno Schmidt and Arne Tarara and Joseph P. {De Veaugh-Geiss} and Sebastian Weber and Max Westing and Volker Wohlgemuth and Stefan Naumann}
}

@article{DARABI2026101307,
title = {CGI-SRAM: Memory cell with data-aware write bitline-free and inner readout features for energy-efficient bitwise logic-in-memory operations},
journal = {Sustain. Comput. Inf. Syst.},
volume = {50},
pages = {101307},
year = {2026},
issn = {2210-5379},
doi = {https://doi.org/10.1016/j.suscom.2026.101307},
url = {https://www.sciencedirect.com/science/article/pii/S221053792600017X},
author = {Abdolreza Darabi and Ebrahim Abiri}
}

@article{ANDRESLARRACOECHEA2026130046,
title = {Energy-aware software design: A hardware and behavioural consumption scoring and labelling algorithm},
journal = {Expert Systems with Applications},
volume = {299},
pages = {130046},
year = {2026},
issn = {0957-4174},
doi = {https://doi.org/10.1016/j.eswa.2025.130046},
url = {https://www.sciencedirect.com/science/article/pii/S0957417425036620},
author = {Jorge {Andrés Larracoechea} and Philippe Roose and Sergio Ilarri}
}

@article{MENCARONI2025146787,
title = {Towards net-zero manufacturing: Carbon-aware scheduling for GHG emissions reduction},
journal = {J. Clean. Product.},
volume = {529},
pages = {146787},
year = {2025},
issn = {0959-6526},
doi = {https://doi.org/10.1016/j.jclepro.2025.146787},
url = {https://www.sciencedirect.com/science/article/pii/S0959652625021377},
author = {Andrea Mencaroni and Pieter Leyman and Birger Raa and Stijn {De Vuyst} and Dieter Claeys}
}

@article{VERMA2025101115,
title = {Sustainable cost-energy aware load balancing in cloud environment using intelligent optimization},
journal = {Sustain. Comput. Inf. Syst.},
volume = {46},
pages = {101115},
year = {2025},
issn = {2210-5379},
doi = {https://doi.org/10.1016/j.suscom.2025.101115},
url = {https://www.sciencedirect.com/science/article/pii/S2210537925000356},
author = {Garima Verma}
}

@article{DACOSTA2025101106,
title = {Hardware and application aware performance, power and energy models for modern HPC servers with DVFS},
journal = {Sustain. Comput. Inf. Syst.},
volume = {46},
pages = {101106},
year = {2025},
issn = {2210-5379},
doi = {https://doi.org/10.1016/j.suscom.2025.101106},
url = {https://www.sciencedirect.com/science/article/pii/S2210537925000265},
author = {Georges {Da Costa}}
}

@article{LI201763,
title = {Energy cost minimization with job security guarantee in Internet data center},
journal = {Future Gener. Comput. Syst.},
volume = {73},
pages = {63-78},
year = {2017},
issn = {0167-739X},
doi = {https://doi.org/10.1016/j.future.2016.12.017},
url = {https://www.sciencedirect.com/science/article/pii/S0167739X16307634},
author = {Zhongjin Li and Jidong Ge and Chuanyi Li and Hongji Yang and Haiyang Hu and Bin Luo and Victor Chang}
}

@article{HOSSEINISHIRVANI2023100856,
title = {An energy-efficient topology-aware virtual machine placement in Cloud Datacenters: A multi-objective discrete JAYA optimization},
journal = {Sustain. Comput. Inf. Syst.},
volume = {38},
pages = {100856},
year = {2023},
issn = {2210-5379},
doi = {https://doi.org/10.1016/j.suscom.2023.100856},
url = {https://www.sciencedirect.com/science/article/pii/S2210537923000112},
author = {Mirsaeid {Hosseini Shirvani}}
}

@article{LIN2024100989,
title = {A systematic review of green-aware management techniques for sustainable data center},
journal = { Sustain. Comput. Inf. Syst.},
volume = {42},
pages = {100989},
year = {2024},
issn = {2210-5379},
doi = {https://doi.org/10.1016/j.suscom.2024.100989},
url = {https://www.sciencedirect.com/science/article/pii/S2210537924000349},
author = {Weiwei Lin and Jianpeng Lin and Zhiping Peng and Huikang Huang and Wenjun Lin and Keqin Li}
}

@article{KASSAB2021100590,
title = {Green power aware approaches for scheduling independent tasks on a multi-core machine},
journal = {Sustain. Comput. Inf. Syst.},
volume = {31},
pages = {100590},
year = {2021},
issn = {2210-5379},
doi = {https://doi.org/10.1016/j.suscom.2021.100590},
url = {https://www.sciencedirect.com/science/article/pii/S2210537921000792},
author = {Ayham Kassab and Jean-Marc Nicod and Laurent Philippe and Veronika Rehn-Sonigo}
}

@article{WANG2026129008,
title = {An energy-aware scheduling method for parallel tasks based on an adaptive differential evolution algorithm in a multi-cloud environment},
journal = {Expert Systems with Applications},
volume = {296},
pages = {129008},
year = {2026},
issn = {0957-4174},
doi = {https://doi.org/10.1016/j.eswa.2025.129008},
url = {https://www.sciencedirect.com/science/article/pii/S0957417425026259},
author = {Qin Wang and Yongsheng Hao and Yue Xu and Tinhuai Ma and Xin Zhang}
}

@article{AMINOROAYA2026101327,
title = {I\_MOAFT: An intelligent multi-layered framework for energy-efficient and low-latency task scheduling in fog-cloud computing},
journal = {Sustain. Comput. Inf. Syst.},
volume = {50},
pages = {101327},
year = {2026},
issn = {2210-5379},
doi = {https://doi.org/10.1016/j.suscom.2026.101327},
url = {https://www.sciencedirect.com/science/article/pii/S2210537926000375},
author = {Leila Aminoroaya and Reihaneh Khorsand and Mahdi Mosleh}
}

@article{BAYDOUN2025101258,
title = {HAPSO: An ACO-initialized, discretization-aware PSO for energy- and carbon-efficient VM consolidation in green cloud datacenters},
journal = {Sustain. Comput. Inf. Syst.},
volume = {48},
pages = {101258},
year = {2025},
issn = {2210-5379},
doi = {https://doi.org/10.1016/j.suscom.2025.101258},
url = {https://www.sciencedirect.com/science/article/pii/S2210537925001799},
author = {Ali M. Baydoun and Ahmed S. Zekri}
}

@article{KHODAYARSERESHT2023100888,
title = {Energy and carbon-aware initial VM placement in geographically distributed cloud data centers},
journal = {Sustain. Comput. Inf. Syst.},
volume = {39},
pages = {100888},
year = {2023},
issn = {2210-5379},
doi = {https://doi.org/10.1016/j.suscom.2023.100888},
url = {https://www.sciencedirect.com/science/article/pii/S2210537923000434},
author = {Ehsan Khodayarseresht and Alireza Shameli-Sendi and Quentin Fournier and Michel Dagenais}
}

@article{LASO2025101088,
title = {Energy consumption and workload prediction for edge nodes in the Computing Continuum},
journal = {Sustain. Comput. Inf. Syst.},
volume = {46},
pages = {101088},
year = {2025},
issn = {2210-5379},
doi = {https://doi.org/10.1016/j.suscom.2025.101088},
url = {https://www.sciencedirect.com/science/article/pii/S2210537925000083},
author = {Sergio Laso and Pablo Rodriguez and Juan Luis Herrera and Javier Berrocal and Juan M. Murillo}
}

@article{DONG2023100926,
title = {ETNAS: An energy consumption task-driven neural architecture search},
journal = {Sustain. Comput. Inf. Syst.},
volume = {40},
pages = {100926},
year = {2023},
issn = {2210-5379},
doi = {https://doi.org/10.1016/j.suscom.2023.100926},
url = {https://www.sciencedirect.com/science/article/pii/S2210537923000811},
author = {Dong Dong and Hongxu Jiang and Xuekai Wei and Yanfei Song and Xu Zhuang and Jason Wang}
}

@article{OLIVEIRAFILHO2025101172,
title = {Phoeni6: A systematic approach for evaluating the energy consumption of neural networks},
journal = {Sustain. Comput. Inf. Syst.},
volume = {47},
pages = {101172},
year = {2025},
issn = {2210-5379},
doi = {https://doi.org/10.1016/j.suscom.2025.101172},
url = {https://www.sciencedirect.com/science/article/pii/S2210537925000939},
author = {Antônio Oliveira-Filho and Wellington Silva-de-Souza and Carlos Alberto Valderrama Sakuyama and Samuel Xavier-de-Souza}
}

@article{SPILLO2026101286,
title = {Balancing carbon footprint and algorithm performance in recommender systems: A comprehensive benchmark},
journal = {Sustain. Comput. Inf. Syst.},
volume = {49},
pages = {101286},
year = {2026},
issn = {2210-5379},
doi = {https://doi.org/10.1016/j.suscom.2025.101286},
url = {https://www.sciencedirect.com/science/article/pii/S2210537925002070},
author = {Giuseppe Spillo and Allegra {De Filippo} and Cataldo Musto and Michela Milano and Giovanni Semeraro}
}

@article{AZIMI2020100412,
title = {PowerCoord: Power capping coordination for multi-CPU/GPU servers using reinforcement learning},
journal = {Sustain. Comput. Inf. Syst.},
volume = {28},
pages = {100412},
year = {2020},
issn = {2210-5379},
doi = {https://doi.org/10.1016/j.suscom.2020.100412},
url = {https://www.sciencedirect.com/science/article/pii/S2210537920301396},
author = {Reza Azimi and Chao Jing and Sherief Reda}
}

@article{SINGHAL2024100985,
title = {Rock-hyrax: An energy efficient job scheduling using cluster of resources in cloud computing environment},
journal = {Sustain. Comput. Inf. Syst.},
volume = {42},
pages = {100985},
year = {2024},
issn = {2210-5379},
doi = {https://doi.org/10.1016/j.suscom.2024.100985},
url = {https://www.sciencedirect.com/science/article/pii/S2210537924000301},
author = {Saurabh Singhal and Shabir Ali and Mohan Awasthy and Dhirendra Kumar Shukla and Rajesh Tiwari}
}

@article{JIANG2019311,
title = {Energy efficiency comparison of hypervisors},
journal = { Sustain. Comput. Inf. Syst.},
volume = {22},
pages = {311-321},
year = {2019},
issn = {2210-5379},
doi = {https://doi.org/10.1016/j.suscom.2017.09.005},
url = {https://www.sciencedirect.com/science/article/pii/S2210537917300963},
author = {Congfeng Jiang and Yumei Wang and Dongyang Ou and Youhuizi Li and Jilin Zhang and Jian Wan and Bing Luo and Weisong Shi}
}

@article{VERDECCHIA2022100767,
title = {The future of sustainable digital infrastructures: A landscape of solutions, adoption factors, impediments, open problems, and scenarios},
journal = { Sustain. Comput. Inf. Syst.},
volume = {35},
pages = {100767},
year = {2022},
issn = {2210-5379},
doi = {https://doi.org/10.1016/j.suscom.2022.100767},
url = {https://www.sciencedirect.com/science/article/pii/S2210537922000889},
author = {Roberto Verdecchia and Patricia Lago and Carol {de Vries}}
}

@article{BELGHACHI2026108452,
title = {Adaptive Federated Learning for energy-aware wireless edge networks in 6G environments},
journal = {Computer Communications},
volume = {250},
pages = {108452},
year = {2026},
issn = {0140-3664},
doi = {https://doi.org/10.1016/j.comcom.2026.108452},
url = {https://www.sciencedirect.com/science/article/pii/S0140366426000423},
author = {Mohammed Belghachi}
}




\end{document}